\documentclass[prx,aps,letterpaper,twocolumn,allowfontchageintitle=true,accepted=2019-02-01]{quantumarticle}

\usepackage{bbm}
\usepackage{overpic}
\usepackage{amsmath}
\usepackage{epsfig,psfrag}
\usepackage{dcolumn}
\usepackage{bm}
\usepackage{graphicx}
\usepackage{color}
\usepackage{version}
\usepackage[allcolors=quantumviolet,pdftitle={A Game of Surface Codes: Large-Scale Quantum Computing with Lattice Surgery},pdfauthor={Daniel Litinski}]{hyperref}
\usepackage{natbib}
\usepackage{multicol}

\usepackage{float}
\usepackage{wasysym}
\usepackage{algorithm2e}
\SetAlgoCaptionSeparator{}
\SetAlCapSkip{0.5em}
\NoCaptionOfAlgo

\usepackage{xcolor}
\usepackage{braket}

\begin{document}

\title{\centering \color{quantumviolet} A Game of Surface Codes: \newpage {\centering Large-Scale Quantum Computing with Lattice Surgery} \newpage}
\author{\vspace{-5ex}}
\affiliation{Daniel Litinski @ Dahlem Center for Complex Quantum Systems, Freie Universit\"at Berlin, Arnimallee 14, 14195 Berlin, Germany}

\date{\vspace{-5ex}}

{\centering \maketitle}

\begin{abstract}

Given a quantum gate circuit, how does one execute it in a fault-tolerant architecture with as little overhead as possible? In this paper, we discuss strategies for surface-code quantum computing on small, intermediate and large scales. They are strategies for space-time trade-offs, going from slow computations using few qubits to fast computations using many qubits. Our schemes are based on surface-code patches, which not only feature a low space cost compared to other surface-code schemes, but are also conceptually simple~--~simple enough that they can be described as a tile-based game with a small set of rules. Therefore, no knowledge of quantum error correction is necessary to understand the schemes in this paper, but only the concepts of qubits and measurements.

\end{abstract}


The field of quantum computing is fuelled by the promise of fast solutions to classically intractable problems, such as simulating large quantum systems or factoring large numbers. Already ${\sim}100$ qubits can be used to solve useful problems that are out of reach for classical computers~\cite{Reiher2017,Babbush2018}. Despite the exponential speed-up, the actual time required to solve these problems is orders of magnitude above the coherence times of any physical qubit. In order to store and manipulate quantum information on large time scales, it is necessary to actively correct errors by combining many physical qubits into logical qubits using a quantum error-correcting code~\cite{Preskill1998,TerhalRMP,Campbell2016}. Of particular interest are codes that are compatible with the locality constraints of realistic devices such as superconducting qubits, which are limited to operations that are local in two dimensions. The most prominent such code is the surface code~\cite{Kitaev2003,Fowler2012}.

Working with logical qubits introduces additional overhead to the computation. Not only is the space cost drastically increased as physical qubits are replaced by logical qubits, but also the time cost increases due to the restricted set of accessible logical operations. Surface codes, in particular, are limited to a set of 2D-local operations, which means that arbitrary gates in a quantum circuit may require several time steps instead of just one. To keep the cost of surface-code quantum computing low, it is important to find schemes that translate quantum circuits into surface-code layouts with a low space-time overhead. This is also necessary to benchmark how well quantum algorithms perform in a surface-code architecture.  

There exist several encoding schemes for surface codes, among others, defect-based~\cite{Fowler2012}, twist-based~\cite{Bombin2010} and patch-based~\cite{Horsman2012} encodings. In this work, we focus on the latter. Surface-code patches have a low space overhead compared to other schemes, and offer low-overhead Clifford gates~\cite{Brown2017,Litinski2017b}. In addition, they are conceptually less difficult to understand, as they do not directly involve braiding of topological defects. Designing computational schemes with surface-code patches only requires the concepts of qubits and measurements. To this end, we describe the operations of surface-code patches as a tile-based game. This is helpful to design protocols and determine their space-time cost. The exact correspondence between this game and surface-code patches is specified in Appendix~\ref{app:surfacecode}, but it is not crucial for understanding this paper. Readers who are interested in the detailed surface-code operations may read Appendix~\ref{app:surfacecode} in parallel to the following section.

\begin{figure}[b!]
\centering
\def\svgwidth{\linewidth}
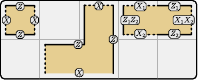
\caption{Examples of one-qubit (a/b) and two-qubit (c) patches in a $5\times 2$ grid of tiles.}
\label{fig:patches}
\end{figure}

\textbf{Surface codes as a game.}
The game is played on a board partitioned into a number of tiles. An example of a $5 \times 2$ grid of tiles is shown in Fig.~\ref{fig:patches}. The tiles can be used to host \textit{patches}, which are representations of qubits. We denote the Pauli operators of each qubit as $X$, $Y$ and $Z$. Patches have dashed and solid edges representing Pauli operators. We consider two types of patches: one-qubit and two-qubit patches. One-qubit patches represent one qubit and consist of two dashed and two solid edges. Each of the two dashed (solid) edges represent the qubit's $X$ ($Z$) operator. While the square patch in Fig.~\ref{fig:patches}a only occupies one tile, a one-qubit patch can also be shaped to, e.g., occupy three tiles (b). A two-qubit patch (c) consists of six edges and represents two qubits. The first qubit's Pauli operators $X_1$ and $Z_1$ are represented by the two top edges, while the second qubit's operators $X_2$ and $Z_2$ are found in the two bottom edges. The remaining two edges represent the operators $Z_1 \cdot Z_2$ and $X_1 \cdot X_2$.

 In the following, we specify the operations that can be used to manipulate the qubits represented by patches. Some of these operations take one time step to complete (denoted by 1\clock), whereas others can be performed instantly, requiring 0\clock. The goal is to implement quantum algorithms using as few tiles and time steps as possible.
There are three types of operations: qubit initialization, qubit measurement and patch deformation.

\begin{figure}[t]
\centering
\def\svgwidth{\linewidth}
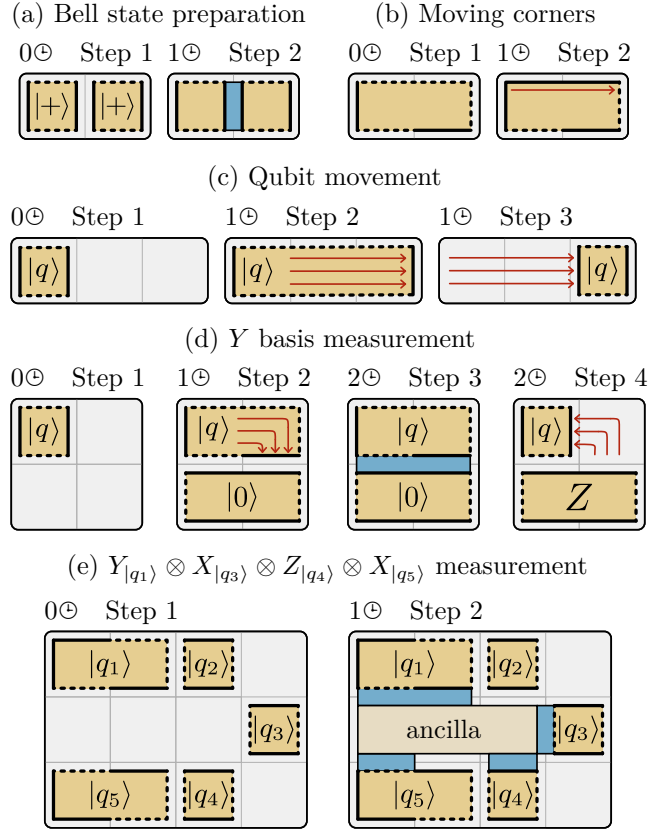
\caption{Examples of short protocols. (a)~Preparation of a two-qubit Bell state in 1\clock. (b) Moving corners of a four-corner patch to change its shape in 1\clock. (c) Moving a square-patch qubit over long distances in 1\clock. (d) Measurement of a square-patch qubit in the $Y$ basis using an ancilla qubit and 2\clock.  (e) A multi-qubit  $Y_{\ket{q_1}}\otimes X_{\ket{q_3}} \otimes Z_{\ket{q_4}} \otimes X_{\ket{q_5}}$ measurement in 1\clock. }
\label{fig:operations}
\end{figure}

\begin{enumerate}
 \item[\textit{I.}] \textit{Qubit initialization:}
 \begin{itemize}
 	\item[--] One-qubit patches can be initialized in the $X$ and $Z$ eigenstates $\ket{+}$ and $\ket{0}$. (Cost: 0\clock)
 	\item[--] Two-qubit patches can be initialized in the states $\ket{+} \otimes \ket{+}$ and $\ket{0} \otimes \ket{0}$. (Cost: 0\clock)
 	\item[--] One-qubit patches can be initialized in an arbitrary state. Unless  this state is $\ket{+}$ or $\ket{0}$, an undetected random Pauli error may spoil the qubit with probability $p$. (Cost: 0\clock)
 \end{itemize}
  \item[\textit{II.}] \textit{Qubit measurement:}
 \begin{itemize}
 	\item[--] Single-patch measurements: The qubits represented by patches can be measured in the $X$ or $Z$ basis. For two-qubit patches, the two qubits must be measured simultaneously and in the same basis. This measurement removes the patch from the board, freeing up previously occupied tiles. (Cost: 0\clock)
 	\item[--] Two-patch measurements: If edges of two different patches are positioned in adjacent tiles, the product of the operators of the two edges can be measured. For example, the product $Z \otimes Z$ between two neighboring square patches can be measured, as highlighted in step 2 of Fig.~\ref{fig:operations}a by the blue rectangle. If the edge of one patch is adjacent to multiple edges of the other patch, the product of all involved Pauli operators can be measured. For instance, if qubit A's $Z$ edge is adjacent to both qubit B's $X$ edge and $Z$ edge, the operator $Z_{\rm A} \otimes Y_{\rm B}$ can be measured (see step 3 of Fig.~\ref{fig:operations}d), since $Y=iXZ$. (Cost: 1\clock)
 	\item[--] Multi-patch measurements: An arbitrarily-shaped ancilla patch can be initialized. The product of any number of operators adjacent to the ancilla patch can be measured. The ancilla patch is discarded after the measurement. The example of a $Y_{\ket{q_1}}\otimes X_{\ket{q_3}} \otimes Z_{\ket{q_4}} \otimes X_{\ket{q_5}}$ measurement is shown in Fig.~\ref{fig:operations}e. (Cost: 1\clock)
 \end{itemize}
\pagebreak
  \item[\textit{III.}] \textit{Patch deformation:}
 \begin{itemize}
 	\item[--] Edges of a patch can be moved to deform the patch. If the edge is moved onto a free tile to increase the size of the patch, this takes 1\clock ~to complete. If the edge is moved inside the patch to make the patch smaller, the action can be performed instantly.
 	\item[--] Corners of a patch can be moved along the patch boundary to change its shape, as shown in Fig.~\ref{fig:operations}b. (Cost: 1\clock)
 \end{itemize}
\end{enumerate}

To illustrate these operations, we go through three short example protocols in Fig.~\ref{fig:operations}a/c/d. The first example (a) is the preparation of a Bell pair. Two square patches are initialized in the $\ket{+}$ state. Next, the operator $Z\otimes Z$ is measured. Before the measurement, the qubits are in the state $\ket{+} \otimes \ket{+} = (\ket{00} + \ket{01} + \ket{10} + \ket{11})/2$. If the measurement outcome is $+1$, the qubits end up in the state $(\ket{00} + \ket{11})/\sqrt{2}$. For the outcome $-1$, the state is $(\ket{01} + \ket{10})/\sqrt{2}$. In both cases, the two qubits are in a maximally entangled Bell state. This protocol takes 1\clock \ to complete. The second example (c) is the movement of a square patch into a different tile. For this, the square patch is enlarged by patch deformation, which takes 1\clock, and then made smaller again at no time cost.
The third example (d) is the measurement of a square patch in the Y basis. For this, the patch is deformed such that the $X$ and $Z$ edge are on the same side of the patch. An ancillary patch is initialized in the $\ket{0}$ state and the operator $Z \otimes Y$ between the ancilla and the qubit is measured. The ancilla is discarded by measuring it in the $Z$ basis.

\textbf{Translation to surface codes.} As described in Appendix~\ref{app:surfacecode}, protocols designed within this framework can be straightforwardly translated into surface-code operations. Essentially, patches correspond to surface-code patches with dashed and solid edges as rough and smooth boundaries. Thus, for surface codes with a code distance $d$, each tile corresponds to $d^2$ physical data qubits. Each time step roughly corresponds to $d$ code cycles, i.e., measuring all surface-code check operators $d$ times. We associate a time step with all surface-code operations which have a time cost that scales with $d$, but no time step with operations whose time cost is independent of the code distance, but may still be nonzero. For this reason, the correspondence between 1\clock \ and $d$ code cycles is not exact. 

Two-patch and multi-patch measurements correspond to (twist-based) lattice surgery~\cite{Horsman2012,Litinski2017b} and multi-qubit lattice surgery~\cite{Fowler2018}, respectively, which both  require $d$ code cycles to account for measurement errors. Qubit initialization has no time cost, since, in the case of $X$ and $Z$ eigenstates, it can be done simultaneously with the subsequent lattice surgery~\cite{Horsman2012,Landahl2014}. For arbitrary states, initialization corresponds to state injection~\cite{Landahl2014,Li2015}. Its time cost does not scale with $d$. Similarly, single-qubit measurements in the $X$ or $Z$ basis correspond to the simultaneous measurement of all physical data qubits in the corresponding basis and some classical error correction, which does not scale with $d$ either. Patch deformation is code deformation, which requires $d$ code cycles, unless the patch becomes smaller in the process, in which case it corresponds to single-qubit measurements. Note that not all surface-code operations are covered by this framework. An extended set of rules is discussed in Appendix \ref{app:extendedrules}.

\begin{figure*}[t]
\centering
\def\svgwidth{\linewidth}
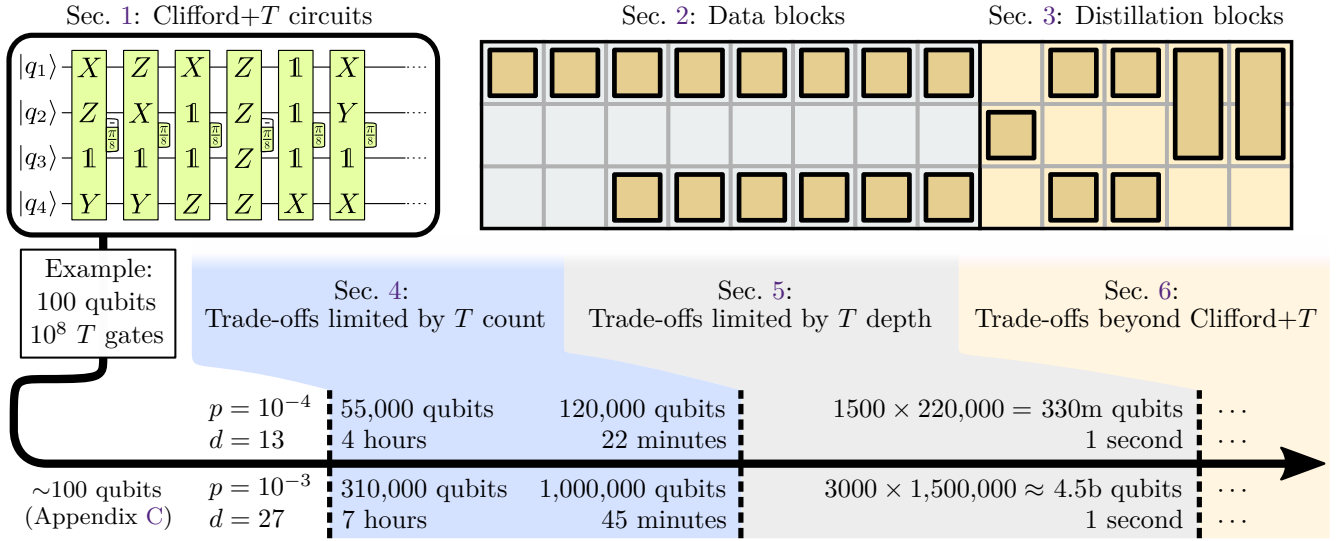
\caption{Overview of the content of this paper. To illustrate the space-time trade-offs discussed in this work, we show the number of physical qubits and the computational time required for a circuit of $10^8$ $T$ gates distributed over $10^6$ $T$ layers. We consider physical error rates of $p=10^{-4}$ and $p=10^{-3}$, for which we need code distances $d=13$ and $d=27$, respectively. We assume that each code cycle takes $1~\mu \mathrm{s}$.}
\label{fig:overview}
\end{figure*}

In essence, the framework can be used to estimate the space-time cost of a computation. The leading-order term of the space-time cost~--~the term that scales with $d^3$~--~of a protocol that uses $s$ tiles for $t$ time steps is $st\cdot d^3$ in terms of (physical data qubits)$\cdot$(code cycles). The space cost is $s \cdot d^2$ physical data qubits. Determining the exact time cost requires special care. In some protocols, the subleading contributions due to state injection and classical processing may need to be taken into account. For these protocols, we will show how they can be adapted to prevent such contributions from increasing the time cost beyond $t \cdot d$ code cycles.

\section*{Overview}

Having established the rules of the game and the correspondence of our framework to surface-code operations, our goal is to implement arbitrary quantum computations. In this work, we discuss strategies to tackle the following problem: Given a quantum circuit, how does one execute it as fast as possible on a surface-code-based quantum computer of a certain size? This is an optimization problem that was shown to be NP-hard~\cite{Herr2017b}, so the focus is on heuristics rather than a general solution. The content of this paper is outlined in Fig.~\ref{fig:overview}.

The input to our problem is an arbitrary gate circuit corresponding to the computation. We refer to the qubits that this circuit acts on as \textit{data qubits}. As we review in Sec.~\ref{sec:circuits}, the natural universal gate set for surface codes is Clifford+$T$, where Clifford gates are cheap and $T$ gates are expensive. In fact, Clifford gates can be treated entirely classically, and $T$ gates require the consumption of a magic state $\ket{0} + e^{i\pi/4} \ket{1}$. Only faulty (\textit{undistilled}) magic states can be prepared in our framework. To generate higher-fidelity magic states for large-scale quantum computation, a lengthy protocol called magic state distillation~\cite{Bravyi2005} is used.

It is therefore natural to partition a quantum computer into a block of tiles that is used to distill magic states (a distillation block) and a block of tiles that hosts the data qubits (a data block) and consumes magic states. The speed of a quantum computer is governed by how fast magic states can be distilled, and how fast they can be consumed by the data block.

In Sec.~\ref{sec:datablock}, we discuss how to design data blocks. In particular, we show three designs: compact, intermediate and fast blocks. The compact block uses $1.5n+3$ tiles to store $n$ qubits, but takes up to 9\clock \ to consume a magic state. Intermediate blocks use $2n+4$ tiles and require up to 5\clock \ per magic state. Finally, the fast block uses $2n+\sqrt{8n}+1$ tiles, but requires only 1\clock \ to consume a magic state. The compact block is an option for early quantum computers with few qubits, where the generation of a single magic state takes longer than 9\clock. The fast block has a better space-time overhead, which makes it more favorable on larger scales.

Data blocks need to be combined with distillation blocks for universal quantum computing. In Sec.~\ref{sec:distillation}, we discuss designs of distillation blocks. Since magic state distillation is the main operation of a surface-code-based quantum computer, it is important to minimize its space-time cost. We discuss distillation protocols based on error-correcting codes with transversal $T$ gates, such as punctured Reed-Muller codes~\cite{Bravyi2005,Haah2018} and block codes~\cite{Bravyi2012,Jones2013a,Fowler2013}. In comparison to braiding-based implementations of distillation protocols, we reduce the space-time cost by up to 90\%.

A data block combined with a distillation block constitutes a quantum computer in which $T$ gates are performed one after the other. At this stage, the quantum computer can be sped up by increasing the number of distillation blocks, effectively decreasing the time it takes to distill a single magic state, as we discuss in Sec.~\ref{sec:tcount}. In order to illustrate the resulting space-time trade-off, we consider the example of a 100-qubit computation with $10^8$ $T$ gates, which can already be used to solve classically intractable problems~\cite{Babbush2018}. Assuming an error rate of $p=10^{-4}$ and a code-cycle time of $1~\mu\mathrm{s}$, a compact data block together with a distillation block can finish the computation in 4 hours using 55,000 physical qubits.\footnote{We will assume that the total number of physical qubits is twice the number of physical data qubits. This is consistent with superconducting qubit platforms, where the use of measurement ancillas doubles the qubit count. If a platform does not require the use of ancilla qubits, the total qubit count is reduced by 50\% compared to the numbers reported in this paper.} Adding 10 more distillation blocks increases the qubit count to 120,000 and decreases the computational time to 22 minutes, using 1\clock \ per $T$ gate. 

For further space-time trade-offs in Sec.~\ref{sec:timeoptimal}, we exploit that the $T$ gates of a circuit are arranged in layers of gates that can be executed simultaneously. This enables linear space-time trade-offs down to the execution of one $T$ layer per qubit measurement time, effectively implementing Fowler's time-optimal scheme~\cite{Fowler2012a}. If the $10^8$~$T$~gates are distributed over $10^6$ layers, and measurements (and classical processing) can be performed in 1~$\mu\mathrm{s}$, up to 1500 \textit{units} of 220,000 qubits can be run in parallel, where each unit is responsible for the execution of one $T$ layer. This way, the computational time can be brought down to 1 second using 330 million qubits. While this is a large number, the units do not necessarily need to be part of the same quantum computer, but can be distributed over up to 1500 quantum computers with 220,000 qubits each, and with the ability to share Bell pairs between neighboring computers.

\begin{figure*}[t!]
\centering
\def\svgwidth{\linewidth}
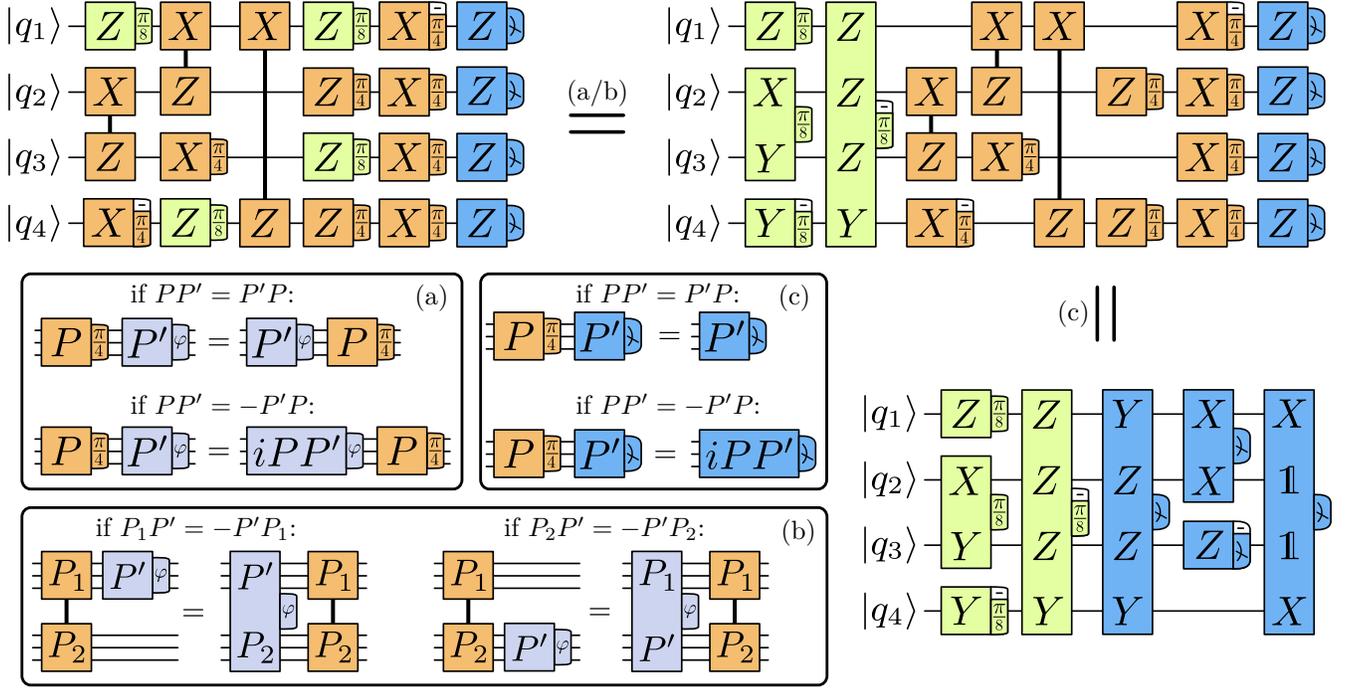
\caption{A generic circuit consists of $\pi/4$ rotations (orange), $\pi/8$ rotations (green) and measurements (blue). The Pauli product in each box specifies the axis of rotation or the basis of measurement. If the Pauli operator is $-P$ instead of $P$, a minus sign is found in the corner of the box, such that, e.g., $Z_{-\pi/4}$ corresponds to an $S^\dagger$ gate. Using the commutation rules in (a/b), all Clifford gates can be moved to the end of the circuit. Using (c), the Clifford gates can be absorbed by the final measurements.}
\label{fig:circuitsimplify}
\end{figure*}

In Sec.~\ref{sec:beyond}, we discuss further space-time trade-offs that are beyond the parallelization of Clifford+$T$ circuits. In particular, we discuss the use of Clifford+$\varphi$ circuits, i.e., circuits containing arbitrary-angle rotations beyond $T$ gates. These require the use of additional resources, but can speed up the computation. We also discuss the possibility of hardware-based trade-offs by using higher code distances, but in turn shorter measurements with a decreased measurement fidelity. Ultimately, the speed of a quantum computer is limited by classical processing, which can only be improved upon by faster classical computing.

Finally, we note that while the number of qubits
required for useful quantum computing is orders of magnitude above what is currently available, a proof-of-principle two-qubit device demonstrating all necessary operations using undistilled magic states can be built with 48 physical data qubits, see Appendix~\ref{app:proofofprinciple}.

\section{Clifford+$T$ quantum circuits}
\label{sec:circuits}

\begin{figure}[b!]
\centering
\def\svgwidth{0.9\linewidth}
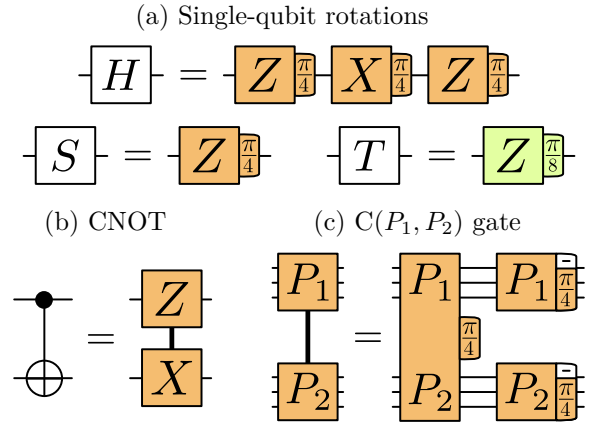
\caption{Clifford+$T$ gates in terms of Pauli rotations. (a)~Single-qubit Clifford gates are $\pi/4$ rotations, and the $T$ gate is a $\pi/8$ rotation. (b/c) $P_1$-controlled-$P_2$ gates are Clifford gates, where $\mathrm{C}(Z,X)$ is the CNOT gate.}
\label{fig:gateset}
\end{figure}

Our goal is to implement full quantum algorithms with surface codes. The input to our problem is the algorithm's quantum circuit. The universal gate set Clifford+$T$ is well-suited for surface codes, since it separates easy operations from difficult ones. Often, this set is generated using the Hadamard gate $H$, phase gate $S$, controlled-NOT (CNOT) gate, and the $T$ gate. Instead, we choose to write our circuits using Pauli product rotations $P_\varphi$ (see Fig.~\ref{fig:gateset}), because it simplifies circuit manipulations. Here, $P_\varphi = \exp(-iP\varphi)$, where $P$ is a Pauli product operator (such as $Z$, $Y\otimes X$, or $X \otimes \mathbbm{1} \otimes X$) and $\varphi$ is an angle. In this sense, $S=Z_{\pi/4}$, $T=Z_{\pi/8}$, and $H=Z_{\pi/4}\cdot X_{\pi/4} \cdot Z_{\pi/4}$. The CNOT gate can also be written in terms of Pauli product rotations as $\mathrm{CNOT}=(Z\otimes X)_{\pi/4} \cdot (\mathbbm{1}\otimes X)_{-\pi/4} \cdot (Z\otimes \mathbbm{1})_{-\pi/4}$. In fact, we can more generally define $P_1$-controlled-$P_2$ gates as $\mathrm{C}(P_1,P_2) = (P_1\otimes P_2)_{\pi/4} \cdot (\mathbbm{1}\otimes P_2)_{-\pi/4} \cdot (P_1\otimes \mathbbm{1})_{-\pi/4}$. The CNOT gate is the specific case of $\mathrm{C}(Z,X)$.

\begin{figure*}[t!]
\centering
\def\svgwidth{\linewidth}
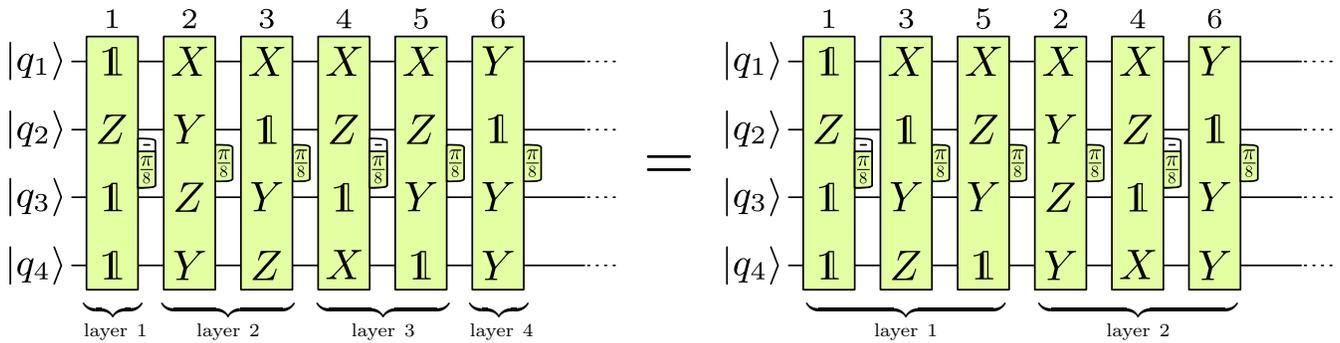
\caption{Clifford+$T$ circuits can be written as a number of consecutive $\pi/8$ rotations. These gates are grouped into layers of mutually commuting rotations. A simple greedy algorithm can be used to reduce the number of layers, i.e., the $T$ depth.}
\label{fig:tlayers}
\end{figure*}

\begin{algorithm}[b!]
 \Repeat{\rm the partitioning no longer changes}{
 \For{\rm each layer $i$}{
 \For{\rm each rotation $j$ in layer $i+1$}{
    \If{\rm (rotation $j$ commutes with all rotations in layer $i$)}{
Move rotation $j$ from  layer $i+1$ to layer $i$\;
   }}
 }
 }
 \caption{Algorithm to reduce the $T$ count and $T$ depth.}
\end{algorithm}

\textbf{Getting rid of Clifford gates.} Clifford gates are considered to be easy, because, by definition, they map Pauli operators onto other Pauli operators~\cite{Gottesman1999}. This can be used to simplify the input circuit. A generic circuit is shown in Fig.~\ref{fig:circuitsimplify}, consisting of Clifford gates, $Z_{\pi/8}$ rotations and $Z$ measurements. If all Clifford gates are commuted to the end of the circuit, the $Z_{\pi/8}$ rotations become Pauli product rotations. The rules for moving $P_{\pi/4}$ rotations past $P'_\varphi$ gates are shown in Fig.~\ref{fig:circuitsimplify}a: If $P$ and $P'$ commute, $P_{\pi/4}$ can simply be moved past $P'_\varphi$. If they anticommute, $P'_\varphi$ turns into $(iPP')_\varphi$ when $P_{\pi/4}$ is moved to the right. Since $\mathrm{C}(P_1,P_2)$ gates consist of $\pi/4$ rotations, similar rules can be derived as shown in Fig.~\ref{fig:circuitsimplify}b: If $P'$ anticommutes with $P_1$, $P'_\varphi$ turns into $(P'P_2)_\varphi$ after commutation. If $P'$ anticommutes with $P_2$, $P'_\varphi$ turns into $(P'P_1)_\varphi$. If $P'$ anticommutes with both $P_1$ and $P_2$, $P'_\varphi$ turns into $(P'P_1P_2)_\varphi$.

After moving the Clifford gates to the right, the resulting circuit consists of three parts: a set of $\pi/8$ rotations, a set of $\pi/4$ rotations, and $Z$ measurements. Because Clifford gates map Pauli operators onto other Pauli operators, the Clifford gates can be absorbed by the final measurements, turning $Z$ measurements into Pauli product measurements. The commutation rules of this final step are shown in Fig.~\ref{fig:circuitsimplify}c and are similar to the commutation of Clifford gates past rotations.

\textbf{$T$ count and $T$ depth.} Thus, every $n$-qubit circuit can be written as a number of consecutive $\pi/8$ rotations and $n$ final Pauli product measurements, as shown in Fig.~\ref{fig:tlayers}. We refer to the number of $\pi/8$ rotations as the $T$ count. An important part of circuit optimization is the minimization of the $T$~count, for which there exist various approaches~\cite{Kliuchnikov2012,Kliuchnikov2013,Gosset2013,Heyfron2017}. The $\pi/8$ rotations of a circuit can be grouped into layers. All $\pi/8$ rotations that are part of a layer need to mutually commute. The number of $\pi/8$ layers of a circuit is strictly speaking not the same quantity as the $T$ depth, but we will still refer to it as the $T$ depth and to $\pi/8$ layers as $T$ layers. Note that, in the usual definition, only up to $n$ $T$ gates can be part of a layer, whereas in our case, there is no limit.

When partitioning $\pi/8$ rotations into layers, the naive approach often yields more layers than are necessary. For instance, a naive partitioning of the first 6 $T$ gates of Fig.~\ref{fig:tlayers} yields 4 layers. A few commutations can bring the number down to 2 layers. There are a number of algorithms for the optimization of the $T$ depth~\cite{Amy2013,Selinger2013,Amy2014}. Here, we use the simple greedy algorithm shown below to reduce the number of layers.

Note that when a reordering puts two equal $\pi/8$ rotations into the same layer, they can be combined into a $\pi/4$ rotation that is commuted to the end of the circuit, thereby decreasing the $T$ count. As we discuss in Sec.~\ref{sec:beyond}, this kind of algorithm can not only be used with $\pi/8$ rotations, but, in principle, with arbitrary Pauli product rotations. The reduction of the circuit depth in terms of non-$\pi/8$ rotations can be useful when going beyond Clifford+$T$ circuits.

\subsection{Pauli product measurements}

\begin{figure}[b!]
\centering
\def\svgwidth{0.93\linewidth}
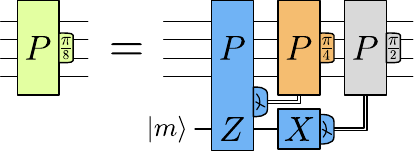
\caption{Circuit to perform a $\pi/8$ rotation by consuming a magic state.}
\label{fig:magicstateuse}
\end{figure}

When implementing circuits like Fig.~\ref{fig:tlayers} with surface codes, one obstacle is that $\pi/8$ rotations are not directly part of the set of available operations. Instead, one uses magic states~\cite{Bravyi2005} as a resource. These states are $\pi/8$-rotated Pauli eigenstates $\ket{m} = \ket{0} + e^{i\pi/4}\ket{1}$. They can be consumed in order to perform $P_{\pi/8}$ rotations. The corresponding circuit~\cite{Litinski2018} is shown in Fig.~\ref{fig:magicstateuse}. A $P_{\pi/8}$ rotation corresponds to a $P \otimes Z$ measurement involving the magic state. If the measurement outcome is $P \otimes Z = -1$, then a corrective $P_{\pi/4}$ operation is necessary. Since this is a Clifford gate, it can be simply commuted to the end of the circuit, changing the axes of the subsequent $\pi/8$ rotations. Finally, in order to discard the magic state, it is disentangled from the rest of the system by an $X$ measurement. Here, an outcome $X=-1$ prompts a $P_{\pi/2}$ correction. $\pi/2$ rotations correspond to Pauli operators, i.e., $P_{\pi/2} = P$. The Pauli correction can also be commuted to the end of the circuit. When $P_{\pi/2}$ is moved past a $P'$ rotation or measurement, it changes the axis of rotation or measurement basis to $-P'$, if $P$ and $P'$ anticommute.

\begin{figure}[t!]
\centering
\def\svgwidth{\linewidth}
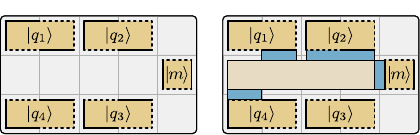
\caption{Example of a $Z_{\ket{q_1}}\otimes Y_{\ket{q_2}} \otimes X_{\ket{q_4}} \otimes Z_{\ket{m}}$ measurement to implement a $(Z\otimes Y \otimes \mathbbm{1} \otimes X)_{\pi/8}$ gate.}
\label{fig:pauliprodmeas}
\end{figure}

In essence, if magic states are available, the only operations required for universal quantum computing are Pauli product measurements. In our framework, such operations can be performed in 1\clock \ via multi-patch measurements, corresponding to multi-qubit lattice surgery. An example is shown in Fig.~\ref{fig:pauliprodmeas}, where a $(Z\otimes Y \otimes \mathbbm{1} \otimes X)_{\pi/8}$ rotation on four qubits $\ket{q_1}$-$\ket{q_4}$ stored in four two-tile one-qubit patches is performed. Using the circuit identity in Fig.~\ref{fig:magicstateuse}, this is done by measuring $Z_{\ket{q_1}}\otimes Y_{\ket{q_2}} \otimes X_{\ket{q_4}} \otimes Z_{\ket{m}}$ between the four qubits and a magic state. 

\textbf{Summary.} Clifford+$T$ circuits can be written in terms of $\pi/8$ rotations, $\pi/4$ rotations and measurements. To convert input circuits into a standard form, $\pi/4$ rotations can be commuted to the end of the circuit and absorbed by the final measurements. Thus, any quantum computation can be written as a sequence of $\pi/8$ rotations grouped into layers of mutually commuting rotations. The number of rotations is the $T$ count and the number of layers is the $T$ depth. Each rotation can be performed by consuming a magic state via a Pauli product measurement. These measurements can be implemented in our framework in 1\clock.

\section{Data blocks}
\label{sec:datablock}

Since Clifford+$T$ circuits are a sequence of $\pi/8$ rotations, each requiring the consumption of a magic state, it is natural to partition a quantum computer into a set of tiles that are used for magic state distillation (distillation blocks) and a set of tiles that host data qubits and consume magic states via Pauli product measurements (data blocks). In this section, we discuss designs for the latter. In principle, the structure shown in Fig.~\ref{fig:pauliprodmeas} is a data block, where each qubit is stored in a two-tile patch and magic states can be consumed every 1\clock. However, this sort of design uses $3n$ tiles to host $n$ data qubits, which is a relatively large space overhead.

\begin{figure}[t!]
\centering
\def\svgwidth{0.85\linewidth}
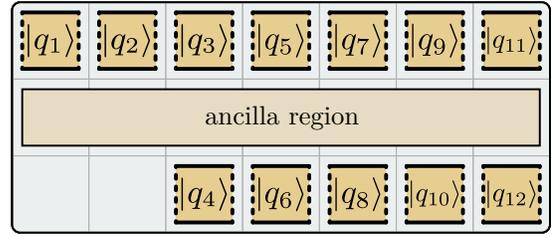
\caption{A compact block stores $n$ data qubits in $1.5n+3$ tiles. The consumption of a magic state can take up to 9\clock.}
\label{fig:compactblock1}
\end{figure}

\begin{figure}[b!]
\centering
\def\svgwidth{0.85\linewidth}
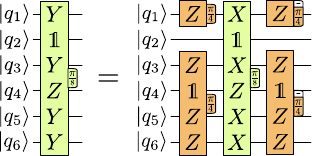
\caption{For compact blocks, the worst-case scenario are Pauli product measurements involving an even number of $Y$ operators, e.g., the measurement required for a $(Y \otimes \mathbbm{1} \otimes Y \otimes Z \otimes Y \otimes Y)_{\pi/8}$ gate. Such measurements require two explicit $\pi/4$ rotations (left), and two $\pi/4$ rotations that are commuted to the end of the circuit (right).}
\label{fig:compactblock2}
\end{figure}

\begin{figure*}[t!]
\centering
\def\svgwidth{0.98\linewidth}
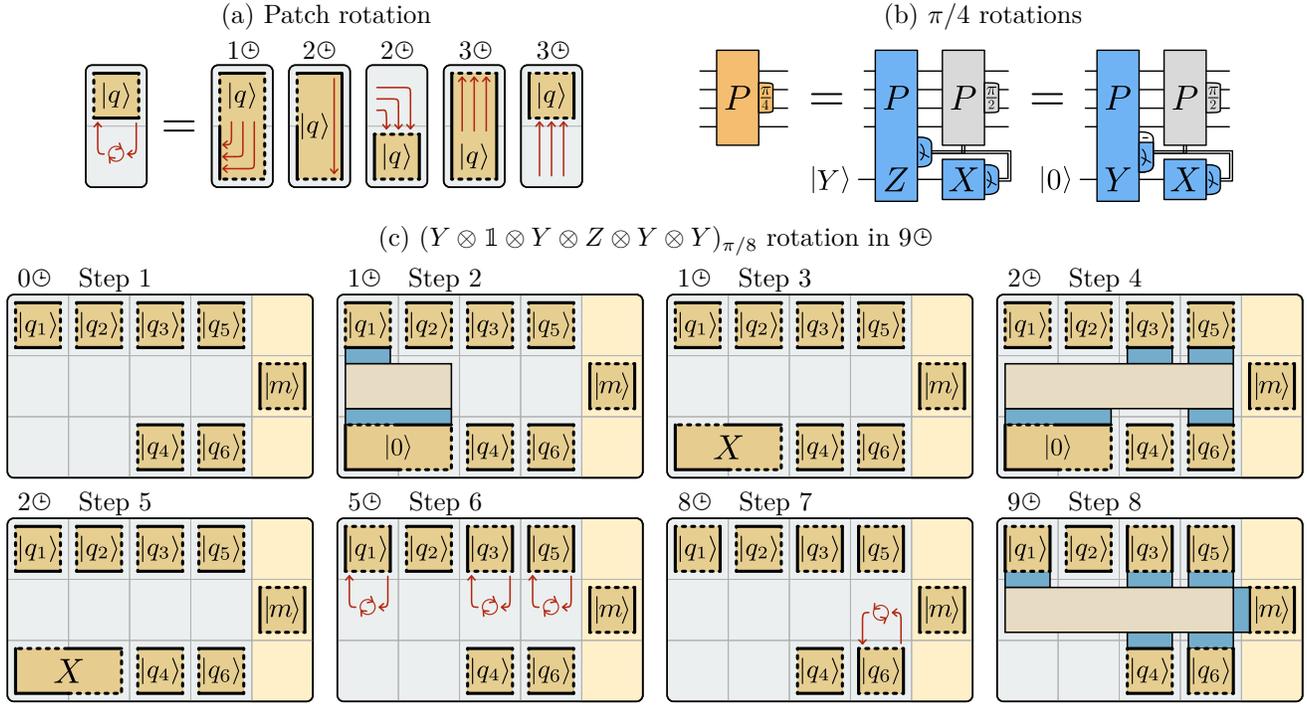
\caption{(a) Patches can be rotated in 3\clock \ to change whether the $X$ or $Z$ operator is adjacent to the compact block's ancilla region.  (b) A $P_{\pi/4}$ gate can be performed explicitly via a $P \otimes Y$ measurement with a $\ket{0}$ ancilla qubit. (c) Six-step protocol to perform the rotation of Fig.~\ref{fig:compactblock2} in a compact block. The magic state is consumed in 9\clock, where steps 2-5 are the two $\pi/4$ rotations in Fig.~\ref{fig:compactblock2}, steps 6 and 7 are patch rotations, and step 8 is the Pauli product measurement consuming the magic state.}
\label{fig:compactblock3}
\end{figure*}

\subsection{Compact block}

The first design that we discuss uses only $1.5n+3$ tiles. This compact block is shown in Fig.~\ref{fig:compactblock1}, where each data qubit is stored in a square patch. This lowers the space cost, but restricts the operators that are accessible by Pauli product measurements, as only the $Z$ operator is free to be measured. Using 3\clock, patches may also be rotated (see Fig.~\ref{fig:compactblock3}a), such that the $X$ operator becomes accessible instead of the $Z$ operator. The problematic operators are $Y$ operators, which are the reason why the consumption of a magic state can take up to 9\clock.

The worst-case scenario is a $\pi/8$ rotation involving an even number of $Y$ operators, such as the one shown in Fig.~\ref{fig:compactblock2}. One possibility to replace $Y$ operators by $X$ or $Z$ operators is via $\pi/4$ rotations, since \linebreak $Y_{\pi/4} = Z_{\pi_4} X_{\pi/4} Z_{-\pi/4}$. Rotations with an even number of $Y$'s require two $\pi/4$ rotations, while an odd number of $Y$'s can be handled by one rotation. Only the left two $\pi/4$ rotations in Fig.~\ref{fig:compactblock2} need to be performed explicitly. The right two rotations can be commuted to the end of the circuit, changing the subsequent $\pi/8$ rotations. Similarly to a $\pi/8$ rotation, a $P_{\pi/4}$ rotation can be executed using a resource state $\ket{Y} = \ket{0} + i\ket{1}$, as shown in Fig.~\ref{fig:compactblock3}b. However, even though this state is a Pauli eigenstate, it cannot be readily prepared in our framework. Instead, we use a $\ket{0}$ state and $Y$ measurements, such that a $P_{\pi/4}$ rotation is performed by a $P\otimes Y$ measurement between the qubits and the $\ket{0}$ state. Afterwards, the $\ket{0}$ state is measured in $X$. If the $-P \otimes Y$ and $X$ measurements in Fig.~\ref{fig:compactblock3}b yield different outcomes, a Pauli correction is necessary.

\begin{figure*}[t!]
\centering
\def\svgwidth{\linewidth}
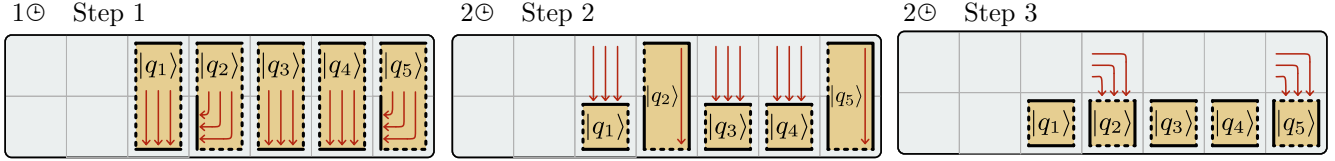
\caption{Patch rotations in preparation of a $Z \otimes X \otimes Z \otimes Z \otimes X$ measurement with an intermediate block.}
\label{fig:intermediateblockmove}
\end{figure*}

In Fig.~\ref{fig:compactblock3}, we go through the steps necessary to perform the $(Y \otimes \mathbbm{1} \otimes Y \otimes Z \otimes Y \otimes Y)_{\pi/8}$ rotation of Fig.~\ref{fig:compactblock2}. In step 1, we start with a 12-tile data block storing 6~qubits in the blue region. The orange region is not part of the data block, but is part of the adjacent distillation block, i.e., it is the source of the magic states. In steps 2-5, we perform the two $\pi/4$ rotations that are necessary to replace the $Y$ operators with $X$'s, i.e., the first two $\pi/4$ rotations in the circuit of Fig.~\ref{fig:compactblock2}. In step 6, we first rotate patches in the upper row, and then, in step 7, in the lower row. Finally, in step 8, we measure the Pauli product involving the magic state.

This general procedure can be used for any $\pi/8$ rotation. First, up to two $\pi/4$ rotations are performed in 2\clock. Next, patches in the upper and lower row are rotated, which takes 3\clock \ per row. Finally, the Pauli product is measured in 1\clock, requiring a total of 9\clock. While this is very slow compared to Fig.~\ref{fig:pauliprodmeas}, the compact block is a valid choice for small quantum computers where the distillation of a magic state takes longer than 9\clock.

\begin{figure}[b!]
\centering
\def\svgwidth{0.84\linewidth}
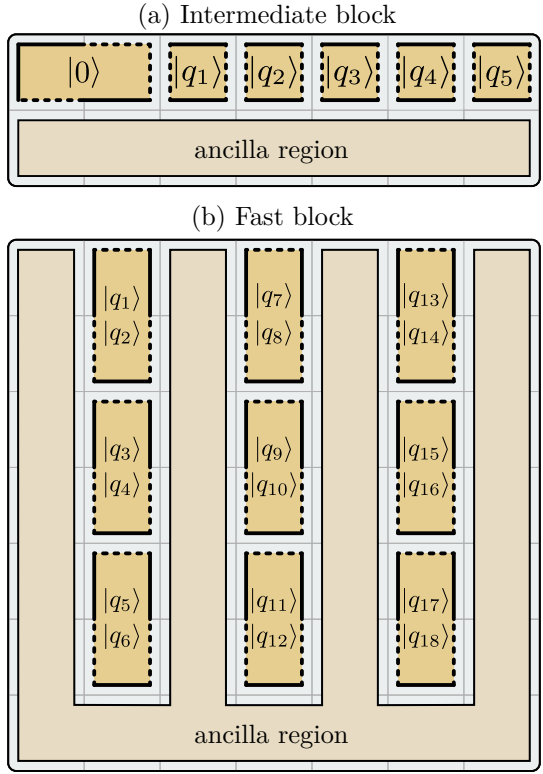
\caption{(a) Intermediate blocks store $n$ data qubits in $2.5n+4$ tiles and require up to 5\clock \ per magic state. (b) Fast blocks  use $2n+\sqrt{8n}+1$ tiles and require 1\clock \ per magic state.}
\label{fig:fasterblocks}
\end{figure}

\subsection{Intermediate block}

One possibility to speed up compact blocks is to store all qubits in one row instead of two. This is the intermediate block shown in Fig.~\ref{fig:fasterblocks}a, which uses $2n+4$ tiles to store $n$ qubits. By eliminating one row, all patch rotations can be done simultaneously. In addition, one can save 1\clock \ by moving all patches to the other side, thereby eliminating the need to move patches back to their row after the rotation. An example is shown in Fig.~\ref{fig:intermediateblockmove}. Suppose we have 5 qubits and need to prepare them for a $Z \otimes X \otimes Z \otimes Z \otimes X$ measurement. The first, third and fourth qubit are moved to the other side, which takes 1\clock. Simultaneously, the second and fifth qubit are rotated, which takes 2\clock. Therefore, the total number of time steps to consume a magic state is at most 5\clock, where 2\clock \ are used for up to two $\pi/4$ rotations, 2\clock \ for the patch rotations, and 1\clock \ for the Pauli product measurement consuming the magic state.

\subsection{Fast block}

The disadvantage of square patches is that only one Pauli operator is adjacent to the data block's ancilla region, i.e., available for Pauli product measurements at any given time. Two-tile one-qubit patches as in Fig.~\ref{fig:pauliprodmeas}, on the other hand, allow for the measurement of any Pauli operator, but use two tiles for each qubit. In order to have both compact storage and access to all Pauli operators, we use two-qubit patches for our fast blocks in Fig.~\ref{fig:fasterblocks}b. These patches use two tiles to represent two qubits (see Fig.~\ref{fig:patches}), where the first qubit's Pauli operators are in the left two edges, and the second qubit's operators are in the right two edges. Therefore, the example in Fig.~\ref{fig:fasterblocks}b is a fast block that stores 18 qubits.

Since all Pauli operators are accessible, the Pauli product measurement protocol of Fig.~\ref{fig:pauliprodmeas} can be used to consume a magic state every 1\clock. $n$ qubits occupy a square arrangement of tiles with a side length of $\sqrt{n/2}+1$, i.e., a total of $2n+\sqrt{8n}+1$ tiles. Even if $\sqrt{n/2}$ is not integer, one should keep the block as square-shaped as possible by picking the closest integer as a side length and shortening the last column.
While the fast block uses more tiles compared to the compact and intermediate blocks, it has a lower space-time cost, making it more favorable for large quantum computers for which the distillation of a magic state takes less than~5\clock.

\begin{figure*}[t!]
\centering
\def\svgwidth{\linewidth}
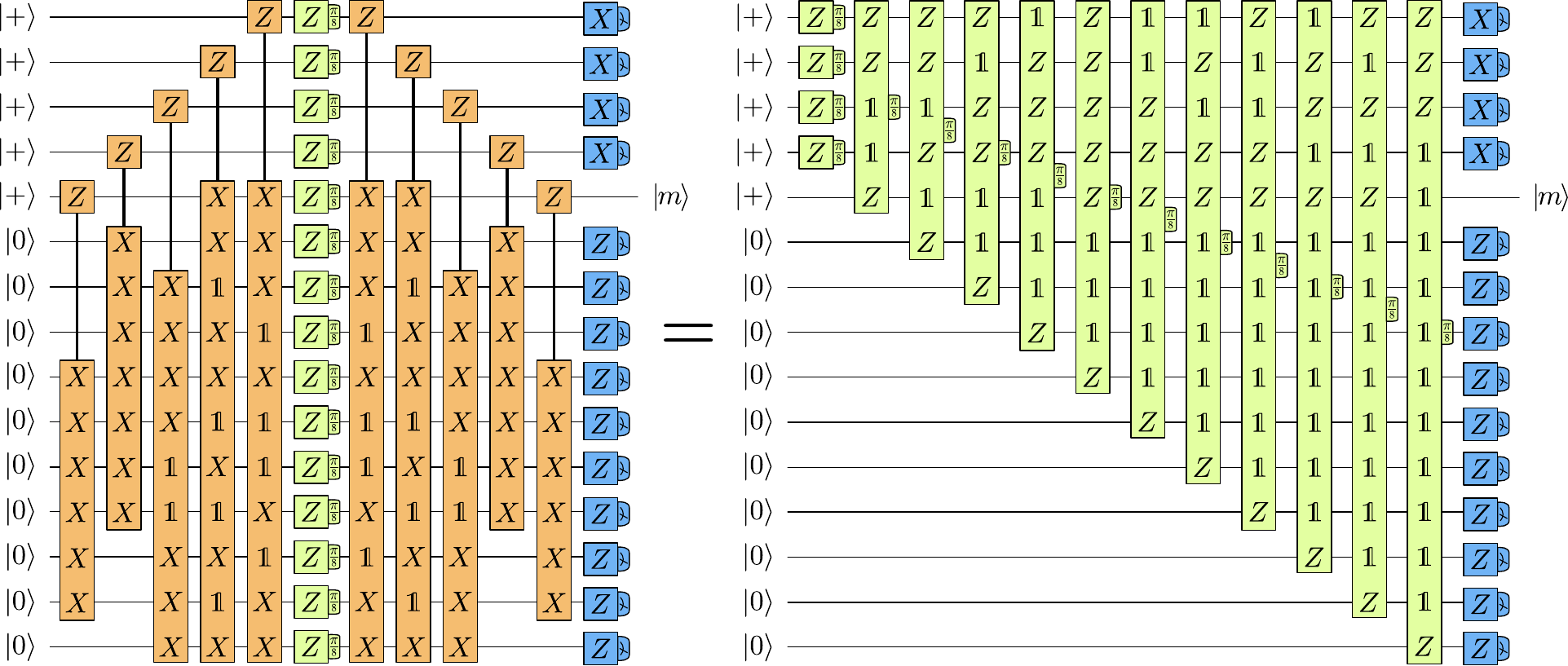
\caption{Encode-$T$-decode circuit of the 15-to-1 distillation protocol. The multi-target CNOTs (orange) can be commuted past the $T$ gates, such that they cancel and leave 15 $Z$-type Pauli product rotations. }
\label{fig:15to1circuit1}
\end{figure*}

Note that if undistilled magic states are sufficient, then any data block can already be used as a full quantum computer. A proof-of-principle two-qubit device in the spirit of Ref.~\cite{Lavasani2018} that constitutes a universal two-qubit quantum computer with undistilled magic states and can demonstrate all the operations that are used in our framework can be realized with six tiles, as shown in Appendix \ref{app:proofofprinciple}. This proof-of-principle device uses $(3d-1)\cdot 2d$ physical data qubits, i.e., 48, 140, or 280 data qubits for distances $d=3$, 5 or 7. If ancilla qubits are used for stabilizer measurements, the number of physical qubits roughly doubles, but it is still within reach of near-term devices. 

\textbf{Summary.} Data blocks store the data qubits of the computation and consume magic states. Compact blocks use $1.5n+3$ tiles for $n$ qubits and require up to 9\clock \ to consume a magic state. Intermediate blocks use $2n+4$ tiles and take up to 5\clock \ per magic state. Fast blocks use $2n+\sqrt{8n}+1$ tiles and take 1\clock \ per magic state. Data blocks need to be combined with distillation blocks for large-scale quantum computation.

\section{Distillation blocks}
\label{sec:distillation}

In this section, we discuss designs of tile blocks that are used for magic state distillation. This is necessary, because with surface codes, the initialization of non-Pauli eigenstates is prone to errors, which means that $\pi/8$ rotations performed using these states may lead to errors. In order to decrease the probability of such an error, magic state distillation~\cite{Bravyi2005} is used to convert many low-fidelity magic states into fewer higher-fidelity states. This requires only Clifford gates (i.e., Pauli product measurements), so, in principle, any of the data blocks discussed in the previous section can be used for this purpose. However, magic state distillation is repeated extremely often for large-scale quantum computation, so it is worth optimizing these protocols.

Here, we discuss a general procedure that can be applied to any distillation protocol based on an error-correcting code with transversal $T$ gates, such as punctured Reed-Muller codes~\cite{Bravyi2005,Haah2018} or block codes~\cite{Bravyi2012,Jones2013a,Fowler2013}. To show the general structure of such a protocol, we go through the example of 15-to-1 distillation~\cite{Bravyi2005}, i.e., a protocol that uses 15 faulty magic states to distill a single higher-fidelity state.

\begin{figure*}[t!]
\centering
\def\svgwidth{0.79\linewidth}
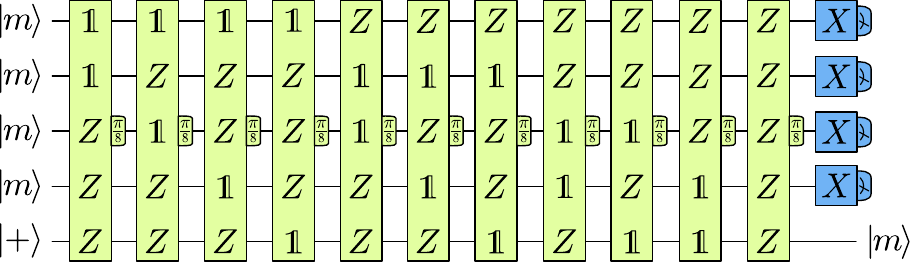
\caption{15-to-1 distillation circuit that uses 5 qubits and 11 $\pi/8$ rotations.}
\label{fig:15to1circuit2}
\end{figure*}

\subsection{15-to-1 distillation}

The 15-to-1 protocol is based on a quantum error-correcting code that uses 15 qubits to encode a single logical qubit with code distance 3. The reason why this can be used for magic state distillation is that, for this code, a physical $T$ gate on every physical qubit corresponds to a logical $T$ gate (actually $T^\dagger$) on the encoded qubit, which is called a transversal $T$ gate. The general structure of a distillation circuit based on a code with transversal $T$ gates is shown in Fig.~\ref{fig:15to1circuit1} for the example of 15-to-1. It consists of four parts: an encoding circuit, transversal $T$ gates, decoding and measurement.

The circuit begins with 5 qubits initialized in the $\ket{+}$ state and 10 qubits in the $\ket{0}$ state. Qubits 1-4, 5 and 6-15 are associated with the four $X$ stabilizers, the logical $X$ operator, and the ten $Z$ stabilizers of the code. The first five operations are multi-target CNOTs that correspond to the code's encoding circuit. They map the $X$ Pauli operators of qubits 1-4 onto the code's $X$ stabilizers, the $X$ Pauli of qubit 5 onto the logical $X$ operator and the $Z$ operators of qubits 6-15 onto the code's $Z$ stabilizers. Because we start out with +1-eigenstates of $X$ and $Z$, this circuit prepares the simultaneous stabilizer eigenstate corresponding to the logical $\ket{+}_L$ state. Next, a transversal $T$ gate is applied, transforming the logical state to $T_L \ket{+}_L$ (actually to $T_L^\dagger \ket{+}_L$). Note that the 15 $Z_{\pi/8}$ rotations are potentially faulty. Finally, the encoding circuit is reverted, shifting the logical qubit information back into qubit 5, and the information about the $X$ and $Z$ stabilizers into qubits 1-4 and 6-15. If no errors occurred, qubit 5 is now a magic state $T\ket{+}$ (actually $T^\dagger\ket{+}$). In order to detect whether any of the 15 $\pi/8$ rotations were affected by an error, qubits 1-4 and 6-15 are measured in the $X$ and $Z$ basis, respectively, effectively measuring the stabilizers of the code. Since the code distance is 3, up to two errors can be detected, which will yield a -1 measurement outcome on some stabilizers. If any error is detected, all qubits are discarded and the distillation protocol is restarted. This way, if the error probability of each of the 15 $T$ gates is $p$, the error probability of the output state is reduced to $35p^3$ to leading order. In other words, this protocol takes 15 magic states with error probability $p$, and outputs a single magic state with an error of $35p^3$.

\textbf{Simplifying the circuit.} Using the commutation rules of Fig.~\ref{fig:circuitsimplify}b, we can commute the first set of multi-target CNOTs to the right. This maps the $Z_{\pi/8}$ rotations onto $Z$-product $\pi/8$ rotations. Since controlled-Pauli gates satisfy $\mathrm{C}(P_1,P_2) = \mathrm{C}(P_1,P_2)^\dagger$, the multi-target CNOTs of the encoding circuit precisely cancel the multi-target CNOTs of the decoding circuit, leaving a circuit of 15 $Z$-type $\pi/8$ rotations in Fig.~\ref{fig:15to1circuit1}.

Note that qubits 6-15 in this circuit are entirely redundant. They are initialized in a $Z$ eigenstate, are then part of a $Z$-type rotation, and are finally measured in the $Z$ basis, trivially yielding the outcome $+1$. Since they serve no purpose, they can simply be removed to yield the five-qubit circuit in Fig.~\ref{fig:15to1circuit2}, where we have absorbed the single-qubit $\pi/8$ rotations into the initial $\ket{+}$ states and rearranged the remaining 11 rotations.

\setcounter{MaxMatrixCols}{25}
\setlength\arraycolsep{2pt}

This kind of circuit simplification is equivalent to the space-time trade-offs mentioned in Ref.~\cite{Haah2018} and can be applied to any protocol that is based on a code with transversal $T$ gates. In general, a code with $m_x$ $X$ stabilizers that uses $n$ qubits to encode $k$ logical qubits yields a circuit of $n-m_x$ $\pi/8$ rotations on $m_x+k$ qubits. Each of the $m_x+k$ qubits are either associated with an $X$ stabilizer or one of the $k$ logical qubits. For each of the $n$ qubits of the code, the circuit contains one $\pi/8$ rotation with an axis that has a $Z$ on each stabilizer or logical $X$ operator that this qubit is part of. In order to more easily determine the $n-m_x$ rotations, it is useful to write down an $n \times (m_x+k)$ matrix that shows the $X$ stabilizers and logical $X$ operators of the code. For 15-to-1, such a matrix could look like this:

\begin{figure*}[t!]
\centering
\def\svgwidth{0.97\linewidth}
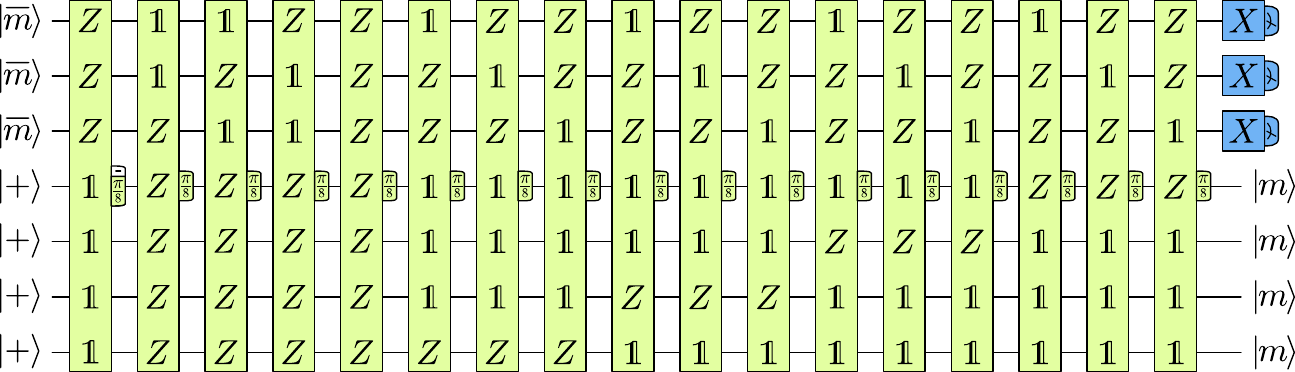
\caption{20-to-4 distillation circuit that uses 7 qubits and 17 $\pi/8$ rotations.}
\label{fig:20to4circuit}
\end{figure*}

\begin{equation}
M_{\text{15-to-1}} = 
\begin{pmatrix}
	0 & 0 & 0 & 1 & 0 & 0 & 0 & 0 & 1 & 1 & 1 & 1 & 1 & 1 & 1 \\
	0 & 0 & 1 & 0 & 0 & 1 & 1 & 1 & 0 & 0 & 0 & 1 & 1 & 1 & 1 \\
	0 & 1 & 0 & 0 & 1 & 0 & 1 & 1 & 0 & 1 & 1 & 0 & 0 & 1 & 1 \\
	1 & 0 & 0 & 0 & 1 & 1 & 0 & 1 & 1 & 0 & 1 & 0 & 1 & 0 & 1 \\
	\hline
	0 & 0 & 0 & 0 & 1 & 1 & 1 & 0 & 1 & 1 & 0 & 1 & 0 & 0 & 1
\end{pmatrix}
\label{eqn:15to1matrix}
\end{equation}

Each of the first four rows describes one of the four $X$ stabilizers of the code, where 0 stands for $\mathbbm{1}$ and 1 stands for $X$. For instance, the first row indicates that the first $X$ stabilizer of this 15-qubit code is $\mathbbm{1} \otimes\mathbbm{1} \otimes\mathbbm{1} \otimes X \otimes \mathbbm{1} \otimes\mathbbm{1} \otimes\mathbbm{1} \otimes\mathbbm{1} \otimes X  \otimes X\otimes X\otimes X\otimes X\otimes X\otimes X$. The rows below the horizontal bar~--~in this case the last row~--~show the logical $X$ operators of the code. The circuit in Fig.~\ref{fig:15to1circuit2} is then obtained by placing a $\ket{+}$ state for each row and a $\pi/8$ rotation for each column, with the axis of rotation determined by the indices in the column~--~a $\mathbbm{1}$ for each 0 and a $Z$ for each 1. Note that, in Fig.~\ref{fig:15to1circuit2}, the first four rotations (columns) of Eq.~\eqref{eqn:15to1matrix} are absorbed by the initial states.

\subsection{Triorthogonal codes}

The aforementioned circuit translation can be applied to any code with transversal $T$ gates. One particularly versatile and simple scheme to generate such codes is based on triorthogonal matrices~\cite{Bravyi2012,Haah2018}, which we briefly review in this section. The first step is to write down a triorthogonal matrix $G$, such as
\begin{equation}
G = \begin{pmatrix}
	1 & 1 & 1 & 1 & 1 & 1 & 1 & 1 & 1 & 1 & 1 & 1 & 1 & 1 & 1 & 1 \\
	0 & 0 & 0 & 0 & 0 & 0 & 0 & 0 & 1 & 1 & 1 & 1 & 1 & 1 & 1 & 1 \\
	0 & 0 & 0 & 0 & 1 & 1 & 1 & 1 & 0 & 0 & 0 & 0 & 1 & 1 & 1 & 1 \\
	0 & 0 & 1 & 1 & 0 & 0 & 1 & 1 & 0 & 0 & 1 & 1 & 0 & 0 & 1 & 1 \\
	0 & 1 & 0 & 1 & 0 & 1 & 0 & 1 & 0 & 1 & 0 & 1 & 0 & 1 & 0 & 1
\end{pmatrix} \, .
\label{eqn:15to1}
\end{equation}
Triorthogonality refers to three criteria: $i)$ The number of 1s in each row is a multiple of 8. $ii)$ For each pair of rows, the number of entries where both rows have a 1 is a multiple of 4. $iii)$ For each set of three rows, the number of entries where all three rows have a 1 is a multiple of 2. In other words,
\begin{equation}
	\begin{split}
	\forall a:& \sum\mathop{}_{i}  G_{a,i} = 0 \pmod 8 \\
	\forall a,b:& \sum\mathop{}_{i} G_{a,i} G_{b,i} = 0 \pmod 4 \\
	\forall a,b,c:& \sum\mathop{}_{i} G_{a,i} G_{b,i} G_{c,i} = 0 \pmod 2
	\end{split}
	\label{eqn:triorth}
\end{equation}
A general procedure based on classical Reed-Muller codes to obtain such matrices is described in Ref.~\cite{Haah2018}.

After obtaining a triorthogonal matrix, such as the one in Eq.~\eqref{eqn:15to1}, the second step is to put it in a row echelon form by Gaussian elimination
\begin{equation}
\tilde{G} = 
\begin{pmatrix}
 0 & 0 & 0 & 0 & 1 & 0 & 0 & 0 & 0 & 1 & 1 & 1 & 1 & 1 & 1 & 1 \\
 0 & 0 & 0 & 1 & 0 & 0 & 1 & 1 & 1 & 0 & 0 & 0 & 1 & 1 & 1 & 1 \\
 0 & 0 & 1 & 0 & 0 & 1 & 0 & 1 & 1 & 0 & 1 & 1 & 0 & 0 & 1 & 1 \\
 0 & 1 & 0 & 0 & 0 & 1 & 1 & 0 & 1 & 1 & 0 & 1 & 0 & 1 & 0 & 1 \\
 1 & 0 & 0 & 0 & 0 & 1 & 1 & 1 & 0 & 1 & 1 & 0 & 1 & 0 & 0 & 1 
\end{pmatrix} \, .
\label{eqn:15to12}
\end{equation}
The last step is to remove one of the columns that contains a single 1, i.e., one of the first five columns, which is also called puncturing.\footnote{Even though this is commonly called puncturing, it would be perhaps more accurate to refer to this process as \emph{shortening} (see, e.g., Ref.~\cite{ModifyingCodes}), as was pointed out to me by a referee.} Puncturing an $a \times b$ triorthogonal matrix $k$ times yields a code encoding $k$ logical qubits with $m_x = b-k$ and $n=a-k$. The rows of the matrix after puncturing that contain an even number of 1s describe $X$ stabilizers, whereas the rows with an odd number of 1s describe $X$ logical operators.
In terms of distillation protocols, a code described by such a matrix can be used for $n$-to-$k$ distillation. Indeed, if we puncture the matrix in Eq.~\eqref{eqn:15to12} once by removing the first column, we retrieve the 15-to-1 protocol of Eq.~\eqref{eqn:15to1matrix}. We can also puncture it twice by removing the first two columns. This yields the matrix
\begin{equation}
M_{\text{14-to-2}} = 
\begin{pmatrix}
 0 & 0 & 1 & 0 & 0 & 0 & 0 & 1 & 1 & 1 & 1 & 1 & 1 & 1 \\
 0 & 1 & 0 & 0 & 1 & 1 & 1 & 0 & 0 & 0 & 1 & 1 & 1 & 1 \\
 1 & 0 & 0 & 1 & 0 & 1 & 1 & 0 & 1 & 1 & 0 & 0 & 1 & 1 \\
 \hline
 0 & 0 & 0 & 1 & 1 & 0 & 1 & 1 & 0 & 1 & 0 & 1 & 0 & 1 \\
 0 & 0 & 0 & 1 & 1 & 1 & 0 & 1 & 1 & 0 & 1 & 0 & 0 & 1
\end{pmatrix} \, ,
\label{eqn:14to2}
\end{equation}
which describes a 14-to-2 protocol. The corresponding circuit can be simply read off from this matrix. It is almost identical to the 15-to-1 protocol of Fig.~\ref{fig:15to1circuit2}, except that the fourth qubit is initialized in the $\ket{+}$ state and is not measured at the end of the circuit, but instead outputs a second magic state. However, because the code of 14-to-2 has a code distance of 2, the output error probability is higher, namely $7p^2$~\cite{Bravyi2012}. Puncturing the matrix $\tilde{G}$ any further would yield codes with a distance lower than 2, precluding them from detecting errors and improving the quality of magic states. In fact, the minimum number of qubits in triorthogonal codes was shown to be 14~\cite{Campbell2018}.

\begin{figure*}[t!]
\centering
\def\svgwidth{0.98\linewidth}
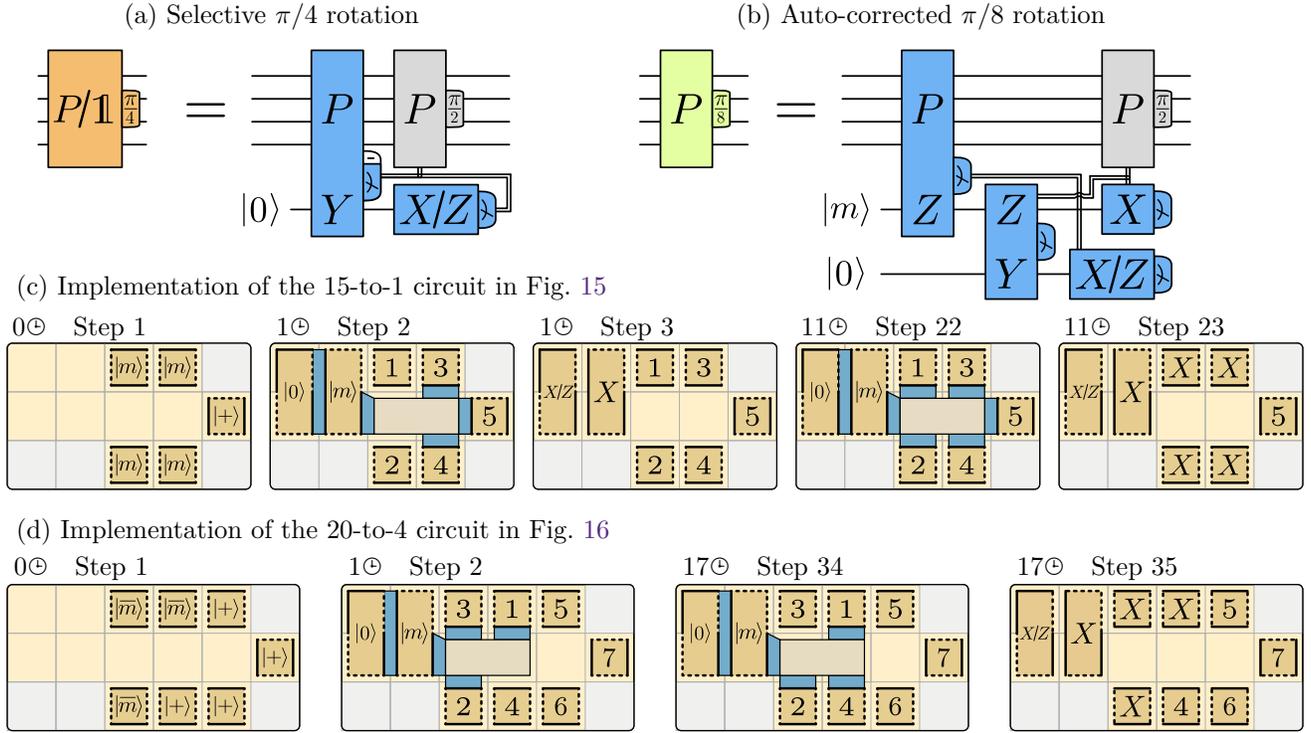
\caption{Implementation of the 15-to-1 and 20-to-4 distillation protocols in our framework. Each time step in (c) and (d) corresponds to an auto-corrected $\pi/8$ rotation (b), which in turn is based on selective $\pi/4$ rotations (a).}
\label{fig:15to1impl}
\end{figure*}

\textbf{Semi-triorthogonal codes.} There are also codes that are based on ``semi-triorthogonal'' matrices, where all three conditions of Eq.~\eqref{eqn:triorth} are only satisfied modulo~2. One example is the matrix

\begin{equation}
\begin{pmatrix}
 0 & 0 & 0 & 0 & 0 & 0 & 1 & 1 & 0 & 0 & 1 & 1 & 0 & 1 & 1 & 0 & 1 & 1 & 0 & 1 & 1 & 0 & 1 & 1 \\
 0 & 0 & 0 & 0 & 0 & 1 & 0 & 1 & 0 & 1 & 0 & 1 & 1 & 0 & 1 & 1 & 0 & 1 & 1 & 0 & 1 & 1 & 0 & 1 \\
 0 & 0 & 0 & 0 & 1 & 0 & 0 & 1 & 1 & 0 & 0 & 1 & 1 & 1 & 0 & 1 & 1 & 0 & 1 & 1 & 0 & 1 & 1 & 0 \\
 0 & 0 & 0 & 1 & 0 & 0 & 0 & 0 & 1 & 1 & 1 & 1 & 0 & 0 & 0 & 0 & 0 & 0 & 0 & 0 & 0 & 1 & 1 & 1 \\
 0 & 0 & 1 & 0 & 0 & 0 & 0 & 0 & 1 & 1 & 1 & 1 & 0 & 0 & 0 & 0 & 0 & 0 & 1 & 1 & 1 & 0 & 0 & 0 \\
 0 & 1 & 0 & 0 & 0 & 0 & 0 & 0 & 1 & 1 & 1 & 1 & 0 & 0 & 0 & 1 & 1 & 1 & 0 & 0 & 0 & 0 & 0 & 0 \\
 1 & 0 & 0 & 0 & 0 & 0 & 0 & 0 & 1 & 1 & 1 & 1 & 1 & 1 & 1 & 0 & 0 & 0 & 0 & 0 & 0 & 0 & 0 & 0 
\end{pmatrix} \, .
\end{equation}
When this matrix is punctured four times, it yields a code that can be used for a 20-to-4 protocol. A scheme to generate such matrices for $3k$+8-to-$k$ distillation is shown in Ref.~\cite{Bravyi2012}. For the case of the 20-to-4 protocol, the matrix that describes the code
\begin{equation}
\underset{\text{20-to-4}}{M} = 
\begin{pmatrix}
 0 & 0 & 1 & 1 & 0 & 0 & 1 & 1 & 0 & 1 & 1 & 0 & 1 & 1 & 0 & 1 & 1 & 0 & 1 & 1 \\
 0 & 1 & 0 & 1 & 0 & 1 & 0 & 1 & 1 & 0 & 1 & 1 & 0 & 1 & 1 & 0 & 1 & 1 & 0 & 1 \\
 1 & 0 & 0 & 1 & 1 & 0 & 0 & 1 & 1 & 1 & 0 & 1 & 1 & 0 & 1 & 1 & 0 & 1 & 1 & 0 \\
 \hline
 0 & 0 & 0 & 0 & 1 & 1 & 1 & 1 & 0 & 0 & 0 & 0 & 0 & 0 & 0 & 0 & 0 & 1 & 1 & 1 \\
 0 & 0 & 0 & 0 & 1 & 1 & 1 & 1 & 0 & 0 & 0 & 0 & 0 & 0 & 1 & 1 & 1 & 0 & 0 & 0 \\
 0 & 0 & 0 & 0 & 1 & 1 & 1 & 1 & 0 & 0 & 0 & 1 & 1 & 1 & 0 & 0 & 0 & 0 & 0 & 0 \\
 0 & 0 & 0 & 0 & 1 & 1 & 1 & 1 & 1 & 1 & 1 & 0 & 0 & 0 & 0 & 0 & 0 & 0 & 0 & 0 
\end{pmatrix} \, ,
\label{eqn:semitriorth1}
\end{equation}
can be straightforwardly translated into the circuit in Fig.~\ref{fig:20to4circuit}. While semi-triorthogonal codes can be used the same way for distillation as properly triorthogonal codes, their caveat is that a Clifford correction may be required. This correction can be obtained by adding columns to the semi-triorthogonal matrix until it becomes properly triorthogonal, e.g., by adding the columns of the matrix
\begin{equation}
\underset{\text{Clifford correction}}{M} = 
\begin{pmatrix}
 0 & 0 & 0 & 0 & 1 & 1 & 1 & 1  \\
 0 & 0 & 1 & 1 & 0 & 0 & 1 & 1  \\
 1 & 1 & 0 & 0 & 0 & 0 & 1 & 1  \\
 0 & 0 & 0 & 0 & 0 & 0 & 0 & 0  \\
 0 & 0 & 0 & 0 & 0 & 0 & 0 & 0  \\
 0 & 0 & 0 & 0 & 0 & 0 & 0 & 0 \\
 0 & 0 & 0 & 0 & 0 & 0 & 0 & 0  
\end{pmatrix} \,
\label{eqn:semitriorth2}
\end{equation}
 to the matrix of Eq.~\eqref{eqn:semitriorth1}. Since the additional columns come in pairs, this Clifford correction always consists of $Z$-type $\pi/4$ rotations~\cite{Bravyi2012}.

In this case, the correction consists of four $\pi/4$ rotations on the first three qubits, effectively changing the first $(Z\otimes Z \otimes Z)_{\pi/8}$ rotation to a $(Z\otimes Z \otimes Z)_{-\pi/8}$ rotation, and the initial magic states to  $\ket{\overline{m}} = \ket{0} + e^{-i\pi/4}\ket{1}$ states. The probability of any of the four output states being affected by an error is $22p^2$. When treating this output error rate as $5.5p^2$ per magic state, one should take into account that, for multiple output states, errors can be correlated. Note that $3k$+8-to-$k$ protocols can be modified to $3k$+4-to-$k$~\cite{Meier2013,Campbell2016a,Campbell2018}.

\subsection{Surface-code implementation}
Having outlined the general structure of distillation protocols, we now discuss their implementation with surface codes. Distillation protocols are particularly simple quantum circuits, since they exclusively consist of $Z$-type $\pi/8$ rotations. Therefore, we can use a construction similar to the compact data block, and still only require 1\clock \ per rotation.

Because distillation circuits are relatively short, it is useful to avoid the Clifford corrections of Fig.~\ref{fig:magicstateuse} that may be required with 50\% probability after a magic state is consumed. These corrections slow down the protocol, because they change the final $X$ measurements to Pauli product measurements. Instead, we use a circuit which consumes a magic state and automatically performs the Clifford correction. It is based on the selective $\pi/4$ rotation circuit in Fig.~\ref{fig:15to1impl}a. To perform a $P_{\pi/4}$ rotation according to the circuit in Fig.~\ref{fig:compactblock3}b, a $\ket{0}$ state is initialized and $P \otimes Y$ is measured, which takes 1\clock. However, the $\pi/4$ rotation is only performed if the $\ket{0}$ qubit is measured in $X$ afterwards. If, instead, it is measured in $Z$, the qubit is simply discarded without performing any operation. In other words, the choice of measurement basis determines whether a $P_{\pi/4}$ or a $\mathbbm{1}$ operation is performed. This can be used to construct the circuit in Fig.~\ref{fig:15to1impl}b. Here, the first step to perform a $P_{\pi/8}$ gate is to measure $P \otimes Z$ between the qubits and a magic state $\ket{m}$, and $Z \otimes Y$ between $\ket{m}$ and $\ket{0}$. These two measurements commute and can be performed simultaneously. If the outcome of the first measurement is +1, no Clifford correction is required and $\ket{0}$ is read out in $Z$. If the outcome is -1, $\ket{0}$ is measured in $X$, yielding the required Clifford correction.

\begin{figure}[b!]
\centering
\def\svgwidth{\linewidth}
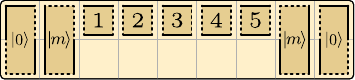
\caption{Two $2\times 2$ ancilla blocks can be used to prevent state injection and classical processing from slowing down the 15-to-1 protocol.}
\label{fig:15to1noslowdown}
\end{figure}

This can be used to implement the 15-to-1 protocol of Fig.~\ref{fig:15to1circuit2} in 11\clock \ using 11 tiles, as shown in Fig.~\ref{fig:15to1impl}c. Four qubits are initialized in $\ket{m}$, and a fifth in $\ket{+}$. A $2\times 2$ block of tiles to the left is reserved for the $\ket{m}$ and $\ket{0}$ qubits of the auto-corrected $\pi/8$ rotations. Two additional tiles are used for the ancilla of the multi-patch measurement. In step 2, the first $\pi/8$ rotation $(\mathbbm{1} \otimes \mathbbm{1} \otimes Z \otimes Z \otimes Z)_{\pi/8}$ is performed. Depending on the measurement outcome of step 2, the $\ket{0}$ ancilla is read out in the $X$ or $Z$ basis. This is repeated 11 times, once for each of the 11 rotations in Fig.~\ref{fig:15to1circuit2}. Finally, in step 23, qubits 1-4 are measured in $X$. If all four outcomes are +1, the distillation protocol yields a distilled magic state in tile 5. Since 11 tiles are used for 11\clock, the space-time cost is $121d^3$ in terms of (physical data qubits)$\cdot$(code cycles) to leading order. Similarly, the 20-to-4 protocol of Fig.~\ref{fig:20to4circuit} is implemented in Fig.~\ref{fig:15to1impl}d using 14 tiles for 17\clock, i..e, with a leading-order space-time cost of $238d^3$.

\textbf{Caveat.} Even though our leading-order estimate of the time cost of $11d$ code cycles for 15-to-1 or $17d$ code cycles for 20-to-4 is correct, the full time cost also contains contributions that do not scale with $d$. The two processes that may require special care in the magic state distillation protocol are state injection and classical processing. Every 1\clock \ requires the initialization of a magic state and a short classical computation to determine whether the $\ket{0}$ state needs to be measured in $X$ or $Z$. While neither of these processes scales with $d$, they can slow down the distillation protocol, depending on the injection scheme and the control hardware that is used. This slowdown can be avoided by using additional $2\times 2$ blocks of $\ket{0}$-$\ket{m}$ pairs, as shown in Fig.~\ref{fig:15to1noslowdown} for 15-to-1 distillation with one additional block. Here, the left and right block can be used in an alternating fashion, i.e., the left block for rotations $1,3,5,\dots$ and the right block for rotations $2,4,6,\dots$ While one block is being used for a rotation, the other one can be used to prepare a new magic state and to process the measurement outcomes of the previous rotation.

\pagebreak

\textbf{General space-time cost.} The scheme of Fig.~\ref{fig:15to1impl} can be used to implement any protocol based on a triorthogonal code. For an $n$-qubit code with $k$ logical qubits and $m_x$ $X$ stabilizers, the protocol uses $1.5(m_x+k)+4$ tiles for $(n-m_x)$ \clock. In this time, it distills $k$ magic states with a success probability of ${\sim}(1-p)^n$, since any error will result in failure. Therefore, such a protocol distills $k$ magic state on average every $(n-m_x)/(1-p)^n$ time steps. Thus, the space-time cost per magic state is
\begin{equation}
	\mathrm{cost}(n,m_x,k,p,d) = \frac{[1.5(m_x+k)+4](n-m_x)}{k(1-p)^n}~d^3 \, .
	\label{eqn:distcost}
\end{equation}
In order to minimize the space-time cost for distillation in our framework, one should pick a distillation protocol that minimizes this quantity for a given input and target error rate.

\begin{figure*}[t!]
\centering
\def\svgwidth{0.9\linewidth}
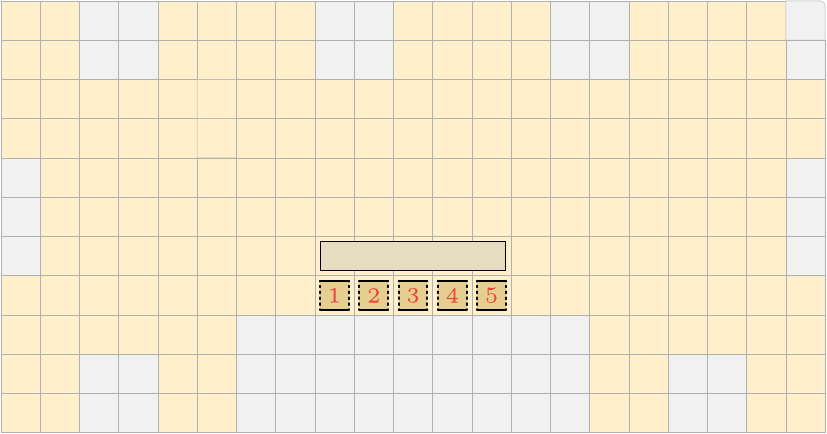
\caption{176-tile block that can be used for 225-to-1 distillation. The qubits highlighted in red are used for the second level of the distillation protocol. The blue ancilla is used to move level-1 magic states into the two $\ket{m}$-$\ket{0}$ blocks of the level-2 distillation.}
\label{fig:225to1}
\end{figure*}

\subsection{Benchmarking}

We can use the previously described 15-to-1 and 20-to-4 schemes to benchmark our implementations. In Ref.~\cite{Herr2017}, these schemes were implemented with lattice surgery and their cost compared to implementations based on braiding of hole defects. In addition, the 7-to-1 scheme was considered, which is a scheme to distill $\ket{Y}$ states. The distillation of these states is not necessary in our framework, but for benchmarking purposes we show the 7-to-1 protocol in Appendix \ref{app:7to1}. It can be implemented using 7 tiles for 4\clock, i.e., with a space-time cost of $28d^3$.

We summarize the leading-order space-time costs of the three protocols in Table \ref{tab:benchmark}. The comparison shows drastic reductions in space-time cost compared to schemes based on braiding of hole defects and compared to other approaches to optimizing lattice surgery. Compared to the braiding-based scheme, the space-time cost of 7-to-1, 15-to-1 and 20-to-4 is reduced by 60\%, 84\% and 90\%, respectively.

\def\arraystretch{1.5}
\begin{table}[b!]
\centering
\begin{tabular}{cccc}
& 7-to-1 & 15-to-1 & 20-to-4 \\
\hline
Hole braiding~\cite{Fowler2012b,Fowler2013} & $70d^3$ & $750d^3$& $2344d^3$ \\
\hline
Lattice surgery~\cite{Herr2017} & $140d^3$ & $540d^3$ & $1134d^3$ \\
\hline
Our framework & $28d^3$ & $121d^3$ & $238d^3$
\end{tabular}
\caption{Comparison of the leading-order space-time cost of 7-to-1, 15-to-1 and 20-to-4 with defect-based schemes, optimized lattice surgery in Ref.~\cite{Herr2017} and our schemes. The space-time cost is in terms of (physical data qubits)$\cdot$(code cycles). }
\label{tab:benchmark}
\end{table}

\subsection{Higher-fidelity protocols}
So far, we have only explicitly discussed protocols that reduce the input error to ${\sim}p^2$ or ${\sim}p^3$. There are two strategies to obtain protocols with a higher output fidelity: concatenation and higher-distance codes.

\textbf{Concatenation.} In the 15-to-1 protocol, we use 15 undistilled magic states to obtain a distilled magic state with an error rate of $35p^3$. If we perform the same protocol, but use 15 distilled magic states from previous 15-to-1 protocols as inputs, the output state will have an error rate of $35(35p^3)^3 = 1500625p^9$. This corresponds to a 225-to-1 protocol obtained from the concatenation of two 15-to-1 protocols. It is also possible to concatenate protocols that are not identical. Strategies to combine high-yield and low-yield protocols are discussed in Ref.~\cite{Bravyi2012}.

In Fig.~\ref{fig:225to1}, we show an unoptimized block that can be used for 225-to-1 distillation. It consists of 11 15-to-1 blocks that are used for the first level of distillation. Since each of these 11 blocks takes 11\clock \ to finish, they can be operated such that exactly one of these blocks finishes in every time step. Therefore, in every time step, one first-level magic state can be used for second-level distillation by moving it into one of the two level-2 $\ket{m}$-$\ket{0}$ blocks via the blue ancilla. The qubits that are used for the second level are highlighted in red. Note that since, for the second level, the single-qubit $\pi/8$ rotations require distilled magic states, the 15-to-1 protocol of Fig.~\ref{fig:15to1circuit2} requires 15 rotations instead of just 11. Therefore, the entire protocol finishes in 15\clock \ using 176 tiles with a total space-time cost of $2640d^3$. It should be noted that, since lower-level distillation blocks produce magic states with low fidelity, there is no benefit in using the full code distance to produce these states. The space-time cost of concatenated protocols can be reduced significantly by running the lower-level distillation blocks at a reduced code distance (see, e.g., Refs.~\cite{Fowler2018,Gidney2018a}), using smaller patches and fewer code cycles. The exact code distance that should be used depends on the protocol and the desired output fidelity.

\begin{figure}[t!]
\centering
\def\svgwidth{0.8\linewidth}
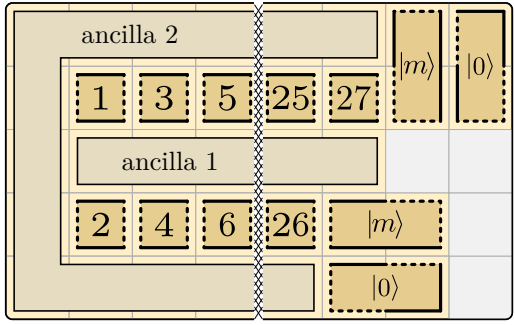
\caption{81-tile block that can be used for the 116-to-12 protocol. Here, two $\pi/8$ rotations can be performed at the same time, where one rotation uses the ancilla space denoted as \textit{ancilla 1}, and the other one uses \textit{ancilla 2}.}
\label{fig:116to12}
\end{figure}

\textbf{Higher-distance codes.} Alternatively, we can use a code that produces higher-fidelity states. In Ref.~\cite{Haah2018}, several protocols based on punctured Reed-Muller codes are discussed. One of these protocols is a 116-to-12 protocol based on a code with $n=116$, $k=12$ and $m_x = 17$. It yields 12 magic states which each have an error rate of  $41.25p^4$. According to Eq.~\eqref{eqn:distcost}, this protocol can be implemented using 44 tiles  for 99\clock \ with a space-time cost of $363d^3$ per output state and a success probability of $(1-p)^{116}$. For protocols with a high space cost such as 116-to-12, the space-time cost can be slightly reduced by introducing additional ancilla space, such that two operations can be performed simultaneously. One possible configuration is shown in Fig.~\ref{fig:116to12}. This increases the space cost to 81 tiles, but reduces the time cost to 50\clock, with a total space-time cost of $337.5d^3$ per output state.

\textbf{Output-to-input ratio is not everything.} A popular figure of merit when comparing $n$-to-$k$ distillation protocols is the ratio $k/n$. One of the protocols in Ref.~\cite{Haah2018} is a 912-to-112 protocol with $n=912$, $k=112$ and $m_x=64$, which yields 112 output state, each with an error rate of $10.63p^6$. While the output fidelity is not as high as for 225-to-1, the output-to-input ratio is much higher. For $p=10^{-3}$, the output fidelity of 225-to-1 is ${\sim} 1.5 \times 10^{-21}$, while it is only ${\sim} 10^{-17}$ for 912-to-112. Therefore, if output-to-input ratio were a good figure of merit, we would expect the 912-to-112 protocol to be considerably less costly compared to 225-to-1. If we use an implementation in the spirit of Fig.~\ref{fig:116to12}, the space cost is roughly $2.5(m_x+k)$ tiles and the protocol takes $(n-m_x)/2$ time steps. Thus, 912-to-112 uses 440 tiles for 424\clock. This would put the space-time cost per state at $1665d^3$, which is indeed lower than that of 225-to-1. However, the success probability of 912-to-112 for $p=10^{-3}$ is only at ${\sim}40\%$, which more than doubles the actual space-time cost. On the other hand, the space-time cost of 225-to-1 is barely affected by the success probability, as each of the level-1 15-to-1 blocks finishes with $98.5\%$ success probability. This means that, with $1.5\%$ probability, a time step of 225-to-1 is skipped, since the necessary level-1 state is missing. This only increases the space-time cost from $2640^3$ to $2680d^3$. Even without further decreasing the space-time cost of 225-to-1 by reducing the code distance of the level-1 distillation blocks, this indicates that the output-to-input ratio is not a good figure of merit in our framework.

\textbf{Summary.} The class of magic state distillation protocols that are based on an $n$-qubit error-correcting code with $m_x$ $X$ stabilizers and $k$ logical qubits can be implemented using $1.5(m_x+k)+4$ tiles and $n-m_x$ time steps. Such protocols output $k$ magic states with a success probability of $(1-p)^n$. Therefore, if the input fidelity and desired output fidelity are known, the distillation protocol should minimize the cost function given in Eq.~\eqref{eqn:distcost}.

\section{Trade-offs limited by $T$ count}
\label{sec:tcount}

Having discussed data blocks and distillation blocks in the previous two sections, we are now ready to piece them together to a full quantum computer. In order to illustrate the steps that are necessary to calculate the space and time cost of a computation and to trade off space against time, we consider an example computation with a $T$ count of $10^8$ and a $T$ depth of $10^6$. We consider two different scenarios: an error rate of $p=10^{-3}$ and an error rate of $p=10^{-4}$. The error rate determines how many physical qubits are required per logical qubit and which distillation protocol should be used. It is only a meaningful number, if we specify an error model for the physical qubits and undistilled magic states. We will assume circuit-level nose for the physical qubits, i.e., faulty qubits, gates and measurements. The error model for undistilled magic states depends on the specific state-injection protocol. We will assume that raw magic states are affected by random Pauli errors with probability $p$. To calculate concrete numbers, we assume that the quantum computer can perform a code cycle every $1~\mu\mathrm{s}$. We want to perform the $10^8$-$T$-gate computation in a way that the probability of any one of the $T$ gates being affected by an error stays below $1\%$. In addition, we require that the probability of an error affecting any of the logical qubits encoded in surface-code patches stays below $1\%$. This results in a $2\%$ chance that the quantum computation will yield a wrong result. In order to exponentially increase the precision of the computation, it can be repeated multiple times or run in parallel on multiple quantum computers.

\subsection{Step 1: Determine distillation protocol}

The first step is to determine which distillation protocol is sufficient for the computation. In order to stay below $1\%$ error probability with $10^8$ $T$ gates, each magic state needs to have an error rate below $10^{-10}$. For $p=10^{-4}$, the 15-to-1 protocol is sufficient, since it yields an output error rate of $35p^3 = 3.5\cdot 10^{-11}$. For $p=10^{-3}$, 15-to-1 is not enough. On the other hand, two levels of 15-to-1, i.e., 225-to-1, yield magic states with an error rate of $1.5 \cdot 10^{-21}$, which is many orders of magnitude above what is required. A less costly protocol is 116-to-12, which yields output states with an error rate of $41.25p^4 = 4.125 \cdot 10^{-11}$, which suffices for our purposes.

\subsection{Step 2: Construct a minimal setup}

In order to determine the necessary code distance, we first construct a minimal setup, i.e., a configuration of tiles that can be used for the computation and uses as little space as possible. The reason why this is useful to determine the code distance is that the initial space-time trade-offs that we discuss significantly improve the overall space-time cost. Therefore, the minimal setup can be used to comfortably upper-bound the required code distance.

\begin{figure}[t!]
\centering
\def\svgwidth{\linewidth}
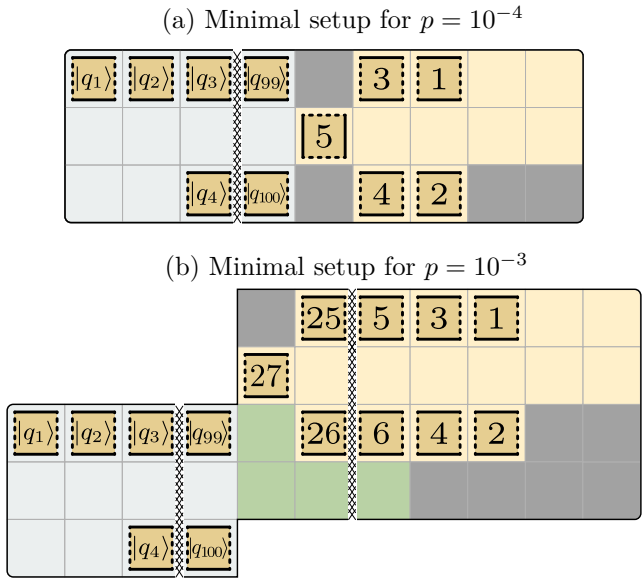
\caption{Minimal setups using compact data blocks for $p=10^{-4}$ (with 15-to-1 distillation) and $p=10^{-3}$ (with 116-to-12 distillation). Blue tiles are data block tiles, orange tiles are distillation block tiles, green tiles are used for magic state storage and gray tiles are unused tiles.}
\label{fig:minimalsetup1}
\end{figure}

For $p=10^{-4}$, a minimal setup consists of a compact data block and a 15-to-1 distillation block, see Fig.~\ref{fig:minimalsetup1}a. The compact block stores 100 qubits in 153 tiles and requires up to 9\clock \ to consume a magic state. The 15-to-1 distillation block uses 11 tiles and outputs a magic state every 11\clock \ with $99.9\%$ success. To ensure that the tile of the distillation block that is occupied by qubit 5 is not blocked during the first time step of the distillation protocol, the first $\pi/8$ rotation of the protocol should be chosen such that it does not involve qubit 5, e.g., the fourth rotation of Fig.~\ref{fig:15to1circuit2}. In total, this minimal setup uses 164 tiles and performs a $T$ gate every 11\clock, i.e., finishes the computation in $11 \cdot 10^8$ time steps.

For $p=10^{-3}$, a minimal setup consists of a compact data block and a 116-to-12 distillation block, as shown in Fig.~\ref{fig:minimalsetup1}b. For the minimal setup, we do not use the larger and faster distillation block shown in Fig.~\ref{fig:116to12}, but instead a block in the spirit of the 15-to-1 block. This 116-to-12 distillation block uses  44 tiles and distills 12 magic states in 99\clock \ with $89\%$ success probability, i.e., on average one state every 9.27\clock. Because this distillation protocol outputs magic states in bursts, i.e., 12 at the same time, these states need to be stored before being consumed. Therefore, we introduce additional storage tiles (green tiles in Fig.~\ref{fig:minimalsetup1}b). Here, we choose the 12 output states to be qubits $6,8,10,\dots,26$ and 27. In the last step of the protocol these states are moved into the green space, where they are consumed by the data block one after the other. This minimal setup uses 153 tiles for the data block, 44 tiles for the distillation block and 13 tiles for storage. In total, it uses 210 tiles and finishes the computation in $9.27 \cdot 10^8$ time steps.

\begin{figure}[t!]
\centering
\def\svgwidth{\linewidth}
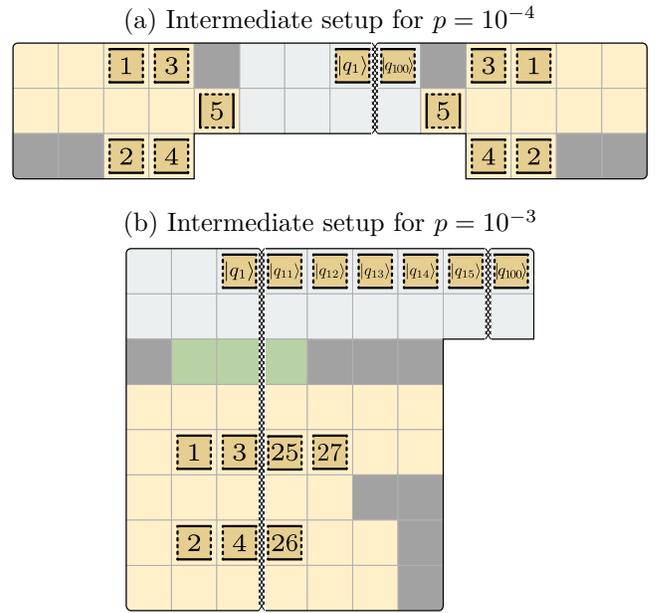
\caption{Intermediate setups using intermediate data blocks and two 15-to-1 distillation blocks for $p=10^{-4}$ or one compact 116-to-12 distillation block for $p=10^{-3}$.}
\label{fig:intermediatesetup}
\end{figure}

\begin{figure*}[t!]
\centering
\def\svgwidth{0.975\linewidth}
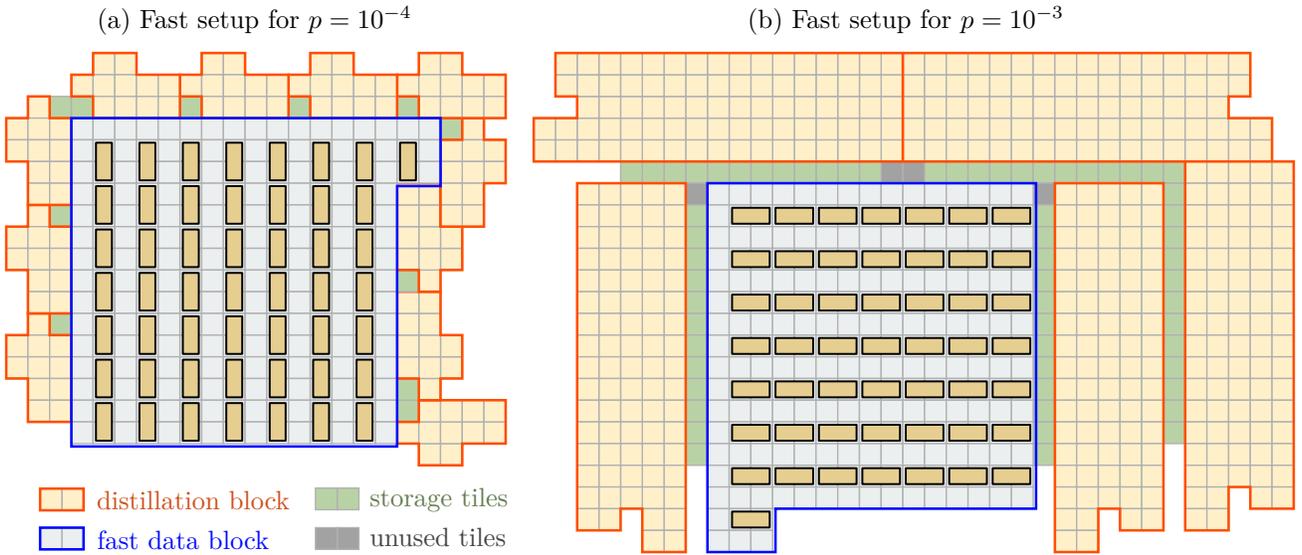
\caption{Fast setups using fast data blocks and 11 15-to-1 distillation blocks for $p=10^{-4}$ or 5 116-to-12 distillation block for $p=10^{-3}$.}
\label{fig:fastsetup}
\end{figure*}

\subsection{Step 3: Determine code distance}

Since each tile corresponds to $d \times d$ physical data qubits and each time step corresponds to $d$ code cycles, 164 encoded logical qubits need to survive for $(11 \cdot 10^8)d$ code cycles for the minimal setup with $p=10^{-4}$. The probability of a single logical error on any of these 164 qubits needs to stay below $1\%$ at the end of the computation. The logical error rate per logical qubit per code cycle can be approximated~\cite{Fowler2018} as
\begin{equation}
	p_L(p,d) = 0.1(100p)^{(d+1)/2} \, 
\end{equation}
for circuit-level noise. Therefore, the condition to determine the required code distance is
\begin{equation}
	164 \cdot 11 \cdot 10^8 \cdot d \cdot p_L(10^{-4},d) < 0.01 \, .
\end{equation}
For distance $d=11$, the final error probability is at $19.8\%$. Therefore, distance $d=13$ is sufficient, with a final error probability of $0.2\%$. The number of physical qubits used in the minimal setup can be calculated as the number of tiles multiplied by $2d^2$, taking measurement qubits into account. The minimal setup for $p=10^{-4}$ uses $164\cdot 2 \cdot 13^2 \approx 55{,}400$ physical qubits and finishes the computation in $13 \cdot 11 \cdot 10^8$ code cycles. With 1$~\mu\mathrm{s}$ per code cycle, this amounts to roughly 4~hours. 

For $p=10^{-3}$, the condition changes to
\begin{equation}
	210 \cdot 9.27 \cdot 10^8 \times d \cdot p_L(10^{-3},d) < 0.01 \, ,
\end{equation}
which is satisfied for $d=27$ with a final error probability of  $0.5\%$. The final error probability for $d=25$ is at $4.9\%$. Thus, the minimal setup uses $210\cdot 2 \cdot 27^2 \approx 306{,}000$ physical qubits and finishes the computation in $27 \cdot 9.27 \cdot 10^8$ code cycles, which amounts to roughly 7~hours.
Note that, in principle, a success probability of less than $50\%$ would be sufficient to reach arbitrary precisions by repeating computations or running them in parallel. This means that the code distances that we consider may be higher than what is necessary. 

\subsection{Step 4: Add distillation blocks}

Only a small fraction of the tiles of the minimal setup is used for magic state distillation, i.e., $6.7\%$ for $p=10^{-4}$ and 21\% for $p=10^{-3}$. On the other hand, adding one additional distillation block doubles the rate of magic state production, potentially doubling the speed of computation. Therefore, in order to speed up the computation and decrease the space-time cost, we add additional distillation blocks to our setup.

For $p=10^{-4}$, adding one more distillation block reduces the time that it takes to distill a magic state to 5.5\clock \ per state. However, the compact block can only consume magic states at 9\clock \ per state. In order to avoid this bottleneck, we can use the intermediate data block instead, which occupies 204 tiles, but consumes one magic state every 5\clock. With 22 tiles for distillation (see Fig.~\ref{fig:intermediatesetup}), this setup uses 226 tiles and finishes the computation after $5.5 \cdot 10^8$ time steps. This increases the number of qubits to 76,400, but reduces the computational time to 2 hours.

For $p=10^{-3}$, the addition of a distillation block reduces the distillation time to 4.64\clock. At this point, one should switch to the more efficient 116-to-12 block of Fig.~\ref{fig:116to12}, which uses 81 tiles and distills a magic state on average every 4.68\clock.
The intermediate data block cannot keep up with this distillation rate, but we can still use it to consume one magic state every 5\clock \ instead of 4.68\clock. Such a configuration uses 228 data tiles, 81 distillation tiles and 13 storage tiles, i.e., a total of 322 tiles corresponding to approximately 469,000 physical qubits. The computational time reduces to $5 \cdot 10^8$ time steps, i.e., 3.75 hours. Note that in Fig.~\ref{fig:intermediatesetup}b, the 12 output states of the 116-to-12 protocol should be chosen as $1,3,5,\dots,25$. They can be moved into the green storage space in the last step of the protocol, since the space denoted as \textit{ancilla 2} in Fig.~\ref{fig:116to12} is not being used in the last step.

\textbf{Trade-offs down to 1\clock \ per $T$ gate.}
Adding additional distillation blocks can reduce the time per $T$ gate down to 1\clock. For $p=10^{-4}$, 11 distillation blocks produce 1 magic state every 1\clock. To consume these magic states fast enough, we need to use a fast data block. This fast block uses 231 tiles and the 11 distillation blocks together with their storage tiles use $11*12=132$ tiles, as shown in Fig.~\ref{fig:fastsetup}a. With a total of 363 tiles, this setup uses 123,000 qubits and finishes the computation in $10^8$\clock, i.e., in 21 minutes and 40 seconds.

For $p=10^{-3}$, parallelizing 5 distillation blocks produces a magic state every 0.936\clock. This is faster than the fast block can consume the states, but allows for the execution of a $T$ gate every 1\clock. With 231 tiles for the fast block, 405 distillation tiles and 60 storage tiles, the total space cost is 696 tiles. The setup shown in Fig.~\ref{fig:116to12}b contains four unused tiles to make sure that all storage lines are connected to the data block. Storage lines need to be connected to the ancilla space of the data block either directly, via other storage lines or via unused tiles. In any case, this corresponds to roughly 1,020,000 physical qubits. The computation finishes after 45 minutes.

\textbf{Avoiding the classical overhead.} Every consumption of a magic state corresponds to a Pauli product measurement, the outcome of which determines whether a Clifford correction is required. This correction is commuted past the subsequent rotations, potentially changing the axis of rotation. Therefore, the computation cannot continue before the measurement outcome is determined. This involves a small classical computation to process the physical measurements (i.e., decoding and feed-forward), which could slow down the quantum computation. In order to avoid this, the magic state consumption can be performed using the auto-corrected $\pi/8$ rotations of Fig.~\ref{fig:15to1impl}b. Here, the classical computation merely determines, whether the ancilla qubit~--~which we refer to as the correction qubit $\ket{c}$~--~is measured in the $X$ or $Z$ basis. While this classical computation is running, the magic state for the subsequent $\pi/8$ rotation can be consumed, as the auto-corrected rotation involves no Clifford correction. This means that distillation blocks should output $\ket{m}-\ket{c}$ pairs, for which we construct modified distillation blocks in the following section. If the classical computation is, on average, faster than 1\clock \ (i.e., $d$ code cycles), then classical processing does not slow down the quantum computation in the $T$-count-limited schemes.

\textbf{Summary.}
Data blocks combined with distillation blocks can be used for large-scale quantum computing. The first step is to determine a sufficiently high-fidelity distillation protocol. Next, one constructs a minimal setup from a compact data block and a single distillation block to upper-bound the required code distance. Finally, one can trade off space against time by using fast data blocks and adding more distillation blocks. This can reduce the time per $T$ gate down to 1\clock. In our example, the trade-off also reduces the space-time cost compared to the minimal setup by a factor of 5 for $p=10^{-4}$ and by a factor of 2.8 for $p=10^{-3}$. In order to fully exploit the space-time trade-offs discussed in this section, the input circuit should be optimized for $T$ count.

\begin{figure}[b!]
\centering
\def\svgwidth{\linewidth}
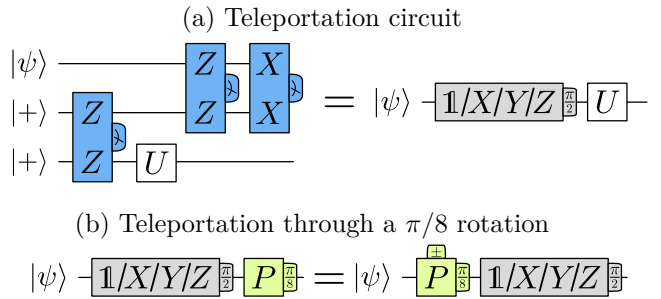
\caption{(a) Circuit for quantum teleportation of $\ket{\psi}$ through a gate $U$. Only if both Bell basis measurement yield +1, the teleported state is $U\ket{\psi}$. If $Z\otimes Z=-1$, the state is $UX\ket{\psi}$. If $X\otimes X=-1$, the state is $UZ \ket{\psi}$. If both measurements yield -1, the state is $UY\ket{\psi}$. (b) If $U$ is a $\pi/8$ rotation, the corrective Paulis change $P_{\pi/8}$ to $P_{-\pi/8}$. }
\label{fig:teleportationcircuit}
\end{figure}

\begin{figure*}[t!]
\centering
\def\svgwidth{\linewidth}
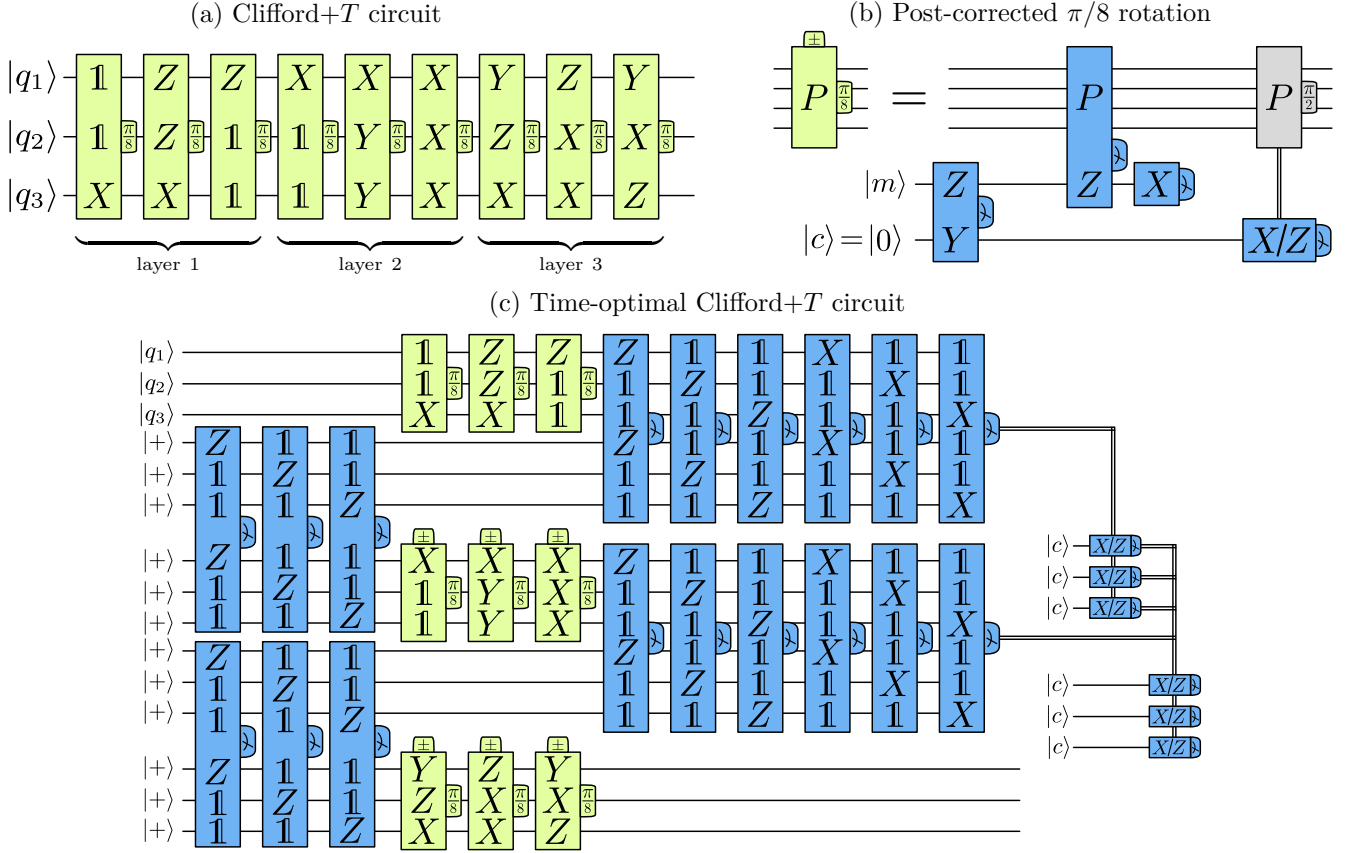
\caption{Time-optimal implementation of a three-qubit quantum computation consisting of 9 $T$ gates in 3 $T$ layers.  Post-corrected $\pi/8$ rotations (b) can be used to decide at a later point, whether the performed operation was a $P_{\pi/8}$ or a $P_{-\pi/8}$ rotation.}
\label{fig:timeoptimalcircuit}
\end{figure*}

\section{Trade-offs limited by $T$ depth}
\label{sec:timeoptimal}

In the previous section, we parallelized distillation blocks to finish computations in a time proportional to the $T$ count. In this section, we combine the previous constructions of data and distillation blocks to what we refer to as \textit{units}.
By parallelizing units, we exploit the fact that, in our example, the $10^8$ $T$ gates are arranged in $10^6$ layers of 100 $T$ gates to finish the computation in a time proportional to the $T$ depth. We first slightly increase the space-time cost compared to the previous section, in order to speed up the computation down to one measurement per $T$ layer. In this sense, we implement Fowler's time-optimal scheme~\cite{Fowler2012a}.

\subsection{$T$ layer parallelization}
The main concept used to parallelize $T$ layers is quantum teleportation. The teleportation circuit is shown in Fig.~\ref{fig:teleportationcircuit}a.  It starts with the generation of a Bell pair $(\ket{00} + \ket{11})/\sqrt{2}$ by the $Z \otimes Z$ measurement of $\ket{+} \otimes \ket{+}$. An arbitrary gate $U$ is performed on the second half of the Bell pair. Next, a qubit $\ket{\psi}$ and the first half of the Bell pair are measured in the Bell basis, i.e., in $X \otimes X$ and $Z \otimes Z$. After the measurement, the first two qubits are discarded and $\ket{\psi}$ is teleported to the third qubit through the gate $U$. This means that the output state is $U\ket{\psi}$, if the teleportation is successful. However, it is only successful, if both Bell basis measurements yield a +1 outcome. In the other three cases, the teleported state is $UX\ket{\psi}$, $UY\ket{\psi}$ or $UZ\ket{\psi}$. Note that the correction operation to recover the state $\ket{\psi}$ is not a Pauli operation $P$, but instead $UPU^\dagger$, which, in general, is as difficult to perform as $U$ itself.

If $U$ is a $P_{\pi/8}$ rotation, as in Fig.~\ref{fig:teleportationcircuit}b, the Pauli errors change $P_{\pi/8}$ to $P_{-\pi/8}$ up to a Pauli correction. Since it is only after the Bell basis measurement that we know, whether we should have performed a $P_{\pi/8}$ or a $P_{-\pi/8}$ gate, we use post-corrected $\pi/8$ rotations in Fig.~\ref{fig:timeoptimalcircuit}b, which are similar to the auto-corrected rotations of Fig.~\ref{fig:15to1impl}b. The post-corrected rotation uses a resource state consisting of two qubits, a magic state $\ket{m}$ and a second qubit that we refer to as a correction qubit $\ket{c}$. The resource state is generated by initializing $\ket{c}$ in $\ket{0}$ and measuring $Z \otimes Y$ between $\ket{m}$ and $\ket{c}$. In order to perform a post-corrected $\pi/8$ rotation, the resource state is consumed by measuring $P \otimes Z$ involving the magic state, and measuring $\ket{m}$ in $X$. The correction qubit $\ket{c}$ is stored for later use. It can be used at a later moment to decide, whether the rotation should have been a $+\pi/8$ or $-\pi/8$ rotation by measuring $\ket{c}$ either in the $Z$ or $X$ basis. Depending on the measurement outcome, a Pauli correction may be required.

\begin{figure*}[t!]
\centering
\def\svgwidth{0.9\linewidth}
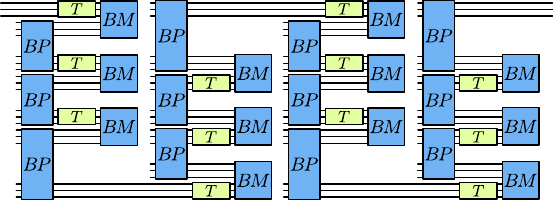
\caption{An example of a time-optimal circuit using four units. In this case, each unit consists of six qubits, i.e., it is a three-qubit quantum computation, where three $T$ layers can be executed simultaneously.}
\label{fig:timeoptimalcircuit2}
\end{figure*}

\textbf{The time-optimal circuit.} This can be used to execute multiple $T$ layers simultaneously. If $U$ is a product of mutually commuting $\pi/8$ rotations, i.e., a $T$ layer, the teleportation corrections replace all $\pi/8$ rotations with post-corrected rotations. An example is shown in Fig.~\ref{fig:timeoptimalcircuit} for a three-qubit computation of three $T$ layers, where all three $T$ layers are executed simultaneously. The reason why we can only group up $T$ gates that are part of the same layer is that otherwise the Pauli corrections of the post-corrected rotation would not commute with the other rotations. The time-optimal circuit consists of three steps: The preparation of Bell pairs for each $T$ layer, the application of $T$ gates, and a set of final Bell measurements. At this point, the computation is not finished, as we still need to measure the correction qubits of the post-corrected rotations. Because these involve potential Pauli corrections, the correction qubits of the different $T$ layers need to be measured one after the other. Thus, every $T$ layer is executed one after the other, where each execution requires the time that it takes to measure the correction qubits and perform the classical processing to determine the next set of measurements from the Pauli corrections. We refer to this time as $t_m$. In other words, any Clifford+$T$ circuit consisting of $n_L$ $T$ layers can be executed in $n_L \cdot t_m$, independent of the code distance, which is the main feature of the time-optimal scheme~\cite{Fowler2012a}.

The circuit in Fig.~\ref{fig:timeoptimalcircuit}c naively requires $2n \cdot n_L$ qubits for an $n$-qubit computation, which scales with the length of the computation. Since we only have a finite number of qubits at our disposal, our goal is to implement the circuit in Fig.~\ref{fig:timeoptimalcircuit2} instead. Here, the qubits form groups of $2n$ qubits. We refer to each of these groups as a \textit{unit}. Using $n_u$ units, $n_u-1$ layers of $T$ gates can be performed at the same time. In the circuit, the steps of Bell state preparation ($BP$), post-corrected $T$ layer execution ($T$) and Bell basis measurement ($BM$) are performed repeatedly until the end of the computation. We refer to the block of operations ($BP$-$T$-$BM$) as \textit{unit preparation}. Every time that unit preparation is finished, all qubits except for the correction qubits (not shown in Fig.~\ref{fig:timeoptimalcircuit2}) and half of the qubits of the last unit are discarded. At this point, the next set of unit preparations begins. Simultaneously, the correction qubits of the recently finished units are measured one after the other, which has a time cost of $(n_u-1) \cdot t_m$. This means that the number of units can be increased to speed up the computation, until $(n_u-1) \cdot t_m$ reaches the time that it takes to prepare a unit $t_u$. At this maximum number of units $n_{\rm max} = t_u/t_m+1$, a $T$ layer is executed every $t_m$ and the computation cannot be sped up any further in the Clifford+$T$ framework.

Note that the first and last unit differ from the other units. While all other units need to execute $n_T$ $T$ gates every $t_u$, the first and last unit need to execute $n_T$ $T$ gates only every $2t_u$, where $n_T$ is the number of $T$ gates per layer. Furthermore, the other blocks need to be able to store up to $2n_T$ correction qubits, since, after the end of a unit preparation, $n_T$ correction qubits are stored, and may need to remain stored until the end of the next unit preparation. For the first and last block, on the other hand, the required storage space is halved.

In the following, we will show how to prepare units in our framework. We find that, for our examples, unit preparation takes 113\clock. If $t_m = 1~\mu\mathrm{s}$, then $n_{\rm max}$ is ${\sim}1500$ for $p=10^{-4}$ and ${\sim}3000$ for $p=10^{-3}$. Independently of the error rate, the computational time drops to one second.

\begin{figure*}[t!]
\centering
\def\svgwidth{0.9\linewidth}
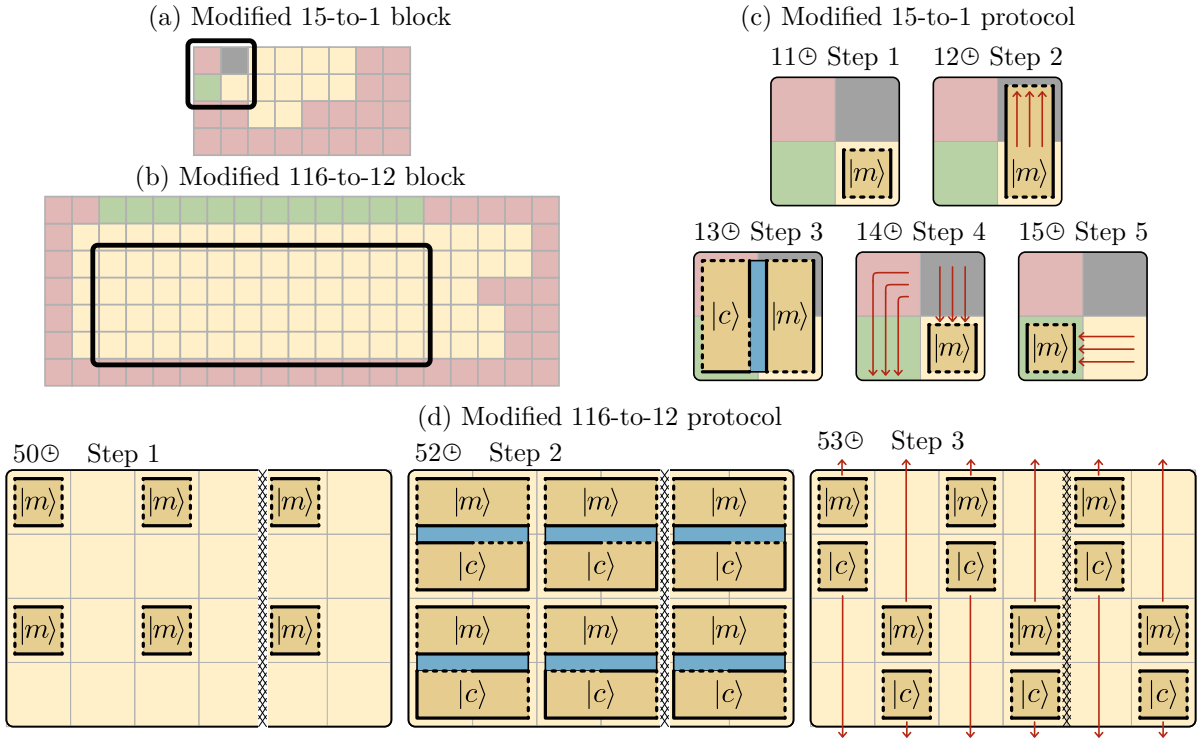
\caption{Modified 15-to-1 distillation blocks (a) output a $\ket{m}$-$\ket{c}$ pair every 11\clock. After the end of the distillation protocol, four additional steps (c) are necessary. The modified 116-to-12 distillation block (b) finishes after 53\clock, due to the three additional steps in (d).}
\label{fig:modifieddistillation}
\end{figure*}

\subsection{Units}

Units differ from the fast setups in Fig.~\ref{fig:fastsetup} in three aspects. First, the number of qubits stored in the data block is doubled. Secondly, the distillation protocols are modified to output $\ket{m}$-$\ket{c}$ pairs, instead of just magic states $\ket{m}$. Thirdly, in order to store correction qubits $\ket{c}$, additional space is required. Contrary to magic-state storage tiles, correction-qubit storage tiles do not need to be connected to the data block's ancilla region.

\textbf{Modified distillation blocks.} In order to have distillation blocks output $\ket{m}$-$\ket{c}$ pairs, extra tiles and operations are required. We show the necessary modifications for the example of 15-to-1 and 116-to-12 distillation. A modified 15-to-1 block is shown in Fig.~\ref{fig:modifieddistillation}a. Apart from the standard 11 distillation tiles (orange) and one magic-state storage tile (green), it also contains 19 correction-qubit storage tiles (purple) and an additional tile (gray) that is used for neither distillation nor storage. The additional steps that modify the protocol are shown in Fig.~\ref{fig:modifieddistillation}c, which zooms into the highlighted region of Fig.~\ref{fig:modifieddistillation}a. In step 1 of the shown protocol, the distillation has just finished after 11\clock. The patch of the output state is deformed in step 2, and an additional qubit $\ket{c}$ is initialized in the $\ket{0}$ state. The $Y\otimes Z$ operator between $\ket{c}$ and $\ket{m}$ is measured in step 3. In step 4, the correction qubit is sent to storage. Finally, in step 5, the magic state $\ket{m}$ is moved to its storage tile. This operation blocks one of the orange tiles that is used for the distillation protocol for 4\clock. Still, this does not slow down 15-to-1 distillation, since the first 4 rotation of the protocol in Fig.~\ref{fig:15to1circuit2} can be chosen, such that the output qubit is not needed. Therefore, the modified distillation block outputs one $\ket{m}$-$\ket{c}$ pair every 11\clock.

For 116-to-12 distillation, a modified block is shown in Fig.~\ref{fig:modifieddistillation}b. We arrange the qubits, such that the 12 output states are found in the positions shown in step 1 of Fig.~\ref{fig:modifieddistillation}d. Using 2\clock, correction qubits are prepared and $Y \otimes Z$ operators are measured. Finally, the patches are deformed back to square patches and all magic states are sent to the green storage, while all correction qubits are sent to the purple storage. This adds 3\clock \ to the protocol, meaning that this block outputs 12 $\ket{m}$-$\ket{c}$ pairs every 53\clock \ with a success probability of $(1-p)^{116}$. For $p=10^{-3}$, this corresponds to one output every 4.96\clock.

As mentioned in Sec.~\ref{sec:tcount}, modified distillation blocks can also be used with setups, in which $T$ gates are performed one after the other, in order to deal with slow classical processing. In this case, only one correction qubit storage tile per magic state is required.

\begin{figure}[b!]
\centering
\def\svgwidth{\linewidth}
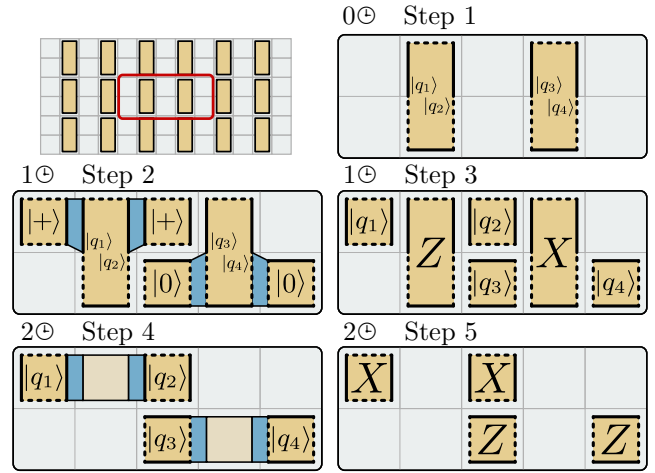
\caption{Bell basis measurement ($BM$) in 2\clock.}
\label{fig:bellmeasurement}
\end{figure}

\begin{figure*}[t!]
\centering
\def\svgwidth{0.97\linewidth}
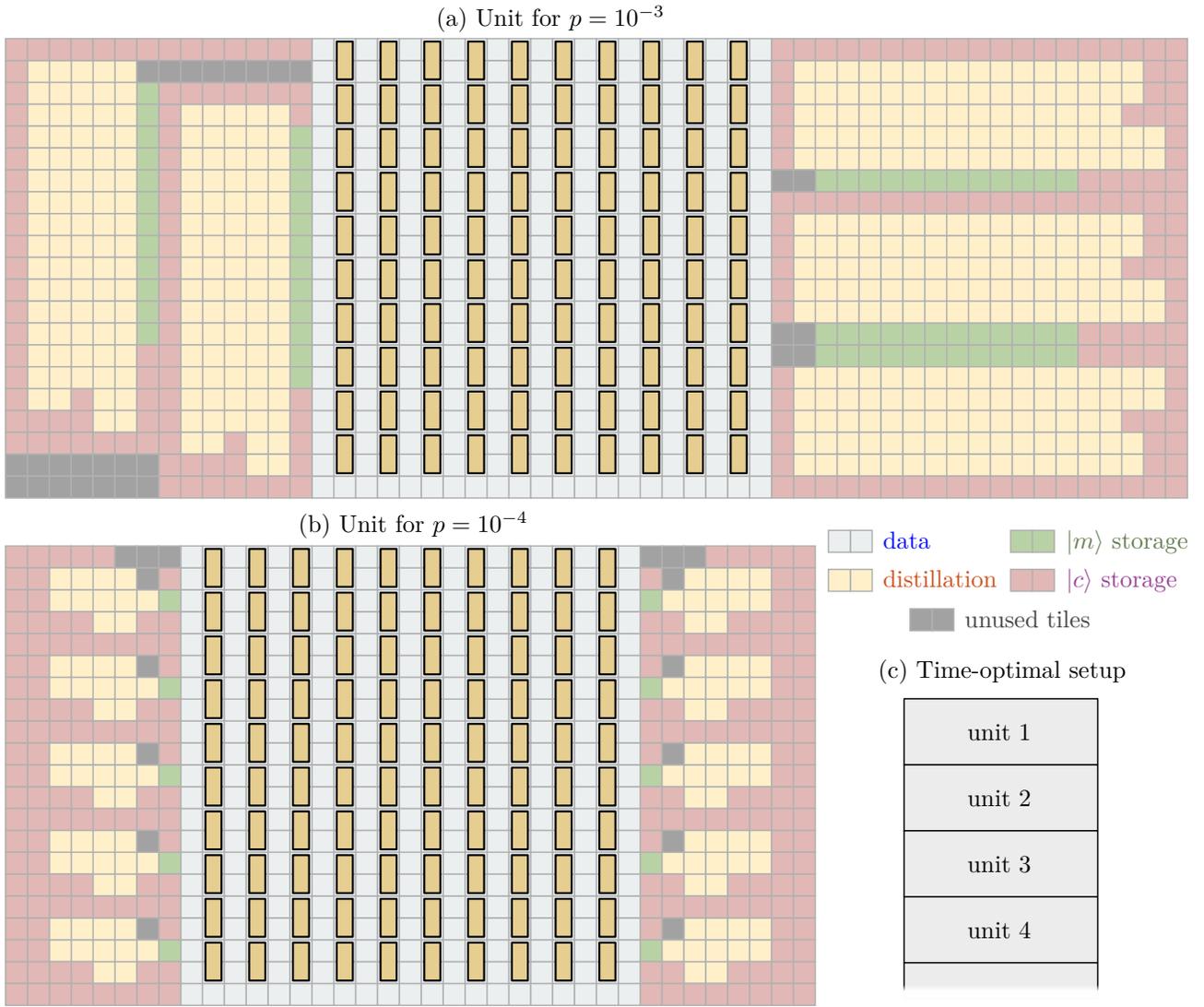
\caption{Units consist of fast data blocks, modified distillation blocks and storage tiles. (a) The unit for $p=10^{-3}$ consists of $54 \times 21 = 1134$ tiles. (b) For $p=10^{-4}$, the number of tiles is $37 \times 21 = 777$. (c) A time-optimal setup consists of a row of multiple units, which means that the space to the bottom and top of the fast data blocks needs to remain free.}
\label{fig:units}
\end{figure*}

\textbf{Units.} Modified distillation blocks together with fast data blocks are what we refer to as units. The units for our example computation for $p=10^{-3}$ and $p=10^{-4}$ are shown in Fig.~\ref{fig:units}a-b. They both consist of a 200-qubit fast data block, 200 correction-qubit storage tiles, and a number of distillation blocks. Since we will show that unit preparation takes 113\clock \ in our case, the number of distillation blocks is chosen such that at least 100 $\ket{m}$-$\ket{c}$ pairs can be distilled in 113\clock. A full time-optimal quantum computer consists of a row of multiple units, see Fig.~\ref{fig:units}c. The units shown in the figure contain some unused tiles. This gives the units a rectangular profiles, even though this is not necessarily required. In our case, the units have a footprint of $54 \times 21$ and $37 \times 21$ tiles, respectively. Note that the first and last unit of a time-optimal setup are smaller, as they only require 100 correction-qubit storage tiles and half the number of distillation blocks.

\textbf{Unit preparation.} In order to implement the time-optimal circuit of Fig.~\ref{fig:timeoptimalcircuit2} with the setup of Fig.~\ref{fig:units}, we show protocols that can be used for the $BP$-$T$-$BM$ operations. The data blocks of every unit store $2n$ qubits in $n$ two-qubit patches. We arrange the qubits in such a way that the the final Bell measurements ($BM$) are $Z\otimes Z$ and $X \otimes X$ measurements of the two qubits of every two-qubit patch. This Bell measurement can be done in 2\clock, as shown in Fig.~\ref{fig:bellmeasurement}.

\begin{figure}[t!]
\centering
\def\svgwidth{\linewidth}
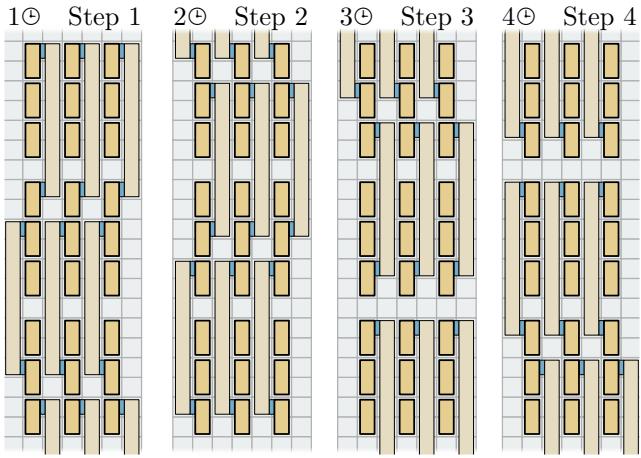
\caption{Bell state preparation ($BP$) for a 9-qubit computation (18 qubits per unit) in 4\clock. All two-qubit patches are initialized in the $\ket{+}^{\otimes 2}$ state. Each measurement ancilla is used for a $Z\otimes Z$ measurement between two qubits in different units. For $n$-qubit computations, this requires $\sqrt{n}+1$ time steps.}
\label{fig:bellprep}
\end{figure}

This arrangement of qubits implies that, for every two-qubit patch, one of the qubits needs to be part of a Bell state preparation ($BP$) with the neighboring unit to the top, and the other with a neighboring unit to the bottom. For an $n$-qubit quantum computation,  this Bell state preparation can be performed in $\sqrt{n}+1$ time steps, as we show in Fig.~\ref{fig:bellprep} for the example of $n=9$. For this, every qubit is initialized in the $\ket{+}$ state. The Bell state preparation requires a series of $Z\otimes Z$ measurements. The protocol in Fig.~\ref{fig:bellprep} shows that, since an $n$-qubit computation implies that the number of rows of the data block is $\sqrt{n}$, these measurements require a total of $\sqrt{n}+1$ time steps.

In total, the unit preparation of an $n$-qubit computation with $n_T$ $T$ gates per layer requires $\sqrt{n}+1$ time steps for the Bell state preparation, $n_T$ time steps for the execution of the $T$ layer, and 2 time steps for the Bell basis measurement, i.e., a total of $n_T+\sqrt{n}+3$ time steps. In our example, this amounts to 113\clock, which corresponds to $t_u = 1469~\mu\mathrm{s}$ for $p=10^{-4}$ and $t_u = 3051~\mu\mathrm{s}$ for $p=10^{-3}$. Thus, time optimality is reached with 1470 units for $p=10^{-4}$ and 3052 units for $p=10^{-3}$.

\textbf{Space-time trade-offs.} Of course, it is also possible to use fewer units than required for time optimality. Using $n_u$ units means that $n_T \cdot (n_u -1)$ $T$ gates are performed every $t_u$. In our example, $100 \cdot (n_u -1)$ $T$ gates are performed every 113\clock. With three units, the computational time drops to 56.5\% of the computational time of the fast setup in Fig.~\ref{fig:fastsetup}. With ten units, it drops to 11\%. The number of qubits per unit is ${\sim}$260,000 for $p=10^{-4}$ and ${\sim}$1,650,000 for $p=10^{-3}$, so going from the fast setup to parallelized units is, initially, not a favorable space-time trade-off. Since the space-time cost has increased compared to the fast setup, it is also useful to check whether the code distance needs to be readjusted. If we use three units~--~ignoring that the first and last unit are, in principle, smaller~--~the space-time cost is still below the space-time cost of the minimal setup in both cases. Adding more units significantly improves the space-time cost. It is also a prescription to linearly speed up the quantum computer down to the time-optimal limit.

\begin{figure}[t!]
\centering
\def\svgwidth{0.99\linewidth}
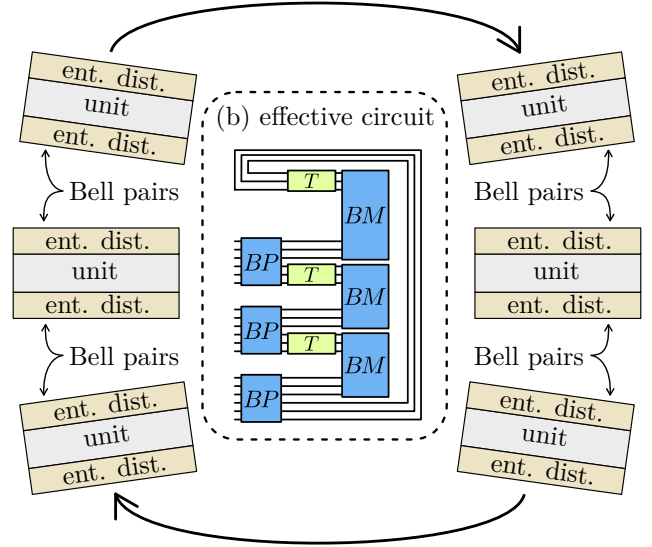
\caption{Scheme for distributed quantum computing in a circular arrangement of quantum computers with the ability to share Bell pairs between nearest neighbors. If the Bell-pair fidelity is low, entanglement distillation (ent.~dist.) can be used to increase the fidelity. This scheme effectively implements the circular time-optimal circuit drawn schematically in (b).}
\label{fig:distributed}
\end{figure}

\subsection{Distributed quantum computing}

Note that, apart from the initial sharing of entangled Bell pairs, the units operate entirely independently of each other. This implies that, if Bell pairs can be shared between different quantum computers, each unit can be located in a separate quantum computer. The shared Bell pairs do not even need to have a high fidelity, as software-based entanglement distillation~\cite{Bennett1996,Bennett1996a} can be used to convert a large number of low-fidelity Bell pairs into fewer high-fidelity Bell pairs. Recent experiments have made progress towards generating entanglement between different superconducting chips~\cite{Dickel2018,Campagne2018,Axline2018}.

For the time-optimal scheme, quantum computers may be arranged in a circle as shown in Fig.~\ref{fig:distributed}a, with the ability to share Bell pairs between neighboring quantum computers. This effectively implements the circuit that is schematically drawn in Fig.~\ref{fig:distributed}b. Note that in this circuit, there is no first and last unit. Here, every unit performs $n_T$ $\pi/8$ rotations every $t_u$. Therefore, time optimality is reached with one fewer unit, and each unit only needs to store $n_T$ correction qubits instead of $2n_T$. With only 100 correction-qubit storage tiles and ignoring the unused tiles, the qubit count of the units in Fig.~\ref{fig:units} drops to ${\sim}$220,000 for $p=10^{-4}$ and ${\sim}$1,470,000 for $p=10^{-3}$, which are the numbers that we report in Fig.~\ref{fig:overview}. Thus, if nearest-neighbor communication between quantum computers is feasible, already fewer than 2 million physical qubits per quantum computer can be used to implement the full time-optimal scheme with 1500-3000 quantum computers.

Entanglement distillation increases the qubit count. Note that it does not slow down the computation, as Bell pairs do not need to be distilled instantly. Entanglement distillation can take up to $t_u$ to distill the $n_T$ Bell pairs required per entanglement distillation block.

\textbf{Summary.}
In order to speed up an $n$-qubit quantum computation beyond 1\clock \ per $T$ gate, we parallelize $T$ layers using units. With an average of $n_T$ $T$ gates per layer, a unit consist of $4n+4\sqrt{n}+1$ tiles for the data block, 2$n_T$ storage tiles for the correction qubits, and enough distillation blocks to distill $n_T$ $\ket{m}$-$\ket{c}$ pairs in the time it takes to prepare a unit, which is $n_T + \sqrt{n}+3$ time steps. If the unit preparation time is $t_u$ and the time for single-qubit measurements and classical processing is $t_m$, a time-optimal setup consists of $t_u/t_m+1$ units, executing one $T$ layer every $t_m$. Using fewer units results in a linear space-time trade-off. With $n_u$ units, $n_T \cdot (n_u-1)$ $T$ gates are performed in $t_u$. A circular arrangement of units can be used for distributed quantum computing. This also reduces the number of correction-qubit storage tiles to $1n_T$ and the number of units in a time-optimal setup to $t_u/t_m$. In order to fully exploit the space-time trade-offs discussed in this section, the input circuit should be optimized for $T$ depth.

\section{Trade-offs beyond Clifford+$T$}
\label{sec:beyond}

Under the assumption that measurements and feed-forward can be done in 1~$\mu\mathrm{s}$, we described how to perform a $10^8$-$T$-gate computation in just 1 second. A more conservative assumption would be a measurement and feed-forward time of $10~\mu\mathrm{s}$, which increases the computation time to 10 seconds. Although this seems fast, many quantum computations have $T$ counts that are significantly higher than $10^8$. While the $T$ count of Hubbard model simulations~\cite{Babbush2018} is indeed in this range, quantum chemistry simulations can be more demanding. In particular, the simulation of FeMoco~\cite{Reiher2017}, a structure that plays an important role in nitrogen fixation, can have a $T$ count of up to $10^{15}$. With a serial execution of one $T$ gate every $10~\mu\mathrm{s}$, the computation takes 317 years to finish. Even if the gates are grouped into 100 $T$ gates per layer, the computation still takes over 3 years.

While Clifford+$T$ is a gate set that is very well suited for surface codes, it is often not the gate set which is natural to the quantum computations in question. In particular, quantum simulation based on Trotterization consists of many small-angle rotations. In the Clifford+$T$ framework, each small-angle rotation is translated into a series of $T$ gates via gate synthesis. Depending on the desired precision, this can require ${\sim}100$ $T$ gates for each rotation~\cite{Ross2014}, which must be executed in series. In order to speed up computations beyond their $T$ count or $T$ depth, it is therefore constructive to consider additional resources for gates other than $T$ gates.

\begin{figure}[t!]
\centering
\def\svgwidth{\linewidth}
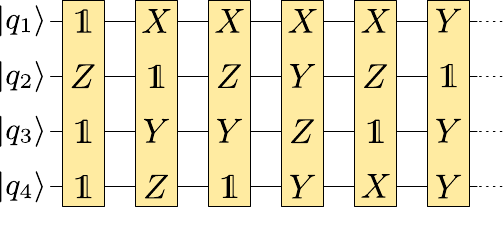
\caption{Clifford+$\varphi$ circuit. The first two rotation layers ($\varphi$ layers) with three rotations per layer are shown.}
\label{fig:rotationlayers}
\end{figure}

\subsection{Clifford+$\varphi$ circuits}

Instead of requiring an input circuit that consists of Clifford gates and $\pi/8$ rotations, we consider circuits that consist of Clifford gates and arbitrary $\varphi$ rotations, which we call Clifford+$\varphi$ circuits. Using the procedure in Sec.~\ref{sec:circuits}, Clifford gates can be commuted to the end of the circuit, such that we end up with a circuit like the one in Fig.~\ref{fig:rotationlayers}. Rotations that mutually commute can be grouped up into layers. The algorithm of Sec.~\ref{sec:circuits} can be used to reduce the number of layers. It can even reduce the number of rotations, since, if two rotations $P_{\varphi_1}$ and $P_{\varphi_2}$ with the same axis of rotation are moved into the same layer, they can be combined into a single rotation $P_{\varphi_1+\varphi_2}$. Clifford+$\varphi$ circuits are characterized by their \textit{rotation count} (or $\varphi$ count) and \textit{rotation depth} (or $\varphi$ depth), rather than $T$ count and $T$ depth.

Each $\varphi$ rotation can be performed using a $\ket{\varphi} = \ket{0} + e^{i(2\varphi)}\ket{1}$ resource state. When this state is consumed to perform a $P_{\varphi}$ rotation, there is a 50\% chance that a $P_{-\varphi}$ rotation is performed instead. For $\pi/8$ rotations, this is not very problematic, since the correction operation is a $\pi/4$ rotation, which can simply be commuted to the end of the circuit. For general $P_{-\varphi}$, the correction is a $P_{2\varphi}$ rotation, which requires the use of a $\ket{2\varphi}$ state. If this fails, the next correction is a $P_{4\varphi}$ rotation requiring a $\ket{4\varphi}$ state and so on. Thus, a wide variety of resource state is required to execute arbitrary-angle rotations. In the case of $\varphi = \pi/2^k$ for an integer $k$, $\ket{\varphi}$ states can be distilled using specialized protocols~\cite{Duclos2015,Campbell2016a}. For other angles, $\ket{\varphi}$ states can be approximated using $\ket{\pi/2^k}$ states, or pieced together from ordinary magic states $\ket{m}$ via circuit synthesis. Ordinary magic states can also generate states that can be used for $V$ gates~\cite{Harrow2002,DuclosCianci2013,Bocharov2013}, which are Pauli rotations with an angle $\theta = \arccos(3/5)$.

\begin{figure}[t!]
\centering
\def\svgwidth{\linewidth}
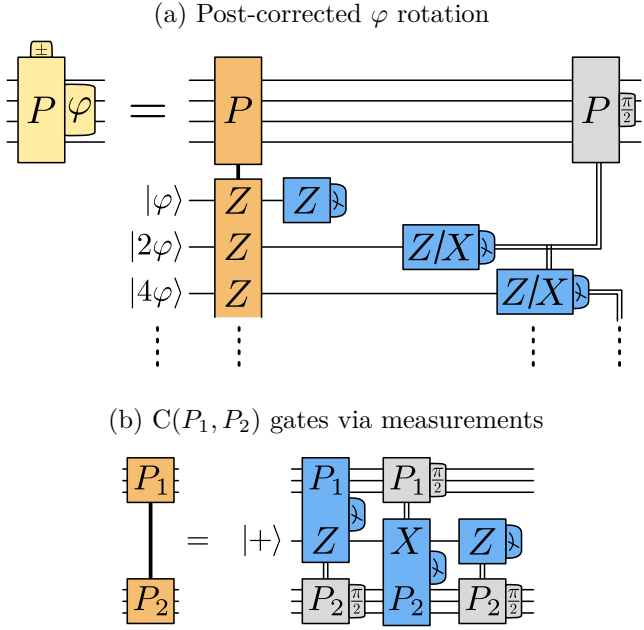
\caption{(a) A post-corrected $\varphi$ rotation can be used to decide at a later point, whether the performed operation was a $P_{\varphi}$ or a $P_{-\varphi}$ gate. (b) A $\mathrm{C}(P_1,P_2)$ gate can be performed explicitly using a $\ket{+}$ ancilla and Pauli product measurements.}
\label{fig:phiconsumption}
\end{figure}

All the schemes discussed in this work can be used with Clifford+$\varphi$ circuits by replacing magic state distillation blocks by distillation blocks that produce resource states for arbitrary-angle rotations. In order to consume these states in a systematic way similar to the post-corrected $\pi/8$ rotations in Fig.~\ref{fig:timeoptimalcircuit}b, we can use the post-corrected version of $\varphi$ rotations shown in Fig.~\ref{fig:phiconsumption}. First, the $n$ resource states are entangled with the data qubits via a $\mathrm{C}(P,Z^{\otimes n})$ gate. Just like magic state consumption, this can be done every 1\clock, since the data qubits are only part of one measurement in the measurement circuit in Fig.~\ref{fig:phiconsumption}b. Next, the $\ket{\varphi}$ state is measured in $Z$. If the outcome of this measurement is +1, then the rotation is successful and all other resource states are discarded by measuring them in $X$. If, instead, the outcome is -1, the $\ket{2\varphi}$ state is measured in $Z$. If the outcome of this $Z$ measurement is +1, the correction is successful, and the remaining resource states are discarded by $X$ measurements. For -1, the corrections continue with a $Z$ measurement of $\ket{4\varphi}$. Note that, in most cases, this cascade of measurements finishes in the second step. Therefore, on average, it takes $2t_m$ to perform these measurements. However, sufficiently many resource state are required in order to be prepared for the most unlikely situations, in which many measurement steps are required. The probability to require $n$ measurement steps (i.e., $n$ resource states down to $\ket{2^n\varphi}$) is exponentially low, $2^{-n}$. Therefore, the number of resource states that need to be generated for each $\varphi$ rotation scales logarithmically with the rotation count of the circuit, if one wants to stay below a certain probability that any of these rotations is slowed down by a missing resource state. If $\ket{\pi/2^k}$ states are used, the cascade of measurements terminates after $k$ steps. This technique of cascading resource state measurements is also referred to as programmable ancilla rotations~\cite{Jones2012}. Note that the cascade of measurements can also be postponed to a later point, such that the post-corrected $\varphi$~rotations can be used in the time-optimal scheme.

\begin{figure}[t!]
\centering
\def\svgwidth{\linewidth}
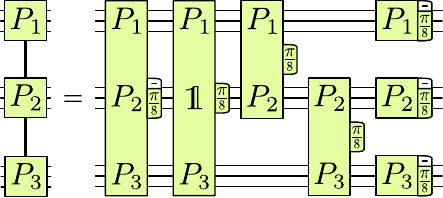
\caption{$C\mathrm(P_1,P_2,P_3)$ gate in terms of seven $\pi/8$ rotations.}
\label{fig:ccc}
\end{figure}

Using the $T$-count-limited scheme of Sec.~\ref{sec:tcount}, we can execute a $\varphi$ rotation every 1\clock. For 100 $T$ gates per $\varphi$ rotation, this speeds up the computation by a factor of 100. Also, the time-optimal setting of Sec.~\ref{sec:timeoptimal} can be used with Clifford+$\varphi$ circuits. However, the execution of a $\varphi$ layer can take more than 2$t_m$, as the measurement cascades for all rotations in the layer need to terminate. For instance, for 100 rotations per layer, each layer execution takes, on average, 8$t_m$. For 100 $T$ gates per rotation, $\varphi$ layer parallelization reduces the computational time by a factor of 12.5 compared to $T$ layer parallelization, i.e., from over 3 years to 3 months. In the specific case of quantum chemistry simulations, their $T$ count can be reduced significantly by using more advanced algorithms~\cite{Low2016,Low2017,Babbush2018a}, which also profit from arbitrary-angle rotations. Thus, if distributed quantum computing is feasible, Clifford+$\varphi$ circuits such as the ones used for quantum chemistry can be executed with qubit counts per quantum computer not far above the numbers reported in Fig.~\ref{fig:overview}. The only difference to Clifford+$T$ units is that larger distillation blocks are required to produce and store the $\ket{\varphi}$ resource states.

\begin{figure*}[t!]
\centering
\def\svgwidth{0.88\linewidth}
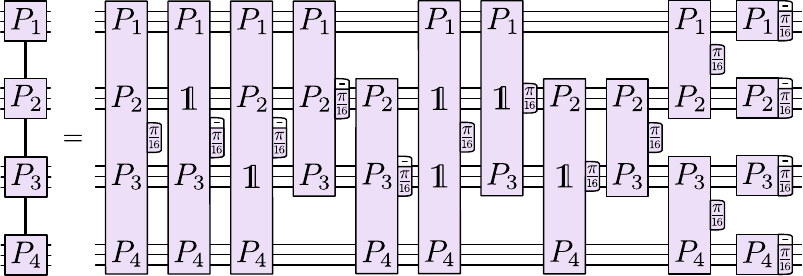
\caption{$C\mathrm(P_1,P_2,P_3,P_4)$ gate in terms of 15 $\pi/15$ rotations.}
\label{fig:cccc}
\end{figure*}

\textbf{Multi-controlled Pauli gates.} Other gates that are used extensively in quantum algorithms are multi-controlled Paulis, such as Toffoli or CCZ gates. In Fig.~\ref{fig:gateset}, we have shown how $\mathrm{C}(P_1,P_2)$ gates can be written in terms of $\pi/4$ rotations. A similar decomposition is possible for multi-controlled Pauli gates. In Fig.~\ref{fig:ccc}, we show how a $\mathrm{C}(P_1,P_2,P_3)$ gate is a product of 7 $\pi/8$ rotations. For instance, $\mathrm{C}(Z,Z,X)$ is the Toffoli gate. From the circuit, it is evident that the $T$ depth of $\mathrm{C}(P_1,P_2,P_3)$ gates is one~\cite{Selinger2013}. In principle, these doubly-controlled Pauli gates can be written with just four T gates~\cite{Jones2013}, but this increases the number of layers and a similar effect can be obtained by cancelling $\pi/8$ rotations from pairs of doubly-controlled gates in a circuit. Reducing the $T$ count by increasing the circuit depth~\cite{Gidney2018} can still be a useful circuit manipulation for $T$-count-limited setups. We also note that the $T$ count can be reduced by combining gate synthesis and magic state distillation (\textit{synthillation})~\cite{Campbell2017,Ogorman2017}.

$\mathrm{C}(P_1,P_2,P_3,P_4)$ gates, i.e., triply-controlled Pauli gates, can be written as 15 $\pi/16$ rotations, as shown in Fig.~\ref{fig:cccc}. While the $T$ depth of this circuit is no longer 1, the \textit{rotation} depth is. In fact, any multi-controlled Pauli gate with $n$ controls can be constructed from $2^n-1$ $P_{\pi/2^n}$ rotations by following the pattern shown in Figs.~\ref{fig:gateset}, \ref{fig:ccc} and \ref{fig:cccc}. The rotation depth of all these gates is 1. Multi-controlled gates can also be pieced together from $\mathrm{C}(P_1,P_2,P_3)$ rotations, but this increases the circuit depth. By using small-angle rotations, any multi-controlled Pauli gate can be executed in one step.

\subsection{Shorter measurements}

If the bottleneck of slow classical processing can be overcome, then the only hardware-based restriction to the speed of quantum computation is the time it takes to measure a physical qubit. In the time-optimal scheme, the execution time of each rotation layer is governed by the measurement time. This measurement time only needs to be high, if the measurement fidelity is required to be sufficiently low. In order to speed up the computation, one can use shorter qubit measurements. This exponentially decreases the measurement fidelity. On the other hand, the measurement fidelity of encoded surface-code qubits increases exponentially with the number of qubits comprising the logical qubit. Thus, by using twice as many physical qubits to encode the measured logical qubit, the measurement time can be decreased by a factor of two, doubling the computational speed of the quantum computer. In fact, not all qubits need to use a higher code distance. Only the correction qubits that are measured to execute each rotation layer need to be larger, and only right before they are measured. The physical qubit measurement does not need to be a quantum non-demolition measurement, but can be a desctructive measurement. Ultimately, however, the speed of quantum computation is limited by the speed of classical computation. Exploring superconducting logic~\cite{Likharev1991} to speed up classical computation may be a viable route to speed up quantum computers.

\textbf{Summary.} All the schemes discussed in this paper can not only be used with Clifford+$T$ circuits, but also with Clifford+$\varphi$ circuits. The only difference is that more and different resource states are required. Their distillation and storage requires more space than ordinary magic state distillation, but their use can speed up the computation by several orders of magnitude.

\begin{figure*}[!t]
\centering
\def\svgwidth{\linewidth}
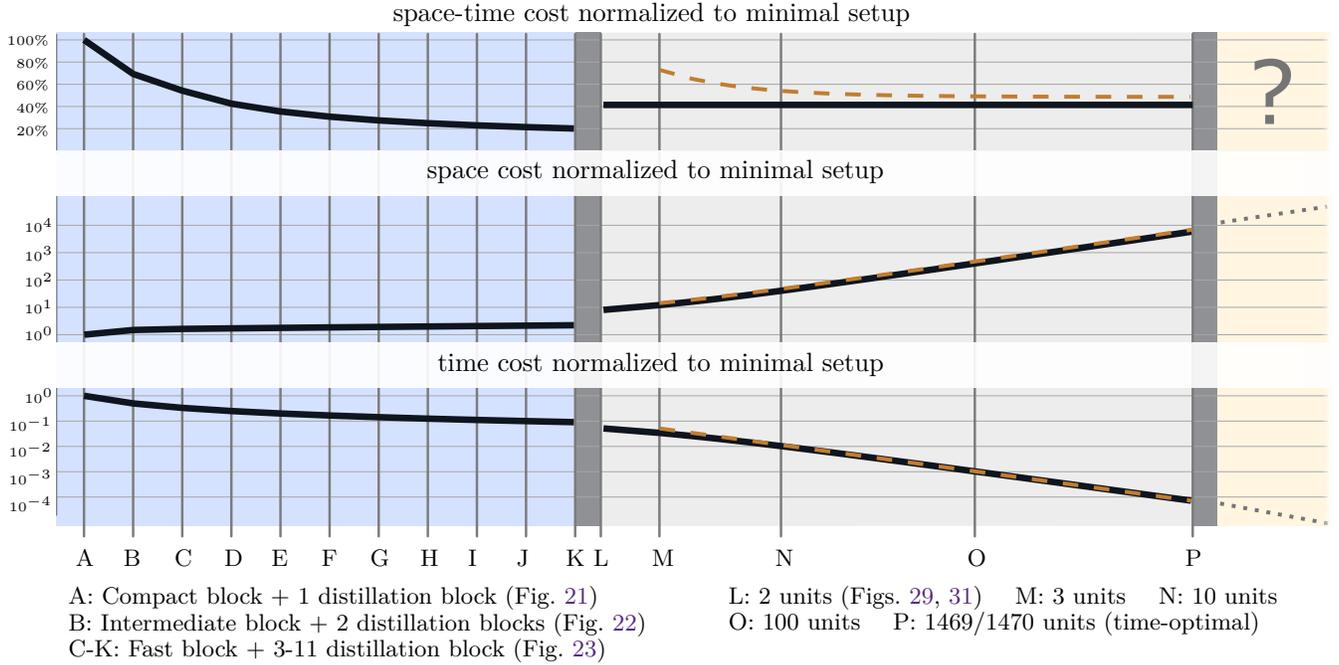
\caption{Space-time, space, and time cost of the schemes discussed in this paper for the example of a 100-qubit quantum computation with $T$ count $10^8$ and $T$ depth $10^6$, under the assumption of a 1~$\mu$s code cycle time, and a 1~$\mu$s measurement and classical processing time. The solid and dashed lines in M-P are for circular (solid) and linear (dashed) arrangements of units.}
\label{fig:stplot}
\end{figure*}

\section{Conclusion}
\label{sec:concl}

In this work, we described how full quantum computations can be performed in surface-code-based architectures of different sizes. Previous works on the translation of quantum computations into surface-code schemes~\cite{Paler2016b,Paler2017,Herr2017,Lao2018} attempted to optimize the logical qubit arrangement via algorithms that take a quantum circuit as an input. Here, we took a different approach by discussing computational schemes that do not require any prior knowledge about the input circuit. This has the advantage that a resource count with our schemes only requires the $T$ count and $T$ depth of the input circuit, and that the schemes consist of modular blocks that can be optimized independently of each other. In addition, the space-time cost is lower compared to earlier works~\cite{Fowler2013,Herr2017}.

\textbf{Big quantum computers are fast.} Starting from the minimal setup in Fig.~\ref{fig:minimalsetup1} that consists of a compact data block and a single distillation block, we traded off space versus time, increasing the size of the quantum computer and, in return, decreasing the computational time. For the example of a computation with a $T$ count of $10^8$ and a $T$ depth of $10^6$ with an error rate of $p=10^{-4}$, the minimal setup consists of 164 tiles and executes one $T$ gate every 11\clock, corresponding to a computational time of 4 hours with 55,400 physical qubits. From here, the space-time cost is drastically reduced by adding more distillation blocks, as shown in Fig.~\ref{fig:stplot} and Tab.~\ref{tab:conclusion}. With this strategy, the computational time is reduced to 1\clock \ per $T$ gate, where the computational cost of a circuit is governed by its $T$ count.

For further space-time trade-offs, we parallelized $T$ layers using units. This is an increase in space-time cost, especially for linear arrangements of units (dashed line in Fig.~\ref{fig:stplot}), but enables further space-time trade-offs. Linearly trading off space versus time, the computational time can be reduced to one measurement per $T$ layer. Units are well-suited for distributed quantum computing, as the sharing of Bell pairs between neighboring units is part of the parallelization scheme.

This exhausts the space-time trade-offs that are possible within the Clifford+$T$ framework. Switching to Clifford+$\varphi$ circuits can provide further trade-offs, as additional resources are introduced for arbitrary-angle rotations. This can be used to execute circuits in a time proportional to their rotation depth, as opposed to their $T$ depth. We have not investigated how this trade-off affects the space-time cost in our scheme.

\begin{table*}[t]
\centering
\scalebox{0.97}{
\begin{tabular}{c|ccc|ccc}
scheme & A & B & C-K & L & M & N - P \\
\hline
physical qubits & 55,400 & 76,400 & 90,200 - 123,000& 447,000  & \begin{tabular}{c}
679,000 \\[-0.6em]
(788,000) \end{tabular} &\begin{tabular}{c}
2,230,000 - 328,000,000 \\[-0.6em]
(2,630,000 - 386,000,000) \end{tabular}  \\
\hline
computational time & 4 h & 2 h & 79-22 min & 12 min & \begin{tabular}{c}
490 sec \\[-0.6em]
(734 sec) \end{tabular} & \begin{tabular}{c}
147 sec - 1 sec \\[-0.6em]
(163 sec - 1 sec) \end{tabular}
\end{tabular}}
\caption{Space and time cost of the schemes plotted in Fig.~\ref{fig:stplot}. The number in parentheses are for linear arrangements of units (dashed lines in Fig.~\ref{fig:stplot}).}\label{tab:conclusion}
\end{table*}

\textbf{Room for optimization.} In our $T$-count-limited schemes and for the preparation of units, one $T$ gate is performed after the other. If the input circuit is known, it is reasonable to assume that qubits can be arranged in a way that allows for the parallel execution of multiple $T$ gates in the same data block. Furthermore, there is a strict separation between tiles used for magic state distillation and tiles used for data blocks in our schemes. By sharing tiles between blocks, the space overhead may be reduced. Moreover, we have only considered a handful of distillation protocols. It would be interesting to see which distillation protocols can be used to optimize the cost function of Eq.~\eqref{eqn:distcost}. Finally, concrete tile layouts that can be used to distill and consume the additional resources necessary for Clifford+$\varphi$ computing are still missing.

\textbf{Beyond surface codes.} Even though we designed our schemes with surface codes in mind, they can, in principle, be applied to other toric-code-based patches, such as Majorana surface-code patches~\cite{Litinski2017b} or color-code patches~\cite{Bombin2006,Landahl2014,Kesselring2018}. Color codes can reduce the number of physical qubits due to more compact encoding, but require more elaborate hardware to measure the higher-weight check operators. The space cost is reduced by replacing all surface-code patches by color-code patches, with the exception of Pauli product measurement ancillas. In order to keep the space cost low, measurement ancillas should remain surface-code patches and color-to-surface code lattice surgery~\cite{Nautrup2016} should be used during the Pauli product measurement protocol, as described in Ref.~\cite{Litinski2017a}.

\textbf{Outlook.}
If the number of qubits continues to double every 8 months~\cite{QubitDouble}, the
60,000 - 300,000 physical qubits necessary for classically intractable Hubbard model simulations with a $T$ count of $10^8$ will be available in 7-9 years, assuming qubit quality improves accordingly. If multiple quantum computers can be connected in a network, time-optimal quantum computing becomes available shortly thereafter, facilitating the implementation of more difficult algorithms such as quantum chemistry simulations or Shor's algorithm. Classical processing in terms of measurements, feed-forward and decoding is expected to be a significant roadblock in speeding up quantum computers. Ultimately, faster classical control hardware will be necessary to build faster quantum computers. I hope that the schemes discussed in this work are a useful roadmap towards large-scale quantum computing, and that the patch-based framework is a valuable toolbox for constructions of surface-code-based implementations of quantum algorithms.

\section*{Acknowledgments}
This work would not have been possible without insightful discussion with Austin Fowler and Craig Gidney about Pauli product measurements and 15-to-1 distillation, with Jens Eisert, Markus Kesselring and Felix von Oppen about Clifford tracking and space-time trade-offs, with Jeongwan Haah and Matthew Hastings about magic state distillation, with Guang Hao Low and Nathan Wiebe about quantum simulation algorithms, and with Ali Lavasani about few-qubit surface-code architectures. This work has been supported by the Deutsche Forschungsgemeinschaft (Bonn) within the network CRC TR 183.

\bibliographystyle{apsrev4-1mod}
\bibliography{biblio}

\begin{thebibliography}{65}%
\makeatletter
\providecommand \@ifxundefined [1]{%
 \@ifx{#1\undefined}
}%
\providecommand \@ifnum [1]{%
 \ifnum #1\expandafter \@firstoftwo
 \else \expandafter \@secondoftwo
 \fi
}%
\providecommand \@ifx [1]{%
 \ifx #1\expandafter \@firstoftwo
 \else \expandafter \@secondoftwo
 \fi
}%
\providecommand \natexlab [1]{#1}%
\providecommand \enquote  [1]{``#1''}%
\providecommand \bibnamefont  [1]{#1}%
\providecommand \bibfnamefont [1]{#1}%
\providecommand \citenamefont [1]{#1}%
\providecommand \href@noop [0]{\@secondoftwo}%
\providecommand \href [0]{\begingroup \@sanitize@url \@href}%
\providecommand \@href[1]{\@@startlink{#1}\@@href}%
\providecommand \@@href[1]{\endgroup#1\@@endlink}%
\providecommand \@sanitize@url [0]{\catcode `\\12\catcode `\$12\catcode
  `\&12\catcode `\#12\catcode `\^12\catcode `\_12\catcode `\%12\relax}%
\providecommand \@@startlink[1]{}%
\providecommand \@@endlink[0]{}%
\providecommand \url  [0]{\begingroup\@sanitize@url \@url }%
\providecommand \@url [1]{\endgroup\@href {#1}{\urlprefix }}%
\providecommand \urlprefix  [0]{URL }%
\providecommand \Eprint [0]{\href }%
\providecommand \doibasemod [0]{http://dx.doi.org/}%
\providecommand \selectlanguage [0]{\@gobble}%
\providecommand \bibinfo  [0]{\@secondoftwo}%
\providecommand \bibfield  [0]{\@secondoftwo}%
\providecommand \translation [1]{[#1]}%
\providecommand \BibitemOpen [0]{}%
\providecommand \bibitemStop [0]{}%
\providecommand \bibitemNoStop [0]{.\EOS\space}%
\providecommand \EOS [0]{\spacefactor3000\relax}%
\providecommand \BibitemShut  [1]{\csname bibitem#1\endcsname}%
\let\auto@bib@innerbib\@empty
\bibitem [{\citenamefont {Reiher}\ \emph {et~al.}(2017)\citenamefont {Reiher},
  \citenamefont {Wiebe}, \citenamefont {Svore}, \citenamefont {Wecker},\ and\
  \citenamefont {Troyer}}]{Reiher2017}%
  \BibitemOpen
  \bibfield  {author} {\bibinfo {author} {\bibfnamefont {M.}~\bibnamefont
  {Reiher}}, \bibinfo {author} {\bibfnamefont {N.}~\bibnamefont {Wiebe}},
  \bibinfo {author} {\bibfnamefont {K.~M.}\ \bibnamefont {Svore}}, \bibinfo
  {author} {\bibfnamefont {D.}~\bibnamefont {Wecker}}, \ and\ \bibinfo {author}
  {\bibfnamefont {M.}~\bibnamefont {Troyer}},\ }\bibfield  {title} {\emph
  {\bibinfo {title} {Elucidating reaction mechanisms on quantum computers},\
  }}\href {\doibasemod 10.1073/pnas.1619152114} {\bibfield  {journal} {\bibinfo
   {journal} {PNAS}\ }\textbf {\bibinfo {volume} {114}},\ \bibinfo {pages}
  {7555} (\bibinfo {year} {2017})}\BibitemShut {NoStop}%
\bibitem [{\citenamefont {Babbush}\ \emph
  {et~al.}(2018{\natexlab{a}})\citenamefont {Babbush}, \citenamefont {Gidney},
  \citenamefont {Berry}, \citenamefont {Wiebe}, \citenamefont {McClean},
  \citenamefont {Paler}, \citenamefont {Fowler},\ and\ \citenamefont
  {Neven}}]{Babbush2018}%
  \BibitemOpen
  \bibfield  {author} {\bibinfo {author} {\bibfnamefont {R.}~\bibnamefont
  {Babbush}}, \bibinfo {author} {\bibfnamefont {C.}~\bibnamefont {Gidney}},
  \bibinfo {author} {\bibfnamefont {D.~W.}\ \bibnamefont {Berry}}, \bibinfo
  {author} {\bibfnamefont {N.}~\bibnamefont {Wiebe}}, \bibinfo {author}
  {\bibfnamefont {J.}~\bibnamefont {McClean}}, \bibinfo {author} {\bibfnamefont
  {A.}~\bibnamefont {Paler}}, \bibinfo {author} {\bibfnamefont
  {A.}~\bibnamefont {Fowler}}, \ and\ \bibinfo {author} {\bibfnamefont
  {H.}~\bibnamefont {Neven}},\ }\bibfield  {title} {\emph {\bibinfo {title}
  {Encoding electronic spectra in quantum circuits with linear {T}
  complexity},\ }}\href {\doibasemod 10.1103/PhysRevX.8.041015} {\bibfield
  {journal} {\bibinfo  {journal} {Phys. Rev. X}\ }\textbf {\bibinfo {volume}
  {8}},\ \bibinfo {pages} {041015} (\bibinfo {year}
  {2018}{\natexlab{a}})}\BibitemShut {NoStop}%
\bibitem [{\citenamefont {Preskill}(1998)}]{Preskill1998}%
  \BibitemOpen
  \bibfield  {author} {\bibinfo {author} {\bibfnamefont {J.}~\bibnamefont
  {Preskill}},\ }\bibfield  {title} {\emph {\bibinfo {title} {Reliable quantum
  computers},\ }}\href {\doibasemod 10.1098/rspa.1998.0167} {\bibfield
  {journal} {\bibinfo  {journal} {Proc. Roy. Soc. Lond. A}\ }\textbf {\bibinfo
  {volume} {454}},\ \bibinfo {pages} {385} (\bibinfo {year}
  {1998})}\BibitemShut {NoStop}%
\bibitem [{\citenamefont {Terhal}(2015)}]{TerhalRMP}%
  \BibitemOpen
  \bibfield  {author} {\bibinfo {author} {\bibfnamefont {B.~M.}\ \bibnamefont
  {Terhal}},\ }\bibfield  {title} {\emph {\bibinfo {title} {Quantum error
  correction for quantum memories},\ }}\href {\doibasemod
  10.1103/RevModPhys.87.307} {\bibfield  {journal} {\bibinfo  {journal} {Rev.
  Mod. Phys.}\ }\textbf {\bibinfo {volume} {87}},\ \bibinfo {pages} {307}
  (\bibinfo {year} {2015})}\BibitemShut {NoStop}%
\bibitem [{\citenamefont {{Campbell}}\ \emph {et~al.}(2017)\citenamefont
  {{Campbell}}, \citenamefont {{Terhal}},\ and\ \citenamefont
  {{Vuillot}}}]{Campbell2016}%
  \BibitemOpen
  \bibfield  {author} {\bibinfo {author} {\bibfnamefont {E.~T.}\ \bibnamefont
  {{Campbell}}}, \bibinfo {author} {\bibfnamefont {B.~M.}\ \bibnamefont
  {{Terhal}}}, \ and\ \bibinfo {author} {\bibfnamefont {C.}~\bibnamefont
  {{Vuillot}}},\ }\bibfield  {title} {\emph {\bibinfo {title} {{Roads towards
  fault-tolerant universal quantum computation}},\ }}\href {\doibasemod
  10.1038/nature23460} {\bibfield  {journal} {\bibinfo  {journal} {Nature}\
  }\textbf {\bibinfo {volume} {549}},\ \bibinfo {pages} {172} (\bibinfo {year}
  {2017})}\BibitemShut {NoStop}%
\bibitem [{\citenamefont {Kitaev}(2003)}]{Kitaev2003}%
  \BibitemOpen
  \bibfield  {author} {\bibinfo {author} {\bibfnamefont {A.~Y.}\ \bibnamefont
  {Kitaev}},\ }\bibfield  {title} {\emph {\bibinfo {title} {{Fault-tolerant
  quantum computation by anyons}},\ }}\href {\doibasemod
  10.1016/S0003-4916(02)00018-0} {\bibfield  {journal} {\bibinfo  {journal}
  {Ann. Phys.}\ }\textbf {\bibinfo {volume} {303}},\ \bibinfo {pages} {2}
  (\bibinfo {year} {2003})}\BibitemShut {NoStop}%
\bibitem [{\citenamefont {Fowler}\ \emph {et~al.}(2012)\citenamefont {Fowler},
  \citenamefont {Mariantoni}, \citenamefont {Martinis},\ and\ \citenamefont
  {Cleland}}]{Fowler2012}%
  \BibitemOpen
  \bibfield  {author} {\bibinfo {author} {\bibfnamefont {A.~G.}\ \bibnamefont
  {Fowler}}, \bibinfo {author} {\bibfnamefont {M.}~\bibnamefont {Mariantoni}},
  \bibinfo {author} {\bibfnamefont {J.~M.}\ \bibnamefont {Martinis}}, \ and\
  \bibinfo {author} {\bibfnamefont {A.~N.}\ \bibnamefont {Cleland}},\
  }\bibfield  {title} {\emph {\bibinfo {title} {Surface codes: Towards
  practical large-scale quantum computation},\ }}\href {\doibasemod
  10.1103/PhysRevA.86.032324} {\bibfield  {journal} {\bibinfo  {journal} {Phys.
  Rev. A}\ }\textbf {\bibinfo {volume} {86}},\ \bibinfo {pages} {032324}
  (\bibinfo {year} {2012})}\BibitemShut {NoStop}%
\bibitem [{\citenamefont {Bombin}(2010)}]{Bombin2010}%
  \BibitemOpen
  \bibfield  {author} {\bibinfo {author} {\bibfnamefont {H.}~\bibnamefont
  {Bombin}},\ }\bibfield  {title} {\emph {\bibinfo {title} {Topological order
  with a twist: Ising anyons from an abelian model},\ }}\href {\doibasemod
  10.1103/PhysRevLett.105.030403} {\bibfield  {journal} {\bibinfo  {journal}
  {Phys. Rev. Lett.}\ }\textbf {\bibinfo {volume} {105}},\ \bibinfo {pages}
  {030403} (\bibinfo {year} {2010})}\BibitemShut {NoStop}%
\bibitem [{\citenamefont {Horsman}\ \emph {et~al.}(2012)\citenamefont
  {Horsman}, \citenamefont {Fowler}, \citenamefont {Devitt},\ and\
  \citenamefont {Meter}}]{Horsman2012}%
  \BibitemOpen
  \bibfield  {author} {\bibinfo {author} {\bibfnamefont {C.}~\bibnamefont
  {Horsman}}, \bibinfo {author} {\bibfnamefont {A.~G.}\ \bibnamefont {Fowler}},
  \bibinfo {author} {\bibfnamefont {S.}~\bibnamefont {Devitt}}, \ and\ \bibinfo
  {author} {\bibfnamefont {R.~V.}\ \bibnamefont {Meter}},\ }\bibfield  {title}
  {\emph {\bibinfo {title} {Surface code quantum computing by lattice
  surgery},\ }}\href {\doibasemod 10.1088/1367-2630/14/12/123011} {\bibfield
  {journal} {\bibinfo  {journal} {New J. Phys.}\ }\textbf {\bibinfo {volume}
  {14}},\ \bibinfo {pages} {123011} (\bibinfo {year} {2012})}\BibitemShut
  {NoStop}%
\bibitem [{\citenamefont {Brown}\ \emph {et~al.}(2017)\citenamefont {Brown},
  \citenamefont {Laubscher}, \citenamefont {Kesselring},\ and\ \citenamefont
  {Wootton}}]{Brown2017}%
  \BibitemOpen
  \bibfield  {author} {\bibinfo {author} {\bibfnamefont {B.~J.}\ \bibnamefont
  {Brown}}, \bibinfo {author} {\bibfnamefont {K.}~\bibnamefont {Laubscher}},
  \bibinfo {author} {\bibfnamefont {M.~S.}\ \bibnamefont {Kesselring}}, \ and\
  \bibinfo {author} {\bibfnamefont {J.~R.}\ \bibnamefont {Wootton}},\
  }\bibfield  {title} {\emph {\bibinfo {title} {Poking holes and cutting
  corners to achieve {C}lifford gates with the surface code},\ }}\href
  {\doibasemod 10.1103/PhysRevX.7.021029} {\bibfield  {journal} {\bibinfo
  {journal} {Phys. Rev. X}\ }\textbf {\bibinfo {volume} {7}},\ \bibinfo {pages}
  {021029} (\bibinfo {year} {2017})}\BibitemShut {NoStop}%
\bibitem [{\citenamefont {Litinski}\ and\ \citenamefont
  {Oppen}(2018)}]{Litinski2017b}%
  \BibitemOpen
  \bibfield  {author} {\bibinfo {author} {\bibfnamefont {D.}~\bibnamefont
  {Litinski}}\ and\ \bibinfo {author} {\bibfnamefont {F.~v.}\ \bibnamefont
  {Oppen}},\ }\bibfield  {title} {\emph {\bibinfo {title} {Lattice {S}urgery
  with a {T}wist: {S}implifying {C}lifford {G}ates of {S}urface {C}odes},\
  }}\href {\doibasemod 10.22331/q-2018-05-04-62} {\bibfield  {journal}
  {\bibinfo  {journal} {{Quantum}}\ }\textbf {\bibinfo {volume} {2}},\ \bibinfo
  {pages} {62} (\bibinfo {year} {2018})}\BibitemShut {NoStop}%
\bibitem [{\citenamefont {Fowler}\ and\ \citenamefont
  {Gidney}(2018)}]{Fowler2018}%
  \BibitemOpen
  \bibfield  {author} {\bibinfo {author} {\bibfnamefont {A.~G.}\ \bibnamefont
  {Fowler}}\ and\ \bibinfo {author} {\bibfnamefont {C.}~\bibnamefont
  {Gidney}},\ }\bibfield  {title} {\emph {\bibinfo {title} {Low overhead
  quantum computation using lattice surgery},\ }}\href
  {https://arxiv.org/abs/1808.06709} {\bibfield  {journal} {\bibinfo  {journal}
  {arXiv:1808.06709}\ } (\bibinfo {year} {2018})}\BibitemShut {NoStop}%
\bibitem [{\citenamefont {Landahl}\ and\ \citenamefont
  {Ryan-Anderson}(2014)}]{Landahl2014}%
  \BibitemOpen
  \bibfield  {author} {\bibinfo {author} {\bibfnamefont {A.~J.}\ \bibnamefont
  {Landahl}}\ and\ \bibinfo {author} {\bibfnamefont {C.}~\bibnamefont
  {Ryan-Anderson}},\ }\bibfield  {title} {\emph {\bibinfo {title} {Quantum
  computing by color-code lattice surgery},\ }}\href
  {https://arxiv.org/abs/1407.5103} {\bibfield  {journal} {\bibinfo  {journal}
  {arXiv:1407.5103}\ } (\bibinfo {year} {2014})}\BibitemShut {NoStop}%
\bibitem [{\citenamefont {Li}(2015)}]{Li2015}%
  \BibitemOpen
  \bibfield  {author} {\bibinfo {author} {\bibfnamefont {Y.}~\bibnamefont
  {Li}},\ }\bibfield  {title} {\emph {\bibinfo {title} {A magic state’s
  fidelity can be superior to the operations that created it},\ }}\href
  {\doibasemod 10.1088/1367-2630/17/2/023037} {\bibfield  {journal} {\bibinfo
  {journal} {New J. Phys.}\ }\textbf {\bibinfo {volume} {17}},\ \bibinfo
  {pages} {023037} (\bibinfo {year} {2015})}\BibitemShut {NoStop}%
\bibitem [{\citenamefont {Herr}\ \emph
  {et~al.}(2017{\natexlab{a}})\citenamefont {Herr}, \citenamefont {Nori},\ and\
  \citenamefont {Devitt}}]{Herr2017b}%
  \BibitemOpen
  \bibfield  {author} {\bibinfo {author} {\bibfnamefont {D.}~\bibnamefont
  {Herr}}, \bibinfo {author} {\bibfnamefont {F.}~\bibnamefont {Nori}}, \ and\
  \bibinfo {author} {\bibfnamefont {S.~J.}\ \bibnamefont {Devitt}},\ }\bibfield
   {title} {\emph {\bibinfo {title} {Optimization of lattice surgery is
  {NP}-hard},\ }}\href {\doibasemod 10.1038/s41534-017-0035-1} {\bibfield
  {journal} {\bibinfo  {journal} {npj Quant. Inf.}\ }\textbf {\bibinfo {volume}
  {3}},\ \bibinfo {pages} {35} (\bibinfo {year}
  {2017}{\natexlab{a}})}\BibitemShut {NoStop}%
\bibitem [{\citenamefont {Bravyi}\ and\ \citenamefont
  {Kitaev}(2005)}]{Bravyi2005}%
  \BibitemOpen
  \bibfield  {author} {\bibinfo {author} {\bibfnamefont {S.}~\bibnamefont
  {Bravyi}}\ and\ \bibinfo {author} {\bibfnamefont {A.}~\bibnamefont
  {Kitaev}},\ }\bibfield  {title} {\emph {\bibinfo {title} {{Universal quantum
  computation with ideal Clifford gates and noisy ancillas}},\ }}\href
  {\doibasemod 10.1103/PhysRevA.71.022316} {\bibfield  {journal} {\bibinfo
  {journal} {Phys. Rev. A}\ }\textbf {\bibinfo {volume} {71}},\ \bibinfo
  {pages} {022316} (\bibinfo {year} {2005})}\BibitemShut {NoStop}%
\bibitem [{\citenamefont {Haah}\ and\ \citenamefont
  {Hastings}(2018)}]{Haah2018}%
  \BibitemOpen
  \bibfield  {author} {\bibinfo {author} {\bibfnamefont {J.}~\bibnamefont
  {Haah}}\ and\ \bibinfo {author} {\bibfnamefont {M.~B.}\ \bibnamefont
  {Hastings}},\ }\bibfield  {title} {\emph {\bibinfo {title} {Codes and
  {P}rotocols for {D}istilling {$T$}, controlled-{$S$}, and {T}offoli
  {G}ates},\ }}\href {\doibasemod 10.22331/q-2018-06-07-71} {\bibfield
  {journal} {\bibinfo  {journal} {{Quantum}}\ }\textbf {\bibinfo {volume}
  {2}},\ \bibinfo {pages} {71} (\bibinfo {year} {2018})}\BibitemShut {NoStop}%
\bibitem [{\citenamefont {Bravyi}\ and\ \citenamefont
  {Haah}(2012)}]{Bravyi2012}%
  \BibitemOpen
  \bibfield  {author} {\bibinfo {author} {\bibfnamefont {S.}~\bibnamefont
  {Bravyi}}\ and\ \bibinfo {author} {\bibfnamefont {J.}~\bibnamefont {Haah}},\
  }\bibfield  {title} {\emph {\bibinfo {title} {Magic-state distillation with
  low overhead},\ }}\href {\doibasemod 10.1103/PhysRevA.86.052329} {\bibfield
  {journal} {\bibinfo  {journal} {Phys. Rev. A}\ }\textbf {\bibinfo {volume}
  {86}},\ \bibinfo {pages} {052329} (\bibinfo {year} {2012})}\BibitemShut
  {NoStop}%
\bibitem [{\citenamefont {Jones}(2013{\natexlab{a}})}]{Jones2013a}%
  \BibitemOpen
  \bibfield  {author} {\bibinfo {author} {\bibfnamefont {C.}~\bibnamefont
  {Jones}},\ }\bibfield  {title} {\emph {\bibinfo {title} {Multilevel
  distillation of magic states for quantum computing},\ }}\href {\doibasemod
  10.1103/PhysRevA.87.042305} {\bibfield  {journal} {\bibinfo  {journal} {Phys.
  Rev. A}\ }\textbf {\bibinfo {volume} {87}},\ \bibinfo {pages} {042305}
  (\bibinfo {year} {2013}{\natexlab{a}})}\BibitemShut {NoStop}%
\bibitem [{\citenamefont {Fowler}\ \emph {et~al.}(2013)\citenamefont {Fowler},
  \citenamefont {Devitt},\ and\ \citenamefont {Jones}}]{Fowler2013}%
  \BibitemOpen
  \bibfield  {author} {\bibinfo {author} {\bibfnamefont {A.~G.}\ \bibnamefont
  {Fowler}}, \bibinfo {author} {\bibfnamefont {S.~J.}\ \bibnamefont {Devitt}},
  \ and\ \bibinfo {author} {\bibfnamefont {C.}~\bibnamefont {Jones}},\
  }\bibfield  {title} {\emph {\bibinfo {title} {Surface code implementation of
  block code state distillation},\ }}\href {\doibasemod 10.1038/srep01939}
  {\bibfield  {journal} {\bibinfo  {journal} {Scientific Rep.}\ }\textbf
  {\bibinfo {volume} {3}},\ \bibinfo {pages} {1939} (\bibinfo {year}
  {2013})}\BibitemShut {NoStop}%
\bibitem [{\citenamefont {Fowler}(2012)}]{Fowler2012a}%
  \BibitemOpen
  \bibfield  {author} {\bibinfo {author} {\bibfnamefont {A.~G.}\ \bibnamefont
  {Fowler}},\ }\bibfield  {title} {\emph {\bibinfo {title} {Time-optimal
  quantum computation},\ }}\href {https://arxiv.org/abs/1210.4626} {\bibfield
  {journal} {\bibinfo  {journal} {arXiv:1210.4626}\ } (\bibinfo {year}
  {2012})}\BibitemShut {NoStop}%
\bibitem [{\citenamefont {Gottesman}(1999)}]{Gottesman1999}%
  \BibitemOpen
  \bibfield  {author} {\bibinfo {author} {\bibfnamefont {D.}~\bibnamefont
  {Gottesman}},\ }\bibfield  {title} {\emph {\bibinfo {title} {{The Heisenberg
  representation of quantum computers}},\ }}\href
  {http://arxiv.org/abs/quant-ph/9807006} {\bibfield  {journal} {\bibinfo
  {journal} {Proc. XXII Int. Coll. Group. Th. Meth. Phys.}\ }\textbf {\bibinfo
  {volume} {1}},\ \bibinfo {pages} {32} (\bibinfo {year} {1999})}\BibitemShut
  {NoStop}%
\bibitem [{\citenamefont {Kliuchnikov}\ \emph
  {et~al.}(2013{\natexlab{a}})\citenamefont {Kliuchnikov}, \citenamefont
  {Maslov},\ and\ \citenamefont {Mosca}}]{Kliuchnikov2012}%
  \BibitemOpen
  \bibfield  {author} {\bibinfo {author} {\bibfnamefont {V.}~\bibnamefont
  {Kliuchnikov}}, \bibinfo {author} {\bibfnamefont {D.}~\bibnamefont {Maslov}},
  \ and\ \bibinfo {author} {\bibfnamefont {M.}~\bibnamefont {Mosca}},\
  }\bibfield  {title} {\emph {\bibinfo {title} {Fast and efficient exact
  synthesis of single-qubit unitaries generated by {C}lifford and {$T$}
  gates},\ }}\href {http://dl.acm.org/citation.cfm?id=2535649.2535653}
  {\bibfield  {journal} {\bibinfo  {journal} {Quantum Info. Comput.}\ }\textbf
  {\bibinfo {volume} {13}},\ \bibinfo {pages} {607} (\bibinfo {year}
  {2013}{\natexlab{a}})}\BibitemShut {NoStop}%
\bibitem [{\citenamefont {Kliuchnikov}\ \emph
  {et~al.}(2013{\natexlab{b}})\citenamefont {Kliuchnikov}, \citenamefont
  {Maslov},\ and\ \citenamefont {Mosca}}]{Kliuchnikov2013}%
  \BibitemOpen
  \bibfield  {author} {\bibinfo {author} {\bibfnamefont {V.}~\bibnamefont
  {Kliuchnikov}}, \bibinfo {author} {\bibfnamefont {D.}~\bibnamefont {Maslov}},
  \ and\ \bibinfo {author} {\bibfnamefont {M.}~\bibnamefont {Mosca}},\
  }\bibfield  {title} {\emph {\bibinfo {title} {Asymptotically optimal
  approximation of single qubit unitaries by {C}lifford and {$T$} circuits
  using a constant number of ancillary qubits},\ }}\href {\doibasemod
  10.1103/PhysRevLett.110.190502} {\bibfield  {journal} {\bibinfo  {journal}
  {Phys. Rev. Lett.}\ }\textbf {\bibinfo {volume} {110}},\ \bibinfo {pages}
  {190502} (\bibinfo {year} {2013}{\natexlab{b}})}\BibitemShut {NoStop}%
\bibitem [{\citenamefont {Gosset}\ \emph {et~al.}(2013)\citenamefont {Gosset},
  \citenamefont {Kliuchnikov}, \citenamefont {Mosca},\ and\ \citenamefont
  {Russo}}]{Gosset2013}%
  \BibitemOpen
  \bibfield  {author} {\bibinfo {author} {\bibfnamefont {D.}~\bibnamefont
  {Gosset}}, \bibinfo {author} {\bibfnamefont {V.}~\bibnamefont {Kliuchnikov}},
  \bibinfo {author} {\bibfnamefont {M.}~\bibnamefont {Mosca}}, \ and\ \bibinfo
  {author} {\bibfnamefont {V.}~\bibnamefont {Russo}},\ }\bibfield  {title}
  {\emph {\bibinfo {title} {An algorithm for the {$T$}-count},\ }}\href
  {https://arxiv.org/abs/1308.4134} {\bibfield  {journal} {\bibinfo  {journal}
  {arXiv:1308.4134}\ } (\bibinfo {year} {2013})}\BibitemShut {NoStop}%
\bibitem [{\citenamefont {Heyfron}\ and\ \citenamefont
  {Campbell}(2018)}]{Heyfron2017}%
  \BibitemOpen
  \bibfield  {author} {\bibinfo {author} {\bibfnamefont {L.~E.}\ \bibnamefont
  {Heyfron}}\ and\ \bibinfo {author} {\bibfnamefont {E.~T.}\ \bibnamefont
  {Campbell}},\ }\bibfield  {title} {\emph {\bibinfo {title} {An efficient
  quantum compiler that reduces {$T$} count},\ }}\href {\doibasemod
  10.1088/2058-9565/aad604} {\bibfield  {journal} {\bibinfo  {journal} {Quantum
  Sci. Technol.}\ }\textbf {\bibinfo {volume} {4}},\ \bibinfo {pages} {015004}
  (\bibinfo {year} {2018})}\BibitemShut {NoStop}%
\bibitem [{\citenamefont {Amy}\ \emph {et~al.}(2013)\citenamefont {Amy},
  \citenamefont {Maslov}, \citenamefont {Mosca},\ and\ \citenamefont
  {Roetteler}}]{Amy2013}%
  \BibitemOpen
  \bibfield  {author} {\bibinfo {author} {\bibfnamefont {M.}~\bibnamefont
  {Amy}}, \bibinfo {author} {\bibfnamefont {D.}~\bibnamefont {Maslov}},
  \bibinfo {author} {\bibfnamefont {M.}~\bibnamefont {Mosca}}, \ and\ \bibinfo
  {author} {\bibfnamefont {M.}~\bibnamefont {Roetteler}},\ }\bibfield  {title}
  {\emph {\bibinfo {title} {A meet-in-the-middle algorithm for fast synthesis
  of depth-optimal quantum circuits},\ }}\href {\doibasemod
  10.1109/TCAD.2013.2244643} {\bibfield  {journal} {\bibinfo  {journal} {IEEE
  Transactions on Computer-Aided Design of Integrated Circuits and Systems}\
  }\textbf {\bibinfo {volume} {32}},\ \bibinfo {pages} {818} (\bibinfo {year}
  {2013})}\BibitemShut {NoStop}%
\bibitem [{\citenamefont {Selinger}(2013)}]{Selinger2013}%
  \BibitemOpen
  \bibfield  {author} {\bibinfo {author} {\bibfnamefont {P.}~\bibnamefont
  {Selinger}},\ }\bibfield  {title} {\emph {\bibinfo {title} {Quantum circuits
  of {$T$}-depth one},\ }}\href {\doibasemod 10.1103/PhysRevA.87.042302}
  {\bibfield  {journal} {\bibinfo  {journal} {Phys. Rev. A}\ }\textbf {\bibinfo
  {volume} {87}},\ \bibinfo {pages} {042302} (\bibinfo {year}
  {2013})}\BibitemShut {NoStop}%
\bibitem [{\citenamefont {Amy}\ \emph {et~al.}(2014)\citenamefont {Amy},
  \citenamefont {Maslov},\ and\ \citenamefont {Mosca}}]{Amy2014}%
  \BibitemOpen
  \bibfield  {author} {\bibinfo {author} {\bibfnamefont {M.}~\bibnamefont
  {Amy}}, \bibinfo {author} {\bibfnamefont {D.}~\bibnamefont {Maslov}}, \ and\
  \bibinfo {author} {\bibfnamefont {M.}~\bibnamefont {Mosca}},\ }\bibfield
  {title} {\emph {\bibinfo {title} {Polynomial-time {$T$}-depth optimization of
  {C}lifford+{$T$} circuits via matroid partitioning},\ }}\href {\doibasemod
  10.1109/TCAD.2014.2341953} {\bibfield  {journal} {\bibinfo  {journal} {IEEE
  Transactions on Computer-Aided Design of Integrated Circuits and Systems}\
  }\textbf {\bibinfo {volume} {33}},\ \bibinfo {pages} {1476} (\bibinfo {year}
  {2014})}\BibitemShut {NoStop}%
\bibitem [{\citenamefont {Litinski}\ and\ \citenamefont {von
  Oppen}(2018)}]{Litinski2018}%
  \BibitemOpen
  \bibfield  {author} {\bibinfo {author} {\bibfnamefont {D.}~\bibnamefont
  {Litinski}}\ and\ \bibinfo {author} {\bibfnamefont {F.}~\bibnamefont {von
  Oppen}},\ }\bibfield  {title} {\emph {\bibinfo {title} {Quantum computing
  with {M}ajorana fermion codes},\ }}\href {\doibasemod
  10.1103/PhysRevB.97.205404} {\bibfield  {journal} {\bibinfo  {journal} {Phys.
  Rev. B}\ }\textbf {\bibinfo {volume} {97}},\ \bibinfo {pages} {205404}
  (\bibinfo {year} {2018})}\BibitemShut {NoStop}%
\bibitem [{\citenamefont {Lavasani}\ and\ \citenamefont
  {Barkeshli}(2018)}]{Lavasani2018}%
  \BibitemOpen
  \bibfield  {author} {\bibinfo {author} {\bibfnamefont {A.}~\bibnamefont
  {Lavasani}}\ and\ \bibinfo {author} {\bibfnamefont {M.}~\bibnamefont
  {Barkeshli}},\ }\bibfield  {title} {\emph {\bibinfo {title} {Low overhead
  {C}lifford gates from joint measurements in surface, color, and hyperbolic
  codes},\ }}\href {\doibasemod 10.1103/PhysRevA.98.052319} {\bibfield
  {journal} {\bibinfo  {journal} {Phys. Rev. A}\ }\textbf {\bibinfo {volume}
  {98}},\ \bibinfo {pages} {052319} (\bibinfo {year} {2018})}\BibitemShut
  {NoStop}%
\bibitem [{\citenamefont {Hall}()}]{ModifyingCodes}%
  \BibitemOpen
  \bibfield  {author} {\bibinfo {author} {\bibfnamefont {J.~I.}\ \bibnamefont
  {Hall}},\ }\href@noop {} {\bibinfo {title} {{Notes on Coding Theory Chapter
  6: Modifying Codes}},\ }\bibinfo {howpublished}
  {\href{https://users.math.msu.edu/users/jhall/classes/codenotes/Mod.pdf}{https://users.math.msu.edu/users/jhall/classes/
  codenotes/Mod.pdf}},\ \bibinfo {note} {accessed: 2019-01-30}\BibitemShut
  {NoStop}%
\bibitem [{\citenamefont {Campbell}\ and\ \citenamefont
  {Howard}(2018)}]{Campbell2018}%
  \BibitemOpen
  \bibfield  {author} {\bibinfo {author} {\bibfnamefont {E.~T.}\ \bibnamefont
  {Campbell}}\ and\ \bibinfo {author} {\bibfnamefont {M.}~\bibnamefont
  {Howard}},\ }\bibfield  {title} {\emph {\bibinfo {title} {Magic state
  parity-checker with pre-distilled components},\ }}\href {\doibasemod
  10.22331/q-2018-03-14-56} {\bibfield  {journal} {\bibinfo  {journal}
  {{Quantum}}\ }\textbf {\bibinfo {volume} {2}},\ \bibinfo {pages} {56}
  (\bibinfo {year} {2018})}\BibitemShut {NoStop}%
\bibitem [{\citenamefont {Meier}\ \emph {et~al.}(2013)\citenamefont {Meier},
  \citenamefont {Eastin},\ and\ \citenamefont {Knill}}]{Meier2013}%
  \BibitemOpen
  \bibfield  {author} {\bibinfo {author} {\bibfnamefont {A.~M.}\ \bibnamefont
  {Meier}}, \bibinfo {author} {\bibfnamefont {B.}~\bibnamefont {Eastin}}, \
  and\ \bibinfo {author} {\bibfnamefont {E.}~\bibnamefont {Knill}},\ }\bibfield
   {title} {\emph {\bibinfo {title} {{Magic-state distillation with the
  four-qubit code}},\ }}\href {http://arxiv.org/abs/1204.4221} {\bibfield
  {journal} {\bibinfo  {journal} {Quant. Inf. Comp.}\ }\textbf {\bibinfo
  {volume} {13}},\ \bibinfo {pages} {195} (\bibinfo {year} {2013})}\BibitemShut
  {NoStop}%
\bibitem [{\citenamefont {Campbell}\ and\ \citenamefont
  {O'Gorman}(2016)}]{Campbell2016a}%
  \BibitemOpen
  \bibfield  {author} {\bibinfo {author} {\bibfnamefont {E.~T.}\ \bibnamefont
  {Campbell}}\ and\ \bibinfo {author} {\bibfnamefont {J.}~\bibnamefont
  {O'Gorman}},\ }\bibfield  {title} {\emph {\bibinfo {title} {An efficient
  magic state approach to small angle rotations},\ }}\href {\doibasemod
  10.1088/2058-9565/1/1/015007} {\bibfield  {journal} {\bibinfo  {journal}
  {Quantum Sci. Technol.}\ }\textbf {\bibinfo {volume} {1}},\ \bibinfo {pages}
  {015007} (\bibinfo {year} {2016})}\BibitemShut {NoStop}%
\bibitem [{\citenamefont {Herr}\ \emph
  {et~al.}(2017{\natexlab{b}})\citenamefont {Herr}, \citenamefont {Nori},\ and\
  \citenamefont {Devitt}}]{Herr2017}%
  \BibitemOpen
  \bibfield  {author} {\bibinfo {author} {\bibfnamefont {D.}~\bibnamefont
  {Herr}}, \bibinfo {author} {\bibfnamefont {F.}~\bibnamefont {Nori}}, \ and\
  \bibinfo {author} {\bibfnamefont {S.~J.}\ \bibnamefont {Devitt}},\ }\bibfield
   {title} {\emph {\bibinfo {title} {Lattice surgery translation for quantum
  computation},\ }}\href {\doibasemod 10.1088/1367-2630/aa5709} {\bibfield
  {journal} {\bibinfo  {journal} {New J. Phys.}\ }\textbf {\bibinfo {volume}
  {19}},\ \bibinfo {pages} {013034} (\bibinfo {year}
  {2017}{\natexlab{b}})}\BibitemShut {NoStop}%
\bibitem [{\citenamefont {Fowler}\ and\ \citenamefont
  {Devitt}(2012)}]{Fowler2012b}%
  \BibitemOpen
  \bibfield  {author} {\bibinfo {author} {\bibfnamefont {A.~G.}\ \bibnamefont
  {Fowler}}\ and\ \bibinfo {author} {\bibfnamefont {S.~J.}\ \bibnamefont
  {Devitt}},\ }\bibfield  {title} {\emph {\bibinfo {title} {A bridge to lower
  overhead quantum computation},\ }}\href {https://arxiv.org/abs/1209.0510}
  {\bibfield  {journal} {\bibinfo  {journal} {arXiv:1209.0510}\ } (\bibinfo
  {year} {2012})}\BibitemShut {NoStop}%
\bibitem [{\citenamefont {Gidney}\ and\ \citenamefont
  {Fowler}(2018)}]{Gidney2018a}%
  \BibitemOpen
  \bibfield  {author} {\bibinfo {author} {\bibfnamefont {C.}~\bibnamefont
  {Gidney}}\ and\ \bibinfo {author} {\bibfnamefont {A.~G.}\ \bibnamefont
  {Fowler}},\ }\bibfield  {title} {\emph {\bibinfo {title} {Efficient magic
  state factories with a catalyzed $\ket{CCZ}$ to $2\ket{T}$ transformation},\
  }}\href {https://arxiv.org/abs/1812.01238} {\bibfield  {journal} {\bibinfo
  {journal} {arXiv:1812.01238}\ } (\bibinfo {year} {2018})}\BibitemShut
  {NoStop}%
\bibitem [{\citenamefont {Bennett}\ \emph
  {et~al.}(1996{\natexlab{a}})\citenamefont {Bennett}, \citenamefont
  {Brassard}, \citenamefont {Popescu}, \citenamefont {Schumacher},
  \citenamefont {Smolin},\ and\ \citenamefont {Wootters}}]{Bennett1996}%
  \BibitemOpen
  \bibfield  {author} {\bibinfo {author} {\bibfnamefont {C.~H.}\ \bibnamefont
  {Bennett}}, \bibinfo {author} {\bibfnamefont {G.}~\bibnamefont {Brassard}},
  \bibinfo {author} {\bibfnamefont {S.}~\bibnamefont {Popescu}}, \bibinfo
  {author} {\bibfnamefont {B.}~\bibnamefont {Schumacher}}, \bibinfo {author}
  {\bibfnamefont {J.~A.}\ \bibnamefont {Smolin}}, \ and\ \bibinfo {author}
  {\bibfnamefont {W.~K.}\ \bibnamefont {Wootters}},\ }\bibfield  {title} {\emph
  {\bibinfo {title} {Purification of noisy entanglement and faithful
  teleportation via noisy channels},\ }}\href {\doibasemod
  10.1103/PhysRevLett.76.722} {\bibfield  {journal} {\bibinfo  {journal} {Phys.
  Rev. Lett.}\ }\textbf {\bibinfo {volume} {76}},\ \bibinfo {pages} {722}
  (\bibinfo {year} {1996}{\natexlab{a}})}\BibitemShut {NoStop}%
\bibitem [{\citenamefont {Bennett}\ \emph
  {et~al.}(1996{\natexlab{b}})\citenamefont {Bennett}, \citenamefont
  {Bernstein}, \citenamefont {Popescu},\ and\ \citenamefont
  {Schumacher}}]{Bennett1996a}%
  \BibitemOpen
  \bibfield  {author} {\bibinfo {author} {\bibfnamefont {C.~H.}\ \bibnamefont
  {Bennett}}, \bibinfo {author} {\bibfnamefont {H.~J.}\ \bibnamefont
  {Bernstein}}, \bibinfo {author} {\bibfnamefont {S.}~\bibnamefont {Popescu}},
  \ and\ \bibinfo {author} {\bibfnamefont {B.}~\bibnamefont {Schumacher}},\
  }\bibfield  {title} {\emph {\bibinfo {title} {Concentrating partial
  entanglement by local operations},\ }}\href {\doibasemod
  10.1103/PhysRevA.53.2046} {\bibfield  {journal} {\bibinfo  {journal} {Phys.
  Rev. A}\ }\textbf {\bibinfo {volume} {53}},\ \bibinfo {pages} {2046}
  (\bibinfo {year} {1996}{\natexlab{b}})}\BibitemShut {NoStop}%
\bibitem [{\citenamefont {Dickel}\ \emph {et~al.}(2018)\citenamefont {Dickel},
  \citenamefont {Wesdorp}, \citenamefont {Langford}, \citenamefont {Peiter},
  \citenamefont {Sagastizabal}, \citenamefont {Bruno}, \citenamefont {Criger},
  \citenamefont {Motzoi},\ and\ \citenamefont {DiCarlo}}]{Dickel2018}%
  \BibitemOpen
  \bibfield  {author} {\bibinfo {author} {\bibfnamefont {C.}~\bibnamefont
  {Dickel}}, \bibinfo {author} {\bibfnamefont {J.~J.}\ \bibnamefont {Wesdorp}},
  \bibinfo {author} {\bibfnamefont {N.~K.}\ \bibnamefont {Langford}}, \bibinfo
  {author} {\bibfnamefont {S.}~\bibnamefont {Peiter}}, \bibinfo {author}
  {\bibfnamefont {R.}~\bibnamefont {Sagastizabal}}, \bibinfo {author}
  {\bibfnamefont {A.}~\bibnamefont {Bruno}}, \bibinfo {author} {\bibfnamefont
  {B.}~\bibnamefont {Criger}}, \bibinfo {author} {\bibfnamefont
  {F.}~\bibnamefont {Motzoi}}, \ and\ \bibinfo {author} {\bibfnamefont
  {L.}~\bibnamefont {DiCarlo}},\ }\bibfield  {title} {\emph {\bibinfo {title}
  {Chip-to-chip entanglement of transmon qubits using engineered measurement
  fields},\ }}\href {\doibasemod 10.1103/PhysRevB.97.064508} {\bibfield
  {journal} {\bibinfo  {journal} {Phys. Rev. B}\ }\textbf {\bibinfo {volume}
  {97}},\ \bibinfo {pages} {064508} (\bibinfo {year} {2018})}\BibitemShut
  {NoStop}%
\bibitem [{\citenamefont {Campagne-Ibarcq}\ \emph {et~al.}(2018)\citenamefont
  {Campagne-Ibarcq}, \citenamefont {Zalys-Geller}, \citenamefont {Narla},
  \citenamefont {Shankar}, \citenamefont {Reinhold}, \citenamefont {Burkhart},
  \citenamefont {Axline}, \citenamefont {Pfaff}, \citenamefont {Frunzio},
  \citenamefont {Schoelkopf},\ and\ \citenamefont {Devoret}}]{Campagne2018}%
  \BibitemOpen
  \bibfield  {author} {\bibinfo {author} {\bibfnamefont {P.}~\bibnamefont
  {Campagne-Ibarcq}}, \bibinfo {author} {\bibfnamefont {E.}~\bibnamefont
  {Zalys-Geller}}, \bibinfo {author} {\bibfnamefont {A.}~\bibnamefont {Narla}},
  \bibinfo {author} {\bibfnamefont {S.}~\bibnamefont {Shankar}}, \bibinfo
  {author} {\bibfnamefont {P.}~\bibnamefont {Reinhold}}, \bibinfo {author}
  {\bibfnamefont {L.}~\bibnamefont {Burkhart}}, \bibinfo {author}
  {\bibfnamefont {C.}~\bibnamefont {Axline}}, \bibinfo {author} {\bibfnamefont
  {W.}~\bibnamefont {Pfaff}}, \bibinfo {author} {\bibfnamefont
  {L.}~\bibnamefont {Frunzio}}, \bibinfo {author} {\bibfnamefont {R.~J.}\
  \bibnamefont {Schoelkopf}}, \ and\ \bibinfo {author} {\bibfnamefont {M.~H.}\
  \bibnamefont {Devoret}},\ }\bibfield  {title} {\emph {\bibinfo {title}
  {Deterministic remote entanglement of superconducting circuits through
  microwave two-photon transitions},\ }}\href {\doibasemod
  10.1103/PhysRevLett.120.200501} {\bibfield  {journal} {\bibinfo  {journal}
  {Phys. Rev. Lett.}\ }\textbf {\bibinfo {volume} {120}},\ \bibinfo {pages}
  {200501} (\bibinfo {year} {2018})}\BibitemShut {NoStop}%
\bibitem [{\citenamefont {Axline}\ \emph {et~al.}(2018)\citenamefont {Axline},
  \citenamefont {Burkhart}, \citenamefont {Pfaff}, \citenamefont {Zhang},
  \citenamefont {Chou}, \citenamefont {Campagne-Ibarcq}, \citenamefont
  {Reinhold}, \citenamefont {Frunzio}, \citenamefont {Girvin}, \citenamefont
  {Jiang} \emph {et~al.}}]{Axline2018}%
  \BibitemOpen
  \bibfield  {author} {\bibinfo {author} {\bibfnamefont {C.~J.}\ \bibnamefont
  {Axline}}, \bibinfo {author} {\bibfnamefont {L.~D.}\ \bibnamefont
  {Burkhart}}, \bibinfo {author} {\bibfnamefont {W.}~\bibnamefont {Pfaff}},
  \bibinfo {author} {\bibfnamefont {M.}~\bibnamefont {Zhang}}, \bibinfo
  {author} {\bibfnamefont {K.}~\bibnamefont {Chou}}, \bibinfo {author}
  {\bibfnamefont {P.}~\bibnamefont {Campagne-Ibarcq}}, \bibinfo {author}
  {\bibfnamefont {P.}~\bibnamefont {Reinhold}}, \bibinfo {author}
  {\bibfnamefont {L.}~\bibnamefont {Frunzio}}, \bibinfo {author} {\bibfnamefont
  {S.}~\bibnamefont {Girvin}}, \bibinfo {author} {\bibfnamefont
  {L.}~\bibnamefont {Jiang}},  \emph {et~al.},\ }\bibfield  {title} {\emph
  {\bibinfo {title} {On-demand quantum state transfer and entanglement between
  remote microwave cavity memories},\ }}\href {\doibasemod
  10.1038/s41567-018-0115-y} {\bibfield  {journal} {\bibinfo  {journal} {Nat.
  Phys.}\ }\textbf {\bibinfo {volume} {14}},\ \bibinfo {pages} {705} (\bibinfo
  {year} {2018})}\BibitemShut {NoStop}%
\bibitem [{\citenamefont {Ross}\ and\ \citenamefont
  {Selinger}(2014)}]{Ross2014}%
  \BibitemOpen
  \bibfield  {author} {\bibinfo {author} {\bibfnamefont {N.~J.}\ \bibnamefont
  {Ross}}\ and\ \bibinfo {author} {\bibfnamefont {P.}~\bibnamefont
  {Selinger}},\ }\bibfield  {title} {\emph {\bibinfo {title} {Optimal
  ancilla-free {C}lifford+{T} approximation of z-rotations},\ }}\href
  {https://arxiv.org/abs/1403.2975} {\bibfield  {journal} {\bibinfo  {journal}
  {arXiv:1403.2975}\ } (\bibinfo {year} {2014})}\BibitemShut {NoStop}%
\bibitem [{\citenamefont {Duclos-Cianci}\ and\ \citenamefont
  {Poulin}(2015)}]{Duclos2015}%
  \BibitemOpen
  \bibfield  {author} {\bibinfo {author} {\bibfnamefont {G.}~\bibnamefont
  {Duclos-Cianci}}\ and\ \bibinfo {author} {\bibfnamefont {D.}~\bibnamefont
  {Poulin}},\ }\bibfield  {title} {\emph {\bibinfo {title} {Reducing the
  quantum-computing overhead with complex gate distillation},\ }}\href
  {\doibasemod 10.1103/PhysRevA.91.042315} {\bibfield  {journal} {\bibinfo
  {journal} {Phys. Rev. A}\ }\textbf {\bibinfo {volume} {91}},\ \bibinfo
  {pages} {042315} (\bibinfo {year} {2015})}\BibitemShut {NoStop}%
\bibitem [{\citenamefont {Harrow}\ \emph {et~al.}(2002)\citenamefont {Harrow},
  \citenamefont {Recht},\ and\ \citenamefont {Chuang}}]{Harrow2002}%
  \BibitemOpen
  \bibfield  {author} {\bibinfo {author} {\bibfnamefont {A.~W.}\ \bibnamefont
  {Harrow}}, \bibinfo {author} {\bibfnamefont {B.}~\bibnamefont {Recht}}, \
  and\ \bibinfo {author} {\bibfnamefont {I.~L.}\ \bibnamefont {Chuang}},\
  }\bibfield  {title} {\emph {\bibinfo {title} {Efficient discrete
  approximations of quantum gates},\ }}\href {\doibasemod 10.1063/1.1495899}
  {\bibfield  {journal} {\bibinfo  {journal} {Journal of Mathematical Physics}\
  }\textbf {\bibinfo {volume} {43}},\ \bibinfo {pages} {4445} (\bibinfo {year}
  {2002})}\BibitemShut {NoStop}%
\bibitem [{\citenamefont {Duclos-Cianci}\ and\ \citenamefont
  {Svore}(2013)}]{DuclosCianci2013}%
  \BibitemOpen
  \bibfield  {author} {\bibinfo {author} {\bibfnamefont {G.}~\bibnamefont
  {Duclos-Cianci}}\ and\ \bibinfo {author} {\bibfnamefont {K.~M.}\ \bibnamefont
  {Svore}},\ }\bibfield  {title} {\emph {\bibinfo {title} {Distillation of
  nonstabilizer states for universal quantum computation},\ }}\href
  {\doibasemod 10.1103/PhysRevA.88.042325} {\bibfield  {journal} {\bibinfo
  {journal} {Phys. Rev. A}\ }\textbf {\bibinfo {volume} {88}},\ \bibinfo
  {pages} {042325} (\bibinfo {year} {2013})}\BibitemShut {NoStop}%
\bibitem [{\citenamefont {Bocharov}\ \emph {et~al.}(2013)\citenamefont
  {Bocharov}, \citenamefont {Gurevich},\ and\ \citenamefont
  {Svore}}]{Bocharov2013}%
  \BibitemOpen
  \bibfield  {author} {\bibinfo {author} {\bibfnamefont {A.}~\bibnamefont
  {Bocharov}}, \bibinfo {author} {\bibfnamefont {Y.}~\bibnamefont {Gurevich}},
  \ and\ \bibinfo {author} {\bibfnamefont {K.~M.}\ \bibnamefont {Svore}},\
  }\bibfield  {title} {\emph {\bibinfo {title} {Efficient decomposition of
  single-qubit gates into $v$ basis circuits},\ }}\href {\doibasemod
  10.1103/PhysRevA.88.012313} {\bibfield  {journal} {\bibinfo  {journal} {Phys.
  Rev. A}\ }\textbf {\bibinfo {volume} {88}},\ \bibinfo {pages} {012313}
  (\bibinfo {year} {2013})}\BibitemShut {NoStop}%
\bibitem [{\citenamefont {Jones}\ \emph {et~al.}(2012)\citenamefont {Jones},
  \citenamefont {Whitfield}, \citenamefont {McMahon}, \citenamefont {Yung},
  \citenamefont {Meter}, \citenamefont {Aspuru-Guzik},\ and\ \citenamefont
  {Yamamoto}}]{Jones2012}%
  \BibitemOpen
  \bibfield  {author} {\bibinfo {author} {\bibfnamefont {N.~C.}\ \bibnamefont
  {Jones}}, \bibinfo {author} {\bibfnamefont {J.~D.}\ \bibnamefont
  {Whitfield}}, \bibinfo {author} {\bibfnamefont {P.~L.}\ \bibnamefont
  {McMahon}}, \bibinfo {author} {\bibfnamefont {M.-H.}\ \bibnamefont {Yung}},
  \bibinfo {author} {\bibfnamefont {R.~V.}\ \bibnamefont {Meter}}, \bibinfo
  {author} {\bibfnamefont {A.}~\bibnamefont {Aspuru-Guzik}}, \ and\ \bibinfo
  {author} {\bibfnamefont {Y.}~\bibnamefont {Yamamoto}},\ }\bibfield  {title}
  {\emph {\bibinfo {title} {Faster quantum chemistry simulation on
  fault-tolerant quantum computers},\ }}\href {\doibasemod
  10.1088/1367-2630/14/11/115023} {\bibfield  {journal} {\bibinfo  {journal}
  {New J. Phys.}\ }\textbf {\bibinfo {volume} {14}},\ \bibinfo {pages} {115023}
  (\bibinfo {year} {2012})}\BibitemShut {NoStop}%
\bibitem [{\citenamefont {Low}\ and\ \citenamefont {Chuang}(2016)}]{Low2016}%
  \BibitemOpen
  \bibfield  {author} {\bibinfo {author} {\bibfnamefont {G.~H.}\ \bibnamefont
  {Low}}\ and\ \bibinfo {author} {\bibfnamefont {I.~L.}\ \bibnamefont
  {Chuang}},\ }\bibfield  {title} {\emph {\bibinfo {title} {Hamiltonian
  simulation by qubitization},\ }}\href {https://arxiv.org/abs/1610.06546}
  {\bibfield  {journal} {\bibinfo  {journal} {arXiv:1610.06546}\ } (\bibinfo
  {year} {2016})}\BibitemShut {NoStop}%
\bibitem [{\citenamefont {Low}\ and\ \citenamefont {Chuang}(2017)}]{Low2017}%
  \BibitemOpen
  \bibfield  {author} {\bibinfo {author} {\bibfnamefont {G.~H.}\ \bibnamefont
  {Low}}\ and\ \bibinfo {author} {\bibfnamefont {I.~L.}\ \bibnamefont
  {Chuang}},\ }\bibfield  {title} {\emph {\bibinfo {title} {Optimal
  {H}amiltonian simulation by quantum signal processing},\ }}\href {\doibasemod
  10.1103/PhysRevLett.118.010501} {\bibfield  {journal} {\bibinfo  {journal}
  {Phys. Rev. Lett.}\ }\textbf {\bibinfo {volume} {118}},\ \bibinfo {pages}
  {010501} (\bibinfo {year} {2017})}\BibitemShut {NoStop}%
\bibitem [{\citenamefont {Babbush}\ \emph
  {et~al.}(2018{\natexlab{b}})\citenamefont {Babbush}, \citenamefont {Berry},
  \citenamefont {McClean},\ and\ \citenamefont {Neven}}]{Babbush2018a}%
  \BibitemOpen
  \bibfield  {author} {\bibinfo {author} {\bibfnamefont {R.}~\bibnamefont
  {Babbush}}, \bibinfo {author} {\bibfnamefont {D.~W.}\ \bibnamefont {Berry}},
  \bibinfo {author} {\bibfnamefont {J.~R.}\ \bibnamefont {McClean}}, \ and\
  \bibinfo {author} {\bibfnamefont {H.}~\bibnamefont {Neven}},\ }\bibfield
  {title} {\emph {\bibinfo {title} {Quantum simulation of chemistry with
  sublinear scaling to the continuum},\ }}\href
  {https://arxiv.org/abs/1807.09802} {\bibfield  {journal} {\bibinfo  {journal}
  {arXiv:1807.09802}\ } (\bibinfo {year} {2018}{\natexlab{b}})}\BibitemShut
  {NoStop}%
\bibitem [{\citenamefont {Jones}(2013{\natexlab{b}})}]{Jones2013}%
  \BibitemOpen
  \bibfield  {author} {\bibinfo {author} {\bibfnamefont {C.}~\bibnamefont
  {Jones}},\ }\bibfield  {title} {\emph {\bibinfo {title} {Low-overhead
  constructions for the fault-tolerant {T}offoli gate},\ }}\href {\doibasemod
  10.1103/PhysRevA.87.022328} {\bibfield  {journal} {\bibinfo  {journal} {Phys.
  Rev. A}\ }\textbf {\bibinfo {volume} {87}},\ \bibinfo {pages} {022328}
  (\bibinfo {year} {2013}{\natexlab{b}})}\BibitemShut {NoStop}%
\bibitem [{\citenamefont {Gidney}(2018)}]{Gidney2018}%
  \BibitemOpen
  \bibfield  {author} {\bibinfo {author} {\bibfnamefont {C.}~\bibnamefont
  {Gidney}},\ }\bibfield  {title} {\emph {\bibinfo {title} {Halving the cost of
  quantum addition},\ }}\href {\doibasemod 10.22331/q-2018-06-18-74} {\bibfield
   {journal} {\bibinfo  {journal} {{Quantum}}\ }\textbf {\bibinfo {volume}
  {2}},\ \bibinfo {pages} {74} (\bibinfo {year} {2018})}\BibitemShut {NoStop}%
\bibitem [{\citenamefont {Campbell}\ and\ \citenamefont
  {Howard}(2017)}]{Campbell2017}%
  \BibitemOpen
  \bibfield  {author} {\bibinfo {author} {\bibfnamefont {E.~T.}\ \bibnamefont
  {Campbell}}\ and\ \bibinfo {author} {\bibfnamefont {M.}~\bibnamefont
  {Howard}},\ }\bibfield  {title} {\emph {\bibinfo {title} {Unified framework
  for magic state distillation and multiqubit gate synthesis with reduced
  resource cost},\ }}\href {\doibasemod 10.1103/PhysRevA.95.022316} {\bibfield
  {journal} {\bibinfo  {journal} {Phys. Rev. A}\ }\textbf {\bibinfo {volume}
  {95}},\ \bibinfo {pages} {022316} (\bibinfo {year} {2017})}\BibitemShut
  {NoStop}%
\bibitem [{\citenamefont {O'Gorman}\ and\ \citenamefont
  {Campbell}(2017)}]{Ogorman2017}%
  \BibitemOpen
  \bibfield  {author} {\bibinfo {author} {\bibfnamefont {J.}~\bibnamefont
  {O'Gorman}}\ and\ \bibinfo {author} {\bibfnamefont {E.~T.}\ \bibnamefont
  {Campbell}},\ }\bibfield  {title} {\emph {\bibinfo {title} {Quantum
  computation with realistic magic-state factories},\ }}\href {\doibasemod
  10.1103/PhysRevA.95.032338} {\bibfield  {journal} {\bibinfo  {journal} {Phys.
  Rev. A}\ }\textbf {\bibinfo {volume} {95}},\ \bibinfo {pages} {032338}
  (\bibinfo {year} {2017})}\BibitemShut {NoStop}%
\bibitem [{\citenamefont {Likharev}\ and\ \citenamefont
  {Semenov}(1991)}]{Likharev1991}%
  \BibitemOpen
  \bibfield  {author} {\bibinfo {author} {\bibfnamefont {K.~K.}\ \bibnamefont
  {Likharev}}\ and\ \bibinfo {author} {\bibfnamefont {V.~K.}\ \bibnamefont
  {Semenov}},\ }\bibfield  {title} {\emph {\bibinfo {title} {{RSFQ}
  logic/memory family: A new {J}osephson-junction technology for
  sub-terahertz-clock-frequency digital systems},\ }}\href {\doibasemod
  10.1109/77.80745} {\bibfield  {journal} {\bibinfo  {journal} {IEEE
  Transactions on Applied Superconductivity}\ }\textbf {\bibinfo {volume}
  {1}},\ \bibinfo {pages} {3} (\bibinfo {year} {1991})}\BibitemShut {NoStop}%
\bibitem [{\citenamefont {Fowler}\ \emph {et~al.}(2017)\citenamefont {Fowler},
  \citenamefont {Devitt},\ and\ \citenamefont {Jones}}]{Paler2016b}%
  \BibitemOpen
  \bibfield  {author} {\bibinfo {author} {\bibfnamefont {A.~G.}\ \bibnamefont
  {Fowler}}, \bibinfo {author} {\bibfnamefont {S.~J.}\ \bibnamefont {Devitt}},
  \ and\ \bibinfo {author} {\bibfnamefont {C.}~\bibnamefont {Jones}},\
  }\bibfield  {title} {\emph {\bibinfo {title} {Synthesis of arbitrary quantum
  circuits to topological assembly: Systematic, online and compact},\ }}\href
  {\doibasemod 10.1038/s41598-017-10657-8} {\bibfield  {journal} {\bibinfo
  {journal} {Scientific Rep.}\ }\textbf {\bibinfo {volume} {7}},\ \bibinfo
  {pages} {10414} (\bibinfo {year} {2017})}\BibitemShut {NoStop}%
\bibitem [{\citenamefont {Paler}\ \emph {et~al.}(2017)\citenamefont {Paler},
  \citenamefont {Polian}, \citenamefont {Nemoto},\ and\ \citenamefont
  {Devitt}}]{Paler2017}%
  \BibitemOpen
  \bibfield  {author} {\bibinfo {author} {\bibfnamefont {A.}~\bibnamefont
  {Paler}}, \bibinfo {author} {\bibfnamefont {I.}~\bibnamefont {Polian}},
  \bibinfo {author} {\bibfnamefont {K.}~\bibnamefont {Nemoto}}, \ and\ \bibinfo
  {author} {\bibfnamefont {S.~J.}\ \bibnamefont {Devitt}},\ }\bibfield  {title}
  {\emph {\bibinfo {title} {Fault-tolerant, high-level quantum circuits: form,
  compilation and description},\ }}\href {\doibasemod 10.1088/2058-9565/aa66eb}
  {\bibfield  {journal} {\bibinfo  {journal} {Quantum Sci. Technol.}\ }\textbf
  {\bibinfo {volume} {2}},\ \bibinfo {pages} {025003} (\bibinfo {year}
  {2017})}\BibitemShut {NoStop}%
\bibitem [{\citenamefont {Lao}\ \emph {et~al.}(2018)\citenamefont {Lao},
  \citenamefont {van Wee}, \citenamefont {Ashraf}, \citenamefont {van Someren},
  \citenamefont {Khammassi}, \citenamefont {Bertels},\ and\ \citenamefont
  {Almudever}}]{Lao2018}%
  \BibitemOpen
  \bibfield  {author} {\bibinfo {author} {\bibfnamefont {L.}~\bibnamefont
  {Lao}}, \bibinfo {author} {\bibfnamefont {B.}~\bibnamefont {van Wee}},
  \bibinfo {author} {\bibfnamefont {I.}~\bibnamefont {Ashraf}}, \bibinfo
  {author} {\bibfnamefont {J.}~\bibnamefont {van Someren}}, \bibinfo {author}
  {\bibfnamefont {N.}~\bibnamefont {Khammassi}}, \bibinfo {author}
  {\bibfnamefont {K.}~\bibnamefont {Bertels}}, \ and\ \bibinfo {author}
  {\bibfnamefont {C.~G.}\ \bibnamefont {Almudever}},\ }\bibfield  {title}
  {\emph {\bibinfo {title} {Mapping of lattice surgery-based quantum circuits
  on surface code architectures},\ }}\href {\doibasemod
  10.1088/2058-9565/aadd1a} {\bibfield  {journal} {\bibinfo  {journal} {Quantum
  Sci. Technol.}\ }\textbf {\bibinfo {volume} {4}},\ \bibinfo {pages} {015005}
  (\bibinfo {year} {2018})}\BibitemShut {NoStop}%
\bibitem [{\citenamefont {Bombin}\ and\ \citenamefont
  {Martin-Delgado}(2006)}]{Bombin2006}%
  \BibitemOpen
  \bibfield  {author} {\bibinfo {author} {\bibfnamefont {H.}~\bibnamefont
  {Bombin}}\ and\ \bibinfo {author} {\bibfnamefont {M.~A.}\ \bibnamefont
  {Martin-Delgado}},\ }\bibfield  {title} {\emph {\bibinfo {title} {Topological
  quantum distillation},\ }}\href {\doibasemod 10.1103/PhysRevLett.97.180501}
  {\bibfield  {journal} {\bibinfo  {journal} {Phys. Rev. Lett.}\ }\textbf
  {\bibinfo {volume} {97}},\ \bibinfo {pages} {180501} (\bibinfo {year}
  {2006})}\BibitemShut {NoStop}%
\bibitem [{\citenamefont {Kesselring}\ \emph {et~al.}(2018)\citenamefont
  {Kesselring}, \citenamefont {Pastawski}, \citenamefont {Eisert},\ and\
  \citenamefont {Brown}}]{Kesselring2018}%
  \BibitemOpen
  \bibfield  {author} {\bibinfo {author} {\bibfnamefont {M.~S.}\ \bibnamefont
  {Kesselring}}, \bibinfo {author} {\bibfnamefont {F.}~\bibnamefont
  {Pastawski}}, \bibinfo {author} {\bibfnamefont {J.}~\bibnamefont {Eisert}}, \
  and\ \bibinfo {author} {\bibfnamefont {B.~J.}\ \bibnamefont {Brown}},\
  }\bibfield  {title} {\emph {\bibinfo {title} {The boundaries and twist
  defects of the color code and their applications to topological quantum
  computation},\ }}\href {\doibasemod 10.22331/q-2018-10-19-101} {\bibfield
  {journal} {\bibinfo  {journal} {{Quantum}}\ }\textbf {\bibinfo {volume}
  {2}},\ \bibinfo {pages} {101} (\bibinfo {year} {2018})}\BibitemShut {NoStop}%
\bibitem [{\citenamefont {Nautrup}\ \emph {et~al.}(2017)\citenamefont
  {Nautrup}, \citenamefont {Friis},\ and\ \citenamefont
  {Briegel}}]{Nautrup2016}%
  \BibitemOpen
  \bibfield  {author} {\bibinfo {author} {\bibfnamefont {H.~P.}\ \bibnamefont
  {Nautrup}}, \bibinfo {author} {\bibfnamefont {N.}~\bibnamefont {Friis}}, \
  and\ \bibinfo {author} {\bibfnamefont {H.~J.}\ \bibnamefont {Briegel}},\
  }\bibfield  {title} {\emph {\bibinfo {title} {Fault-tolerant interface
  between quantum memories and quantum processors},\ }}\href {\doibasemod
  10.1038/s41467-017-01418-2} {\bibfield  {journal} {\bibinfo  {journal} {Nat.
  Commun.}\ }\textbf {\bibinfo {volume} {8}},\ \bibinfo {pages} {1321}
  (\bibinfo {year} {2017})}\BibitemShut {NoStop}%
\bibitem [{\citenamefont {Litinski}\ and\ \citenamefont {von
  Oppen}(2017)}]{Litinski2017a}%
  \BibitemOpen
  \bibfield  {author} {\bibinfo {author} {\bibfnamefont {D.}~\bibnamefont
  {Litinski}}\ and\ \bibinfo {author} {\bibfnamefont {F.}~\bibnamefont {von
  Oppen}},\ }\bibfield  {title} {\emph {\bibinfo {title} {Braiding by
  {M}ajorana tracking and long-range {CNOT} gates with color codes},\ }}\href
  {\doibasemod 10.1103/PhysRevB.96.205413} {\bibfield  {journal} {\bibinfo
  {journal} {Phys. Rev. B}\ }\textbf {\bibinfo {volume} {96}},\ \bibinfo
  {pages} {205413} (\bibinfo {year} {2017})}\BibitemShut {NoStop}%
\bibitem [{Qub()}]{QubitDouble}%
  \BibitemOpen
  \href@noop {} {\bibinfo {title} {{IBM} doubling qubits every 8 months},\
  }\bibinfo {howpublished}
  {\href{https://www.nextbigfuture.com/2018/02/ibm-doubling-qubits-every-8-months-and-ecommerce-cryptography-at-risk-in-7-15-years.html}{https://www.nextbigfuture.com/2018/02/ibm-doubling-qubits-every-8-months-and-ecommerce-cryptography-at-risk-in-7-15-years.html}},\
  \bibinfo {note} {accessed: 2018-08-01}\BibitemShut {NoStop}%
\end{thebibliography}%

\begin{figure}[!b]
\centering
\def\svgwidth{\linewidth}
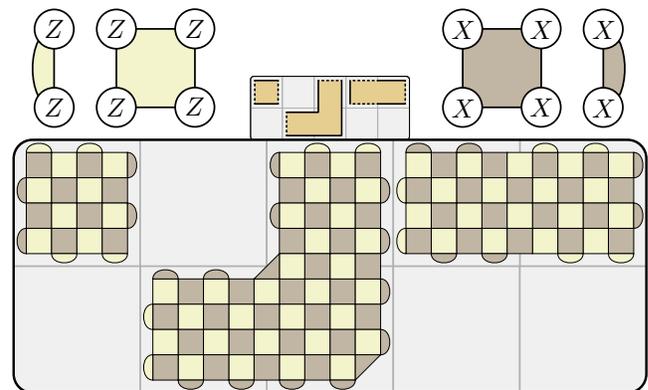
\caption{Surface-code implementation of the patches shown in Fig.~\ref{fig:patches}. Physical qubits are placed on vertices. Bright faces correspond to $Z$ stabilizers and dark faces to $X$ stabilizers.}
\label{fig:scpatches}
\end{figure}

\begin{figure}[!t]
\centering
\def\svgwidth{\linewidth}
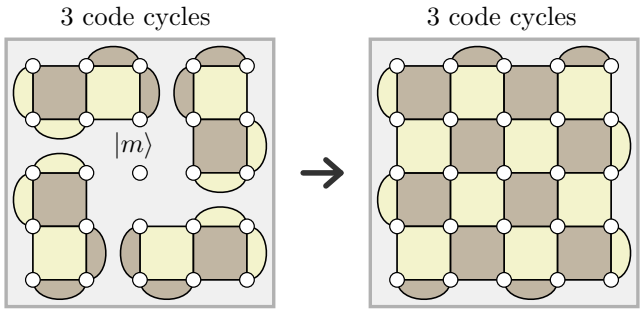
\caption{State-injection protocol of Ref.~\cite{Landahl2014}.}
\label{fig:stateinjection}
\end{figure}

\appendix

\section{Surface-code qubits and lattice-surgery operations}
\label{app:surfacecode}

To illustrate the translation of protocols in our framework into surface-code patches, we show how the patches of Fig.~\ref{fig:patches} and the rules of the game  and protocols of Fig.~\ref{fig:operations} are implemented with surface codes. 

\textbf{Surface-code patches.} Each patch corresponds to a surface-code patch with code distance $d$. Therefore, each tile corresponds to $d^2$ physical data qubits, as shown in Fig.~\ref{fig:scpatches} for $d=5$. In our surface-code patches, physical qubits are placed on the vertices, bright faces correspond to $Z$ stabilizers and dark faces to $X$ stabilizers. Solid and dashed boundaries correspond to $X$ and $Z$ boundaries (also called rough and smooth boundaries). For one-qubit patches, the product of all $d$ physical $X$ ($Z$) operators along any of the $X$ ($Z$) boundaries is the logical $X$ ($Z$) operator of the encoded qubit. For two-qubit patches with six boundaries, the string operators  located at the boundaries correspond to the logical operators shown in Fig.~\ref{fig:patches}, i.e., going clockwise, $X_1$, $Z_1$, $X_1 \cdot X_2$, $Z_2$, $X_2$, and $Z_1 \cdot Z_2$. Note that, in principle, the width of two-tile patches can be $2d-1$ instead of $2d$, potentially reducing the space cost~\cite{Litinski2017b}. Furthermore, the correspondence between solid and dashed, and $X$ and $Z$ boundaries is interchangeable.

\textbf{State initialization.} We now show how the operations and protocols of Fig.~\ref{fig:operations} are implemented with surface codes for $d=5$, and motivate their time cost in the framework, where the reasoning is that 1\clock \ is associated with operations whose time cost scales with $d$. Surface-code patches can be initialized in the logical $\ket{0}$ or $\ket{+}$ state by initializing all physical qubits of the patch in $\ket{0}$ or $\ket{+}$, and then measuring all stabilizers.

Naively, one would expect that there should be a time cost associated with this operation, since the stabilizers need to be measured for $d$ code cycles to account for measurement errors. However, this can be done simultaneously with the subsequent lattice-surgery operation, as will become apparent in the example of the Bell state preparation. For arbitrary states, the logical states are prepared via state injection. This is a non-fault-tolerant procedure with a constant time cost that does not scale with $d$, which is why we do not associate a time step with it. One such state-injection protocol is described in Ref.~\cite{Landahl2014} and is shown in Fig.~\ref{fig:stateinjection} for the preparation of a logical magic state $\ket{m}$. In the left panel, a physical magic state is prepared, along with a stabilizer state by measuring the shown stabilizers for three code cycles. Note that any single-qubit error during these three code cycles will corrupt the logical information. Next, the stabilizer configuration is switched to the ordinary surface code in the right panel. Here, the stabilizers are, again, only measured for three code cycles, independently of $d$, since the state-injection protocol is, in any case, non-fault-tolerant, i.e., produces logical states with an error rate proportional to the physical error rate $p$.

\begin{figure}[!t]
\centering
\def\svgwidth{\linewidth}
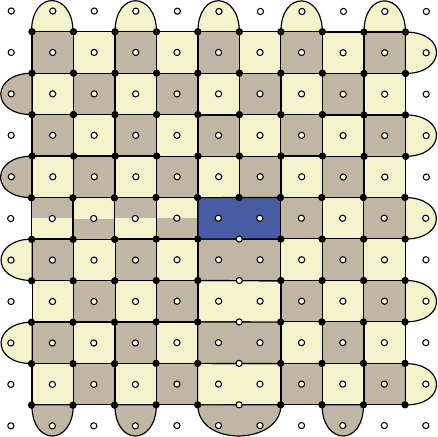
\caption{Twist-based lattice surgery in a square lattice of qubits with nearest-neighbor couplings. The black dots are physical data qubits and the white dots are physical measurement qubits.}
\label{fig:twist}
\end{figure}

\begin{figure*}[!t]
\centering
\def\svgwidth{\linewidth}
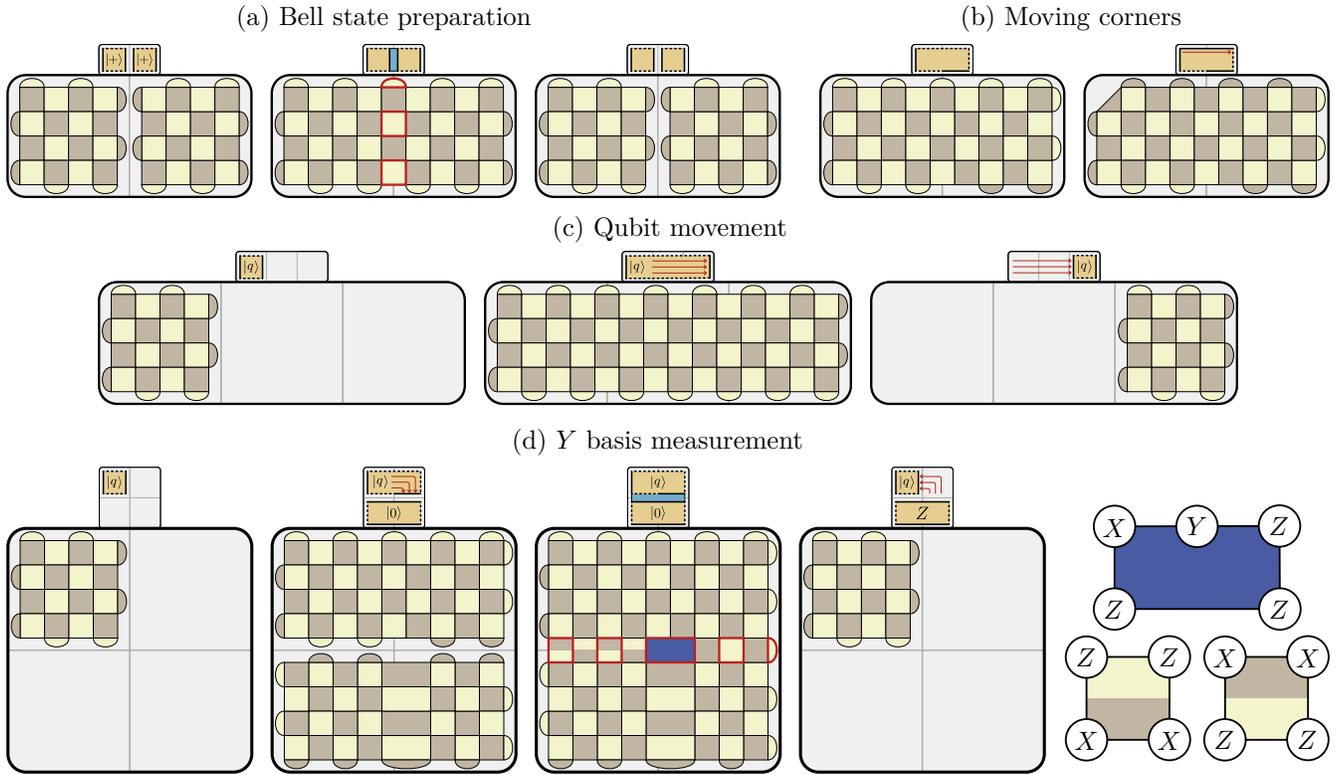
\caption{Surface-code implementation of the protocols in Fig.~\ref{fig:operations}a-d.}
\label{fig:scop1}
\end{figure*}

\textbf{Patch measurement and Bell state preparation.} Surface-code patches are measured in the $X$ or $Z$ basis by measuring all physical qubits in the corresponding basis and performing some classical error correction, where the time cost does not scale with $d$. Two-patch measurements correspond to lattice surgery and can be demonstrated via the preparation of a Bell state, as shown in Fig.~\ref{fig:scop1}a. Two surface-code patches are initialized in the logical $\ket{+}$ state by initializing all physical qubits in $\ket{+}$ and measuring the stabilizers.
Simultaneously, lattice surgery between the two patches is performed, measuring the logical $Z \otimes Z$ operator. The measurement outcome is the product of the newly introduced $Z$ stabilizers highlighted in red, as the product of these stabilizers corresponds to the product of the logical $Z$ operators encoded in the two surface-code $Z$ boundaries. To account for measurement errors, this measurement is repeated for $d$ code cycles. Finally, the patch is split into two patches again, leaving the two logical surface-code qubits in an entangled Bell state.

\textbf{$Y$ measurements.} Two-patch measurements can be used to measure products of two Pauli operators other than $Z\otimes Z$, e.g., operators involving the $Y$ operator, as shown in Fig.~\ref{fig:scop1}d. First, a patch is deformed to a wider patch by initializing physical qubits in the $X$ basis and measuring the new stabilizers, which takes $d$ code cycles. Below the wide patch, a rectangular ancilla patch is initialized in the $\ket{0}$ state. A column of physical qubits in the center is missing, so that, in the next step, the ancilla can be used for twist-based lattice surgery~\cite{Litinski2017b}, measuring the $Y$ operator. The product of the operators highlighted in red in the third step corresponds to the logical $Y \otimes Z$ operator between the two logical qubits. The lattice surgery in the third step involves dislocation operators and a five-qubit twist defect. Even though these stabilizers are irregular, they can still be measured in a square lattice of physical qubits with nearest-neighbor couplings, as we show in Fig.~\ref{fig:twist}. For the measurement of twist operators and wide $X$ and $Z$ stabilizers, up to three measurement ancillas can be used.

\begin{figure*}[!t]
\centering
\def\svgwidth{\linewidth}
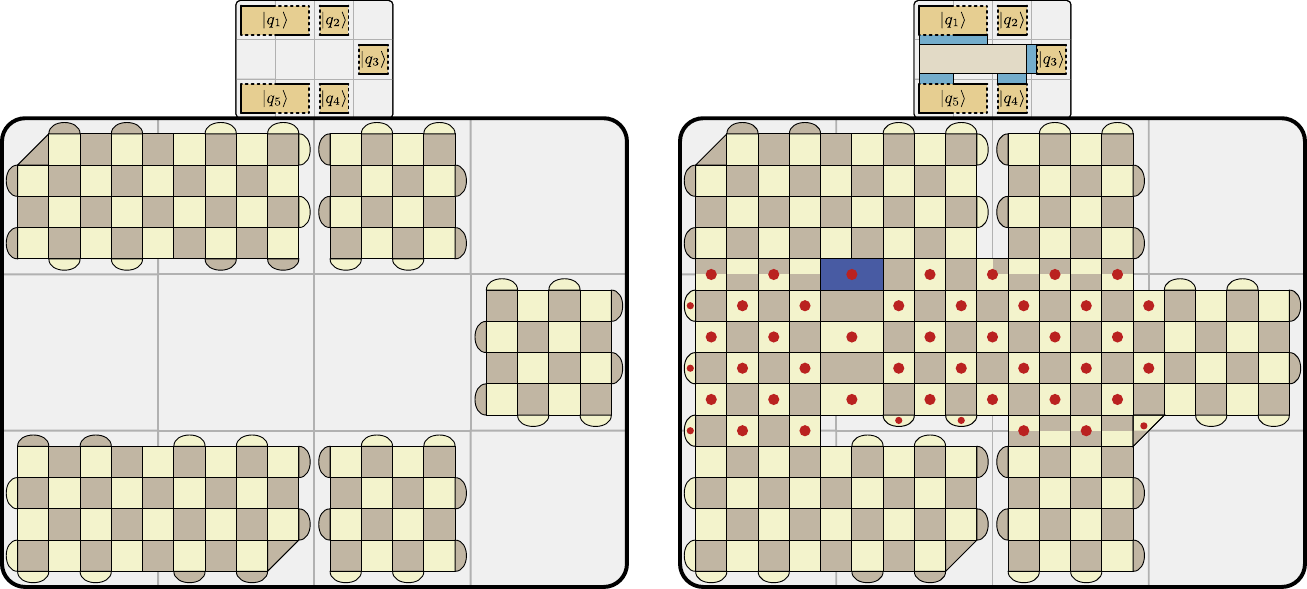
\caption{Surface-code implementation of the multi-patch measurement in Fig.~\ref{fig:operations}e. The measurement outcome is the product of all check operators with a red dot.}
\label{fig:scop2}
\end{figure*}

\textbf{Multi-patch measurements.} For a multi-patch measurement in Fig.~\ref{fig:scop2}, all physical qubits located in the region of the ancilla patch are initialized in the $\ket{+}$ state. Next, new check operators are introduced. The newly introduced $X$-type stabilizers all yield trivial outcomes, since they are products of physical qubits initialized in an $X$ eigenstate and previously measured check operators. The nontrivial operators are highlighted by a red dot in Fig.~\ref{fig:scop2}. Their product is equivalent to the desired operator, i.e., $Y_{\ket{q_1}}\otimes X_{\ket{q_3}} \otimes Z_{\ket{q_4}} \otimes X_{\ket{q_5}}$. The new check operators are measured for $d$ code cycles to account for measurement errors. This procedure corresponds to the multi-body lattice surgery protocol introduced in Ref.~\cite{Fowler2018}. It can be used to measure any product of surface-code-boundary Pauli operators by initializing physical qubits in the $\ket{+}$ state in an ancilla region of width $d$, and then measuring new check operators, where the product of the nontrivial operators yields the outcome of the desired multi-patch measurement. The ancilla region of width $d$ is required to ensure that the code distance of the stabilizer configuration during the multi-body lattice surgery remains $d$.

\textbf{Moving boundaries.} The protocol to move patches is similar to lattice surgery. It is shown in Fig.~\ref{fig:scop1}c. Extending the patch via its $Z$ boundary in the second step is the same operation as a $Z \otimes Z$ lattice surgery between the patch and a rectangular $\ket{+}$ ancilla qubit to the right. This needs to be done for $d$ code cycles to account for measurement errors. Finally, the patch is shortened again by measuring the left two thirds of physical qubits in the $X$ basis.

\textbf{Moving corners.} The movement of corners of a surface-code patch is shown in Fig.~\ref{fig:scop1}b. It corresponds to a change of boundary stabilizers. In order to account for measurement errors of the newly measured stabilizers, this requires $d$ code cycles. The top left physical qubit in the second step of Fig.~\ref{fig:scop1}b is removed from the patch via an $X$ measurement.

\begin{figure}[b!]
\centering
\def\svgwidth{\linewidth}
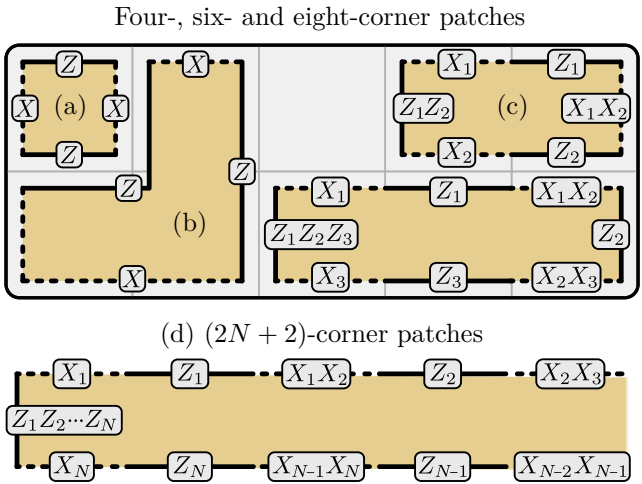
\caption{Patches with $2N+2$ corners represent $N$ qubits. Their $2N+2$ edges represent the shown Pauli operators.}
\label{fig:oldpatches}
\end{figure}

\section{Extended ruleset}
\label{app:extendedrules}

Some surface-code operations are not covered by the rules discussed in the introduction. In particular, we only consider patches with 4 or 6 corners, where we refer to the points where two edges meet as corners. In general, one could also consider patches with a higher number of corners.
A patch with $2N+2$ corners represents $N$ qubits, as shown in Fig.~\ref{fig:oldpatches}. 
The simplest case is a four-corner patch (a/b) representing a single qubit. Six-corner patches (c) are two-qubit patches. The general rule that assigns the operators of $N$ qubits to the edges of a $(2N+2)$-corner patch is given in Fig.~\ref{fig:oldpatches}d. Going clockwise, the dashed boundaries correspond to $X_1, X_1X_2, X_2X_3, \dots, X_{N-1}X_N$ and $X_N$. Starting to the right of $X_1$, the solid edges correspond to $Z_1, Z_2, \dots, Z_N$ and the product $Z_1Z_2\cdots Z_N$.

\begin{figure}[!t]
\centering
\def\svgwidth{\linewidth}
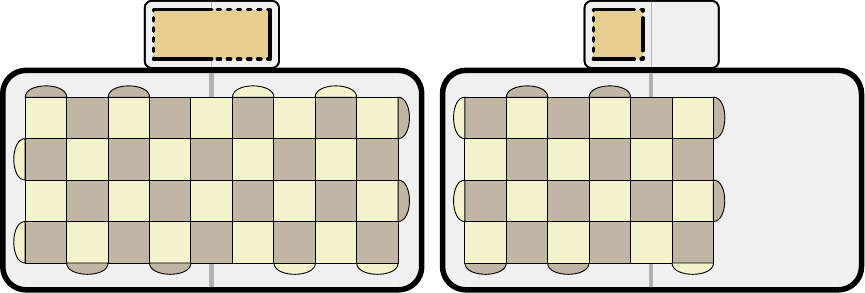
\caption{Surface-code implementation of a six-corner patch with shortened boundaries}
\label{fig:scop5}
\end{figure}

One can also consider patches with shortened edges, such that they occupy fewer tiles. The drawback of this is that in every time step, an error corresponding to the Pauli operator represented by the shortened edge will occur with a certain probability $p_{\rm err}$. An example of a six-corner patch with two shortened $X$ edges is shown in Fig.~\ref{fig:scop5}, meaning that this six-corner patch is susceptible to $X$ errors. In the surface-code implementation, this corresponds to a patch with boundaries that are shorter than $d$ physical data qubits, effectively reducing the code distance of the logical operators encoded by the shortened edges. Note that patches with shortened edges may occupy more than $d^2$ physical data qubits per tile.

With $(2N+2)$-corner patches, the set of operations needs to be modified. The initialization rule for such patches is:
\begin{itemize}
 	\item[--] Qubits can be initialized in the $X$ and $Z$ eigenstates $\ket{+}$ and $\ket{0}$. All qubits that are part of one patch must be initialized in the same state. (Cost: 0\clock)
\end{itemize}
Similarly, the single-patch measurement rule is modified to
\begin{itemize}
\item[--] Qubits can be measured in the $X$ or $Z$ basis. All qubits that are part of the same patch are measured simultaneously and in the same basis. This measurement removes the patch from the board. (Cost: 0\clock)
\end{itemize}

\begin{figure}[!t]
\centering
\def\svgwidth{\linewidth}
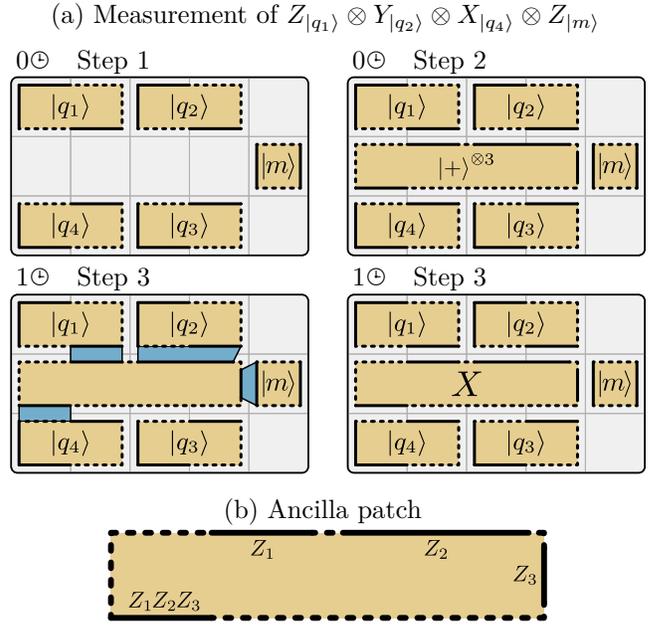
\caption{Pauli product measurement protocol. (a) Example of a measurement of the operator $Z\otimes Y \otimes \mathbbm{1} \otimes X \otimes Z$ of the qubits $\ket{q_1}$, $\ket{q_2}$, $\ket{q_3}$, $\ket{q_4}$ and $\ket{m}$. (b) Ancilla patch used during the measurement.}
\label{fig:pauliprodmeasold}
\end{figure}

\textbf{Pauli product measurements.} Using multi-corner patches with shortened boundaries, the multi-patch measurement rule is, in principle, redundant. For instance, the Pauli product measurement of Fig.~\ref{fig:pauliprodmeas} can be equivalently performed in 1\clock \ via the protocol shown in Fig.~\ref{fig:pauliprodmeasold}. An 8-corner ancilla patch is initialized in the $\ket{+}^{\otimes 3}$ state. The shape of this patch is chosen, such that each of the four $Z$ edges is adjacent to one of the four operators that are part of the measurement. Note that this means that some of the $X$ edges are shortened, such that the qubits are susceptible to $X$ errors. In this case, this is not a problem, since the qubits are initialized in $X$ eigenstates and random $X$ errors will cause no change to the states. Next, in step 3, we measure the four Pauli products $Z_{\ket{q_1}} \otimes Z_1$, $Y_{\ket{q_2}} \otimes Z_2$, $Z_{\ket{m}} \otimes Z_3$ and $X_{\ket{q_4}} \otimes (Z_1 \cdot Z_2 \cdot Z_3)$. Because the ancilla is initialized in an $X$ eigenstate, the operators $Z_1$, $Z_2$ and $Z_3$ are unknown, and the outcome of each of the four aforementioned measurements is entirely random. However, multiplying the four measurement outcomes yields $Z_{\ket{q_1}}\otimes Y_{\ket{q_2}} \otimes X_{\ket{q_4}} \otimes Z_{\ket{m}} \otimes (Z_1 \cdot Z_2 \cdot Z_3 \cdot Z_1 \cdot Z_2 \cdot Z_3)$, which is precisely the operator $Z_{\ket{q_1}}\otimes Y_{\ket{q_2}} \otimes X_{\ket{q_4}} \otimes Z_{\ket{m}}$ that we wanted to measure. Finally, to discard the ancilla patch we measure its three qubits in the $X$ basis. Again, $X$ errors will have no effect, as they commute with the measurement basis. Measurement outcomes of $X_i=-1$ prompt a Pauli correction. If in the previous step, the $Z_i$ edge was measured together with a Pauli operator $P$, the correction is a $P_{\pi/2}$ gate. For instance, if in Fig.~\ref{fig:pauliprodmeas} the final measurements yield $X_2 = -1$ and $X_3 = -1$, the corrections are a $Y_{\pi/2}$ rotation on $\ket{q_2}$ and a $Z_{\pi/2}$ rotation on $\ket{m}$.

\begin{figure*}[!t]
\centering
\def\svgwidth{\linewidth}
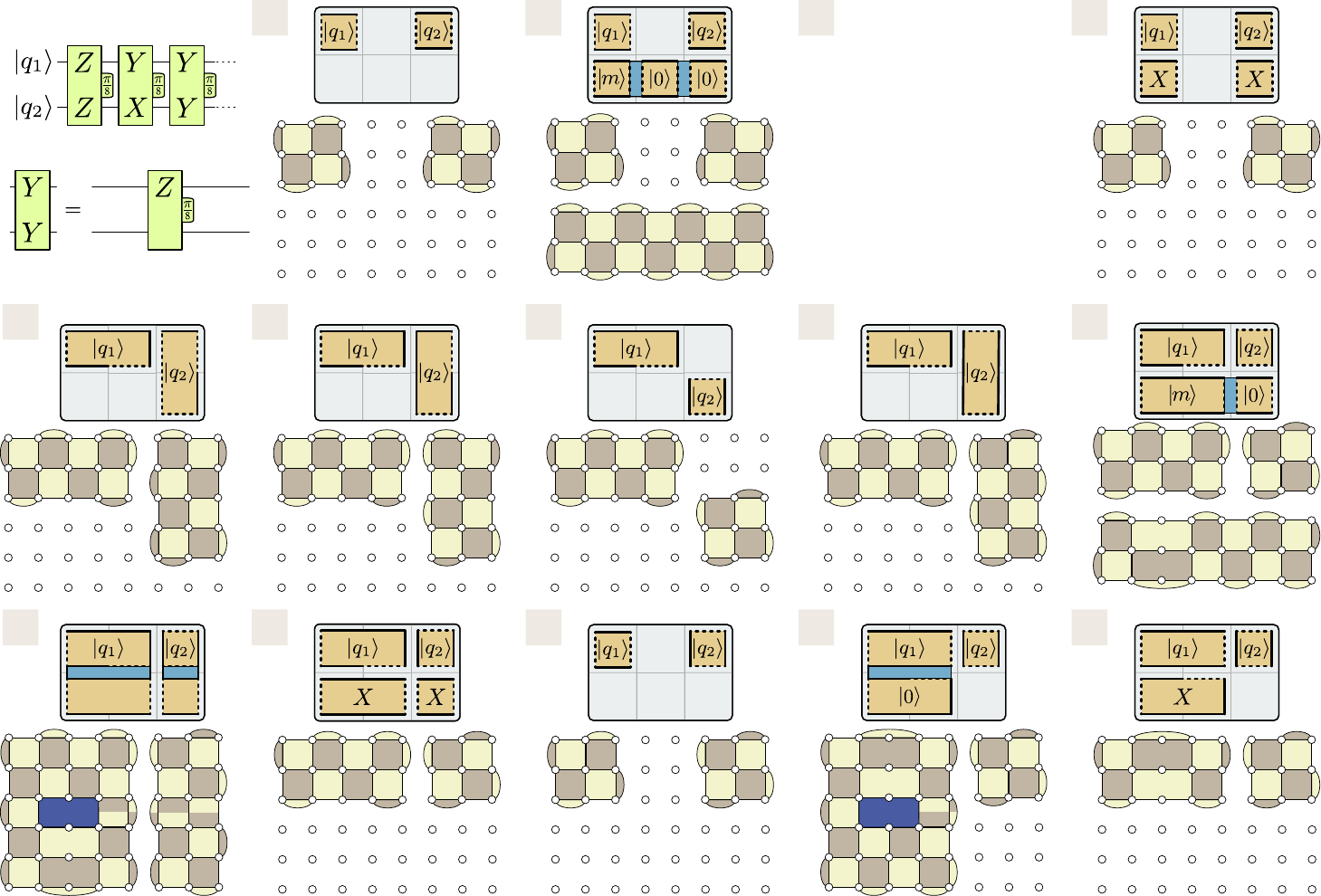
\caption{Proof-of-principle two-qubit device implemented with 48 physical data qubits.}
\label{fig:proofofprinciple}
\end{figure*}

This type of protocol can be used to measure any product of $n$ Pauli operators. An ancilla patch needs to be initialized in the $\ket{+}^{\otimes n}$ state with $Z$ edges adjacent to the $n$ operators part of the measurement. The surface-code implementation of this protocol is identical to the surface-code implementation of multi-patch measurements in Fig.~\ref{fig:scop2}.

While multi-corner patches and shortened edges increase the number of surface-code operations that are covered by the framework, there are still rules that can be added to the ruleset to account for more operations, such as, e.g., the movement of corners inside a patch~\cite{Brown2017}. Also, for the initialization of non-Pauli eigenstates, error models other than random Pauli errors can be considered.

\section{Proof-of-principle device}
\label{app:proofofprinciple}

Here, we discuss how $(3d-1)\cdot 2d$ physical data qubits can be used to build a proof-of-principle device that is a universal two-qubit error-corrected quantum computer that uses undistilled magic states and can demonstrate all the operations required for large-scale quantum computing. We go through the example of a computation that starts with three $\pi/8$ rotations around $Z\otimes Z$, $Y \otimes X$ and $Y \otimes Y$ in Fig.~\ref{fig:proofofprinciple}. For the first rotation, we need to measure $Z_1 \otimes Z_2 \otimes Z_{\ket{m}}$. A magic state is initialized in a long patch in step 2, which is equivalent to initializing a magic state and measuring $X \otimes X$ between the magic state and neighboring $\ket{0}$ ancillas. This effectively encodes the magic state in a three-qubit repetition code with a logical $Z$ operator $Z_L = Z \otimes Z \otimes Z$. To consume the magic state, $Z_1 \otimes Z_2 \otimes Z_L$ is measured in step 3. This consumes a magic state for the $Z\otimes Z$ rotation.

The next rotation is a $Y \otimes X$ rotation. Here, we first need to deform $\ket{q_1}$, such that both the $X$ and $Z$ boundaries of the qubit are accessible. Qubit $\ket{q_2}$ is rotated in steps 5-8 using the protocol in Fig.~\ref{fig:compactblock3}a. In step 9, again, a magic state is initialized in a two-qubit repetition code with $Z_L = Z_{a1} \otimes Z_{a2}$. In step 10, the magic state is consumed via a $Y_1 \otimes Z_{a_1}$ and a $X_1 \otimes Z_{a_2}$ measurement.

This kind of protocol consisting of patch deformations and patch rotations can be used to perform any $\pi/8$ rotation with the exception of $(Y\otimes Y)_{\pi/8}$, since there is not enough space to make both $Y$ operators accessible for lattice surgery. For this rotation, we first explicitly execute a Clifford gate to change $(Y\otimes Y)_{\pi/8}$ to any other rotation. Any Clifford gate that does not commute with $Y\otimes Y$ will suffice. In our example, we choose a $Z_{\pi/4}$ rotation. It is performed by initializing a $\ket{0}$ state in step 13, and measuring $Z_1 \otimes Y$ between $\ket{q_1}$ and the ancilla, following the protocol of Fig.~\ref{fig:compactblock3}b.

This demonstrates that a proof-of-principle experiment can be built with 48 physical data qubits. In general, this requires $6d^2-2d$ qubits, i.e., 48 for $d=3$, 140 for $d=5$ and 280 for $d=7$. If measurement qubits are required for syndrome readout, the number of physical qubits roughly doubles.

\begin{figure*}[!t]
\centering
\def\svgwidth{0.9\linewidth}
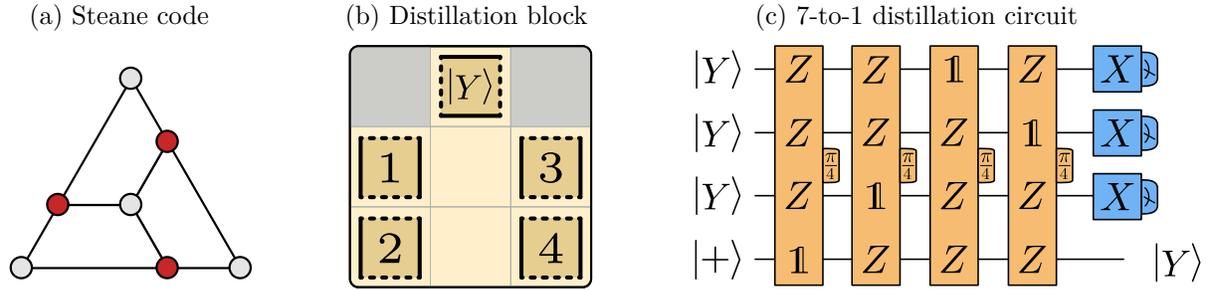
\caption{The Steane code (a) is the basis of 7-to-1 distillation (c). In our framework, the corresponding distillation block (b) uses 7 tiles for 4\clock.}
\label{fig:7to1}
\end{figure*}

\section{Implementation of the 7-to-1 protocol}
\label{app:7to1}

Even though the distillation of $\ket{Y} = \ket{0} + i \ket{1}$ states has no use in our framework, we show how to implement the 7-to-1 distillation protocol for benchmarking purposes in Fig.~\ref{fig:7to1}. The protocol is based on the 7-qubit Steane code. Its $X$ stabilizers are the faces shown in Fig.~\ref{fig:7to1}a, and its logical $X$ operator can be chosen as the $X\otimes X \otimes X$ operator with support on the three qubits drawn in red.

Following the procedure in Sec.~\ref{sec:distillation}, the distillation circuit is obtained by initializing $m_x+k=4$ qubits in the $\ket{+}$ state, where the first three qubits are associated with the three $X$ stabilizers, and the last qubit is associated with the logical $X$ operator. For each qubit of the Steane code, the circuit contains a $\pi/4$ rotation with $Z$'s on each stabilizer and logical operator that the qubit is part of. The three qubits in the corner of the triangle are only part of a single stabilizer and no logical operator, therefore they contribute with  single-qubit  $Z_{\pi/4}$ rotations, which can be absorbed into  \linebreak \newpage \noindent  the  initial state. The remaining four rotations are shown in  Fig.~\ref{fig:7to1}c.

A distillation block that can be used for this protocol  is shown in Fig.~\ref{fig:7to1}b. Since the consumption of $\ket{Y}$   resource states requires no Clifford correction, this block consists of only 7 tiles. With four rotations, the leading order of the space-time cost of this protocol is $7d^2\cdot 4 d = 28d^3$.

\end{document}